\documentclass[twocolumn]{aastex7}
\usepackage{amsmath}
\shorttitle{BayeSED-GALAXIES II.}
\shortauthors{Han et al. 2025}

\begin{document}

\title{BayeSED-GALAXIES II. Bayesian full spectrum analysis of galaxies and application in the CSST wide-field slitless spectroscopy survey}

\correspondingauthor{Yunkun Han}
\email{hanyk@ynao.ac.cn}

\newcommand{\AffilYNAO}{Yunnan Observatories, Chinese Academy of Sciences, 396 Yangfangwang, Guandu District, Kunming 650216, P. R. China}
\newcommand{\AffilCAMS}{Center for Astronomical Mega-Science, Chinese Academy of Sciences, 20A Datun Road, Chaoyang District, Beijing, 100012, P. R. China}
\newcommand{\AffilICESUN}{International Centre of Supernovae (ICESUN), Yunnan Key Laboratory of Supernova Research, Yunnan Observatories, Chinese Academy of Sciences (CAS), Kunming 650216, China}
\newcommand{\AffilPMO}{Purple Mountain Observatory, Chinese Academy of Sciences, 10 Yuanhua Road, Nanjing 210023, China}
\newcommand{\AffilUSTC}{School of Astronomy and Space Science, University of Science and Technology of China, Hefei 230026, China}
\newcommand{\AffilUSTCpeople}{School of Astronomy and Space Sciences, University of Science and Technology of China, Hefei, Anhui 230026, People's Republic of China} 
\newcommand{\AffilGZU}{College of Physics, Guizhou University, Guiyang 550025, China}
\newcommand{\AffilNAOCKeyLab}{Key Laboratory of Optical Astronomy, National Astronomical Observatories, Chinese Academy of Sciences, Beijing 100101, China}
\newcommand{\AffilUCAS}{School of Astronomy and Space Science, University of Chinese Academy of Sciences, Beijing 101408, China}
\newcommand{\AffilSJTUAstro}{Department of Astronomy, School of Physics and Astronomy, Shanghai Jiao Tong University, Shanghai 200240, China}
\newcommand{\AffilSJTUTDLI}{Tsung-Dao Lee Institute and State Key Laboratory of Dark Matter Physics, Shanghai Jiao Tong University, Shanghai 201210, China}
\newcommand{\AffilSHAO}{Shanghai Astronomical Observatory, Chinese Academy of Sciences, 80 Nandan Road, Shanghai 200030, China}
\newcommand{\AffilTsinghua}{Department of Astronomy, Tsinghua University, Beijing 100084, China}

\newcommand{\AffilUCASnew}{University of Chinese Academy of Sciences, Beijing 100049, China}
\newcommand{\AffilRadioKeyLab}{State Key Laboratory of Radio Astronomy and Technology, National Astronomical Observatories, Chinese Academy of Sciences, Beijing 100101, China}

\newcommand{\AffilNAOCInfrared}{Infrared Astrophysics Research Group, National Astronomical Observatories, Chinese Academy of Sciences, 20A Datun Road, Chaoyang District, Beijing 100101, China}

\newcommand{\AffilNAOCGen}{National Astronomical Observatories, Chinese Academy of Sciences, 20A Datun Road, Chaoyang District, Beijing 100101, China}

\newcommand{\AffilXidianAero}{School of Aerospace Science and Technology, Xidian University, Xi'an 710126, China}
\newcommand{\AffilXidianShaanxiKeyLab}{Shaanxi Key Laboratory of Space Extreme Detection, Xidian University, Xi'an 710126, China}

\newcommand{\AffilNAOCPRC}{National Astronomical Observatories, Chinese Academy of Sciences, 20A Datun Road, Beijing 100101, People's Republic of China}
\newcommand{\AffilNAOCCSSTSC}{Science Center for China Space Station Telescope, National Astronomical Observatories, Chinese Academy of Sciences, 20A Datun Road, Beijing 100101, People's Republic of China}
\newcommand{\AffilUCASPRC}{University of Chinese Academy of Sciences, Beijing, 100049, People's Republic of China}

\author[0000-0002-2547-0434]{Yunkun Han}
\email{hanyk@ynao.ac.cn}
\affiliation{\AffilICESUN}
\affiliation{\AffilCAMS}

\author[0000-0003-3728-9912]{Xian Zhong Zheng}
\email{xzzheng@pmo.ac.cn}
\affiliation{\AffilSJTUTDLI}

\author{Xiaohu Yang}
\email{xyang@sjtu.edu.cn}
\affiliation{\AffilSJTUTDLI}
\affiliation{\AffilSJTUAstro}

\author[0000-0002-8705-6327]{Run Wen}
\email{wenrun1214@sjtu.edu.cn}
\affiliation{\AffilSJTUTDLI}
\affiliation{\AffilSJTUAstro}

\author{F. S. Liu}
\email{fsliu@nao.cas.cn}
\affiliation{\AffilNAOCGen}
\affiliation{\AffilNAOCInfrared}
\affiliation{\AffilNAOCKeyLab}

\author[0000-0002-6684-3997]{Hu Zou}
\email{zouhu@nao.cas.cn}
\affiliation{\AffilNAOCKeyLab}
\affiliation{\AffilUCAS}

\author{Jin-Ming Bai}
\email{baijinming@ynao.ac.cn}
\affiliation{\AffilICESUN}
\affiliation{\AffilCAMS}

\author[0000-0002-9128-818X]{Yinghe Zhao}
\email{zhaoyinghe@ynao.ac.cn}
\affiliation{\AffilYNAO}
\affiliation{\AffilRadioKeyLab}

\author[0000-0003-4200-4432]{Lulu Fan}
\email{llfan@ustc.edu.cn}
\affiliation{Department of Astronomy, University of Science and Technology of China, Hefei 230026, China}
\affiliation{\AffilUSTC}
\affiliation{Deep Space Exploration Laboratory, Hefei 230088, China}
\affiliation{\AffilGZU}
\author{Fenghui Zhang}
\email{zhangfh@ynao.ac.cn}
\affiliation{\AffilICESUN}
\affiliation{\AffilCAMS}

\author{Xiaoyu Kang}
\email{kxyysl@ynao.ac.cn}
\affiliation{\AffilICESUN}
\affiliation{\AffilCAMS}

\author{Xiejin Li}
\email{lixiejin@ynao.ac.cn}
\affiliation{\AffilYNAO}
\affiliation{\AffilUCASnew}

\author[0000-0003-4936-8247]{Hong Guo}
\email{guohong@shao.ac.cn}
\affiliation{\AffilSHAO}

\author{Pengjie Zhang}
\email{zhangpj@sjtu.edu.cn}
\affiliation{\AffilSJTUAstro}
\affiliation{\AffilSJTUTDLI}

\author{Hu Zhan}
\email{zhanhu@nao.cas.cn}
\affiliation{\AffilNAOCKeyLab}
\affiliation{\AffilUCAS}

\author[0000-0003-4726-6714]{Gongbo Zhao}
\email{gbzhao@nao.cas.cn}
\affiliation{\AffilNAOCKeyLab}
\affiliation{\AffilUCAS}

\author[0000-0002-8711-8970]{Cheng Li}
\email{cli2015@tsinghua.edu.cn}
\affiliation{\AffilTsinghua}

\author[0000-0003-0709-0101]{Yan Gong}
\email{gongyan@bao.ac.cn}
\affiliation{\AffilNAOCPRC} 
\affiliation{\AffilNAOCCSSTSC} 
\affiliation{\AffilUCASPRC} 

\author[0000-0003-3196-7938]{Yizhou Gu}
\email{guyizhou@sjtu.edu.cn}
\affiliation{\AffilSJTUTDLI}

\author[0000-0002-9968-2894]{Feng Shi}
\email{fshi@xidian.edu.cn}
\affiliation{\AffilXidianAero}
\affiliation{\AffilXidianShaanxiKeyLab}

\author{Xingchen Zhou}
\email{xczhou@nao.cas.cn}
\affiliation{\AffilNAOCPRC} 
\affiliation{\AffilNAOCCSSTSC} 

\author{Jipeng Sui}
\email{suijp@bao.ac.cn}
\affiliation{\AffilNAOCKeyLab}
\affiliation{\AffilUCAS}

\author{Yipeng Jing}
\email{ypjing@sjtu.edu.cn}
\affiliation{\AffilSJTUAstro}
\affiliation{\AffilSJTUTDLI}

\author[0000-0001-9204-7778]{Zhanwen Han}
\email{zhanwenhan@ynao.ac.cn}
\affiliation{\AffilICESUN}
\affiliation{\AffilCAMS}


\begin{abstract}
The China Space Station Telescope (CSST) will simultaneously conduct wide-field multiband photometric imaging and spectroscopic slitless surveys, poised to significantly advance cosmology and galaxy evolution research. Meeting CSST's cosmological goals requires precise redshifts ($\sigma_{\mathrm{NMAD}}\lesssim 0.002-0.005$) from its low-resolution ($R\sim200$), and potentially blended slitless spectra. We present BayeSED3, extended for Bayesian full spectrum analysis, incorporating detailed nebular emission modeling (via \textsc{Cloudy}) and a robust Bayesian treatment of model scaling factor, enhancing reliability over optimization techniques for low Signal-to-Noise Ratio (SNR) spectra. Validated on realistic mock data generated with the CESS emulator (median SNR of 1.65, instrumental and self-blending effects included), our method achieves excellent redshift precision with three-band (GU+GV+GI) spectroscopy, exceeding requirements: $\sigma_{\mathrm{NMAD}}=0.0008$ ($\sim$80\% success rate) for star-forming (SF) and $\sigma_{\mathrm{NMAD}}=0.0015$ ($\sim$50\% success rate) for quiescent galaxies. Stellar mass ($M_{*}$; $\sigma_{\mathrm{NMAD}}\approx 0.015$ dex for SF, $\approx 0.016$ dex for quiescent) and SFR ($\sigma_{\mathrm{NMAD}}\approx 0.05$ dex for SF galaxies, especially at $SNR>1$) can also be reliably recovered. While spectral self-blending increases scatter ($\sigma_{\mathrm{NMAD}}$) by $\gtrsim 30\%$, combining spectroscopy with CSST's seven-band photometry crucially improves accuracy, especially for quiescent galaxies and in data-limited scenarios. Combining photometry with single-band spectroscopy leads to reasonable redshift estimation. While GU+photometry shows limited performance, GI+photometry yields success rates $>60\%$ (SF) and $>40\%$ (quiescent) with $\sigma_{\mathrm{NMAD}}\lesssim 0.002$, and GV+photometry achieves success rates $>35\%$ (SF) and $\sim$40\% (quiescent) at similar precision. Our comprehensive Bayesian framework thus provides a robust tool for precise galaxy characterization, enhancing scientific returns from CSST's unique dataset despite inherent slitless spectroscopy challenges.
\end{abstract}

\section{Introduction} \label{sec:intro}

Large galaxy surveys have been pivotal in mapping the cosmos, with successive generations like the Two-Degree Field Galaxy Redshift Survey (2dFGRS; \citealt{CollessM2001a}), the Sloan Digital Sky Survey (SDSS; \citealt{YorkD2000a}), and the Dark Energy Spectroscopic Instrument (DESI; \citealt{DESI-Collaboration2016a,DESI-Collaboration2016b}) providing foundational datasets for understanding large-scale structure and galaxy evolution. The next frontier involves wide-field space-based missions employing slitless spectroscopy to efficiently survey vast cosmic volumes. Among these, the China Space Station Telescope \citep[CSST;][]{ZhanH2011i,ZhanH2021j,CaoY2018a,GongY2019a} is poised to conduct an unprecedented survey, combining deep imaging in seven photometric bands (NUV, $u, g, r, i, z$, and $y$) with low-resolution ($R = 241$, $263$, and $270$ for GU, GV, and GI bands respectively) slitless spectroscopy over a sky coverage of approximately 17,500 deg$^2$. This unique combination surpasses contemporary missions like Euclid (15,000 deg$^2$) \citep{Euclid-Collaboration2022a} and the upcoming Roman Space Telescope (2,000 deg$^2$) \citep{AkesonR2019a} in survey area and complements them in wavelength coverage, band numbers and depth, holding immense potential for transformative studies in cosmology and galaxy formation.
However, like other slitless spectroscopic surveys, CSST faces unique observational challenges, particularly from spectral contamination due to overlapping spectra along dispersed directions.

To address these challenges, significant effort has been invested in developing mock galaxy redshift surveys that can help evaluate selection effects and validate analysis techniques \citep{GuY2024a}.
These mock catalogs are essential for understanding how observational systematics might impact cosmological measurements.
Recent forecasts by \citet{MiaoH2023a, MiaoH2024a} demonstrate CSST's exceptional potential for cosmological studies using multiple probes (including cosmic shear, galaxy-galaxy lensing, photometric and spectroscopic galaxy clustering, and galaxy cluster counts), highlighting the telescope's capability to significantly advance cosmology by achieving sub-percent accuracy for key parameters like $\Omega_m$ and $\sigma_8$ through multi-probe analysis \citep{MiaoH2023a} and enabling high-precision Baryon Acoustic Oscillation measurements \citep{MiaoH2024a, ShiF2025a}. However, realizing this potential hinges critically on the ability to extract accurate information, particularly precise spectroscopic redshifts with an accuracy of approximately $\sigma_{\mathrm{NMAD}} \lesssim 0.001-0.005$ required for cosmological analyses \citep{MiaoH2023a, MiaoH2024a}, from the vast number of low-resolution, potentially blended slitless spectra that CSST will deliver. This necessitates robust analysis techniques capable of handling the unique challenges of CSST data.

The measurement of galaxy redshifts and inference of their physical properties through spectral energy distribution (SED) fitting are fundamental to our understanding of the formation and evolution of galaxies and cosmology.
Photometric SED fitting uses a small number of broadband flux measurements to constrain galaxy properties, offering computational efficiency but suffering from significant parameter degeneracies.
In contrast, full spectrum analysis utilizes hundreds or thousands of spectral elements, providing much tighter constraints on physical parameters through detailed spectral features.
However, full spectrum analysis faces its own challenges, including increased computational complexity, sensitivity to spectral calibration errors, and in the case of slitless spectroscopy, contamination from overlapping sources.

Modern spectrum fitting codes generally follow either minimum $\chi^2$ or Bayesian approaches, each with distinct advantages and limitations.
Minimum $\chi^2$ methods, exemplified by codes like STARLIGHT \citep{Cid-FernandesR2005a} and FIREFLY \citep{WilkinsonD2017y}, are computationally efficient and can quickly find best-fit solutions, though they may underestimate parameter uncertainties and can be trapped in local minima.
While both STARLIGHT and FIREFLY decomposes galaxy spectra using a basis of simple stellar populations (SSPs) to find the best-fit $\chi^2$ solution, FIREFLY additionally employs an iterative $\chi^2$ minimization approach and guided by the Bayesian Information Criterion.
Penalized Pixel-Fitting (pPXF; \citealt{CappellariM2004a,CappellariM2017a,CappellariM2023a}) takes a hybrid approach, exploiting the quadratic nature of the template weights fitting problem to optimize linear and nonlinear parameters simultaneously, with recent versions supporting full spectrum fitting with photometry, though its regularization approach may sometimes lead to over-smoothing of the star formation history.

In contrast, Bayesian approaches like BAGPIPES \citep{CarnallA2018a} and Prospector \citep{JohnsonB2021m} provide full posterior probability distributions, naturally account for parameter degeneracies, and can incorporate prior knowledge, though at the cost of increased computational complexity.
BAGPIPES combines highly customizable models with nested sampling through MultiNest (or nautilus), while Prospector offers a modular framework that can forward-model data calibration parameters.
Both codes provide robust parameter estimation but with computational demands that can be challenging for large survey applications.
Some codes like pyPipe3D \citep{SanchezS2016a} take a multi-step approach, first fitting basic parameters with simplified models before proceeding to more detailed analysis.
Recent deep learning approaches have shown promise in specific applications - for example, StarNet \citep{FabbroS2018a} achieves remarkable computational efficiency (processing a spectrum in $\sim$0.0004s compared to 11-99s for traditional methods).
A notable recent advancement comes from \citet{ZhouX2024a}, who developed a specialized deep learning framework for CSST slitless spectroscopy that combines a 1D Convolutional Neural Network (CNN) with a Bayesian Neural Network (BNN) approach.
They report an impressive accuracy for redshift estimation ($\sigma_{\rm NMAD} = 0.00063$, outlier percentage 0.92\%) and demonstrates particular robustness in handling low signal-to-noise ratio (SNR) data.
However, these methods require extensive training data, cannot easily incorporate new physical models without retraining, and may provide limited insight into parameter degeneracies.
Their performance also heavily depends on the quality and representativeness of the training data.
Other specialized techniques, such as emission-line fitting, have also demonstrated high precision ($\sigma_{\rm NMAD} \lesssim 0.001$) and purity ($>85\%$) for specific galaxy populations like ELGs \citep{SuiJ2025a}.
Beyond these examples, a rich ecosystem of SED fitting codes has been developed over the past two decades, each with distinct strengths and modeling choices (see comprehensive reviews by \citealt{PacificiC2023a} and \citealt{IyerK2025a} or  check the full list at \url{https://sites.google.com/view/sed-fitting-forum/sed-fitting-codes} and \url{http://www.sedfitting.org/Fitting.html}).

BayeSED has evolved through three formal versions since its prototype in 2012.
The prototype version introduced a novel approach combining artificial neural networks for model interpolation with principal component analysis (PCA) for library dimensionality reduction, establishing a core framework for Bayesian parameter estimation and model comparison for analyzing complex systems like dust-obscured starburst-AGN composite galaxies \citep{HanY2012a}.
The first formal version, BayeSED1 \citep{HanY2014a}, established a general approach to galaxy SED fitting by integrating the MultiNest algorithm for efficient parameter space exploration, implementing K-nearest neighbors algorithms as an alternative interpolation method, and adding MPI parallelization for simultaneous analysis of multiple galaxies.
Through systematic testing on mock galaxies and application to a large Ks-selected sample in the COSMOS/UltraVISTA field, it demonstrated the first Bayesian comparison of different stellar population synthesis models and showed advantages over traditional grid-based methods in parameter distribution estimation.
BayeSED2 \citep{HanY2019a} introduced a comprehensive Bayesian framework for simultaneously dealing with multiple uncertain components in galaxy SED modeling.
Using this framework, it enabled objective discrimination among 16 different SSP models, five forms of star formation history, and four types of dust attenuation laws using Bayesian evidence as a quantitative implementation of Occam's razor, validated through analysis of a $z<1$ sample of galaxies.
The latest version, BayeSED3 \citep{HanY2023a}, is optimized for large survey analysis with more flexible choices of star formation histories and dust attenuation laws.
The code includes improved MPI-based parallel processing with checkpointing capabilities and optimized data I/O for handling large surveys.
For typical SED model assumptions and CSST photometric data, it achieves computational efficiency of approximately 2 seconds per galaxy when using one processor thread on a 2.2GHz CPU.

In \cite{HanY2023a}, we conducted a comprehensive evaluation of BayeSED3's performance for simultaneous photometric redshift and stellar population parameter estimation of galaxies using CSST's seven-band photometric data.
We used two types of mock galaxy samples with different characteristics and findings.
For the empirical statistics-based mock sample with idealized SED modeling, using optimized MultiNest parameters, we achieved photometric redshift estimation with $\sigma_{\rm NMAD} = 0.056$, bias of $-0.0025$, and outlier fraction of $0.215$; stellar mass estimation with $\sigma_{\rm NMAD} = 0.113$, bias of $-0.025$, and outlier fraction of $0.285$; and star formation rate estimation with $\sigma_{\rm NMAD} = 0.08$, bias of $-0.01$, and outlier fraction of $0.255$.
With this sample, we found that random observational errors in photometry were more significant than parameter degeneracies, and simpler SED models often performed better.
For the hydrodynamical simulation-based mock sample without idealized SED modeling, we found that the assumptions about star formation history and dust attenuation law had severe impact on parameter estimation quality.
For this more realistic sample, model errors were more important than observational noise for stellar mass and SFR estimation, and more complex models could improve the results in some cases.
The quality of parameter estimation was found to depend heavily on the discriminative power of the photometric data.
Through extensive testing with both samples, we found that Bayesian model comparison could effectively identify the optimal model complexity for different survey configurations.
This work established the foundation for photometric analysis of CSST data.
However, it left open the question of how to optimally incorporate spectroscopic information and nebular emission modeling.

In this paper, we extend our methodology to full spectrum analysis of CSST slitless spectroscopic data.
Using standard assumptions about star formation history (SFH) and dust attenuation law (DAL), we focus on incorporating nebular emission modeling - a crucial component that was missing in the photometric analysis.
This extension is particularly important as spectroscopic data provides much richer information about galaxy properties through detailed spectral features, especially emission lines that cannot be resolved by broad-band photometry.
We develop a more sophisticated mock data simulation framework that includes detailed spectral features including nebular emission lines and continuum.
It also includes realistic observational effects and comprehensive modeling of spectral contamination from overlapping sources, challenges specific to slitless spectroscopy.
By incorporating nebular emission modeling, we aim to maximize the scientific return from CSST's joint photometric and spectroscopic observations.

This paper is organized as follows. Section~\ref{sec:csst_mock} details the generation of our realistic mock CSST dataset, incorporating nebular emission and simulated slitless spectroscopy effects via the CSST Emulator for Slitless Spectroscopy (CESS; \citealt{WenR2024a}). Section~\ref{sec:bayesed_dec} presents our extended Bayesian full spectrum fitting framework within BayeSED3, including our treatment of model scaling and performance evaluation metrics. Section~\ref{sec:results} establishes the baseline performance by evaluating the recovery of redshift, stellar mass, and star formation rate using the full three-band (GU, GV, GI) CSST spectroscopic data. Building on this, Section~\ref{sec:disc} investigates the impact of spectral self-blending and then explores the synergistic benefits of combining spectroscopy with photometry, particularly focusing on more realistic data-limited scenarios involving only two or one spectroscopic bands. Finally, Section~\ref{sec:summary} summarizes our key findings and conclusions.

\section{CSST Mock Galaxy Catalog and Data Generation} \label{sec:csst_mock}

In this work, we utilize the DESI Legacy Survey (LS) seed galaxies, referred to as the 'parent sample', that form the basis of the CSST mock galaxy redshift surveys developed by \citet{GuY2024a}. This choice allows us to directly leverage their carefully constructed mock catalogs, which are specifically designed to evaluate CSST slitless spectroscopic selection effects and observational systematics. By using the same underlying galaxy sample, our analysis of spectral fitting methods can be directly applicable to understanding the systematic effects in CSST observations and can be integrated into the broader effort to validate CSST analysis techniques.

\subsection{Sample selection} \label{ss:sample}
\begin{figure*}[ht]
    \centering
    \includegraphics[width=1.0\textwidth]{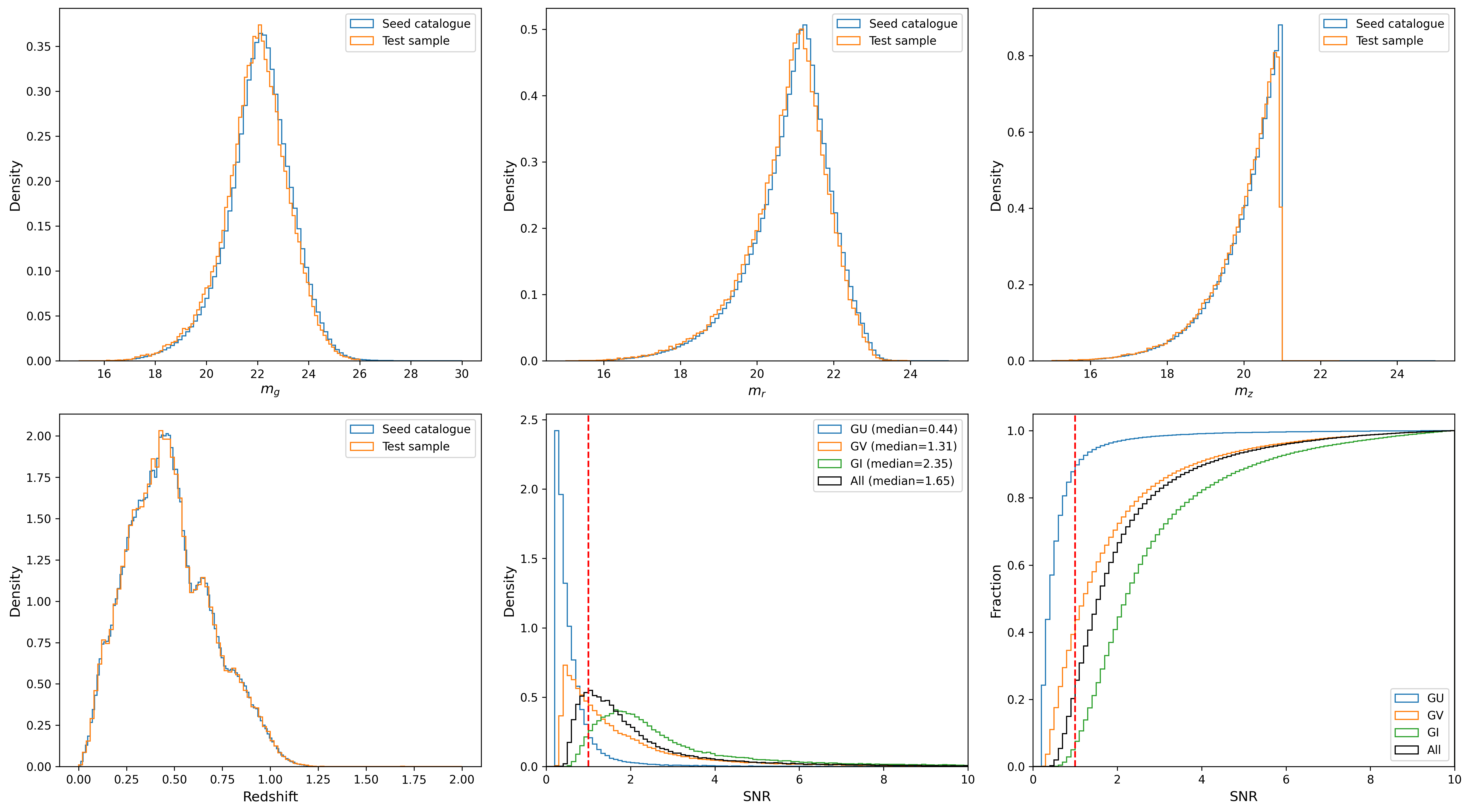}
    \caption{Distributions of DESI magnitudes (top row) and redshift (bottom left) for the test sample (orange) compared to the parent sample (blue), demonstrating the representativeness of our test sample. The bottom middle and right panels show SNR distributions (probability density and cumulative distribution) of CSST slitless spectroscopy in the GU, GV, and GI bands. Median SNR values are 0.44, 1.31, and 2.35 for GU, GV, and GI respectively, with median mean SNR across all bands of 1.65. The GI band shows significantly better SNR characteristics, with $<$10\% of measurements below $SNR=1$, compared to $\sim$40\% for GV and $\sim$90\% for GU.}
    \label{fig:sample_dist}
\end{figure*}

The DESI LS seed galaxies catalog contains approximately 138 million galaxies with photometric measurements in five bands: $g$, $r$, $z$ from DESI imaging, and W1, W2 from WISE.
For performance testing of galaxy redshift and stellar population parameter estimation, we randomly selected approximately 100,000 sources from the parent sample, referred to as the 'test sample'.
The test sample maintains the same redshift and magnitude distributions as the parent sample (see Figure~\ref{fig:sample_dist}), ensuring the representativeness of our test results.
We further divided the test sample into star-forming and quiescent galaxies based on their specific star formation rates (sSFR), with sSFR = $10^{-11}\,{\rm yr}^{-1}$ as the dividing threshold.
Galaxies with sSFR $> 10^{-11}\,{\rm yr}^{-1}$ are classified as star-forming, while those below this threshold are considered quiescent.

The bottom middle and right panels of Figure~\ref{fig:sample_dist} shows the distribution of $SNR$ across the three CSST slitless spectroscopic bands for our test sample.
The GU band exhibits the lowest median $SNR$ of 0.44, while GV and GI bands show progressively higher median values of 1.31 and 2.35, respectively.
The mean $SNR$ across all bands has a median value of 1.65.
About 90\% of GU measurements fall below $SNR=1$, while this fraction is approximately 40\% for GV and less than 10\% for GI measurements.
This systematic difference in $SNR$ distributions reflects the varying sensitivity limits across the slitless spectroscopic bands.

\subsection{Spectral energy distribution modeling} \label{ss:sedm}

Our galaxy spectral energy distribution modeling employs the BC03 \citep{BruzualG2003a} SSP models with a Chabrier Initial Mass Function (IMF), using an exponentially declining (Exp-dec) form for the SFH and the \citet{CalzettiD2000a} law for dust attenuation.
To optimize the computational efficiency for CSST slitless spectroscopic data analysis, we carefully tuned the spectral resolution of our models.
We tested model resolutions of $R=2000$, $500$, and $300$, comparing the resulting metrics for redshift, stellar mass, and SFR estimation.
We found that $R=300$ (comparable to CSST's instrumental $R = 241$, $263$, and $270$ for the GU, GV, and GI bands respectively) shows no appreciable degradation in these metrics compared to higher resolutions, while providing substantial computational savings.
We therefore adopted $R=300$ as the optimal balance between computational efficiency and model fidelity for this analysis.
The intrinsic model spectra at $R=300$ are subsequently convolved to CSST's instrumental resolution during the mock data generation process (Section~\ref{ss:sim}), ensuring that the final mock observations accurately represent CSST's capabilities.

For stellar populations younger than 10\,Myr (typical molecular cloud lifetime), we incorporate nebular emission following the framework described in \citet{BylerN2017a}.
The photoionization modeling utilizes the CLOUDY code \citep{FerlandG2017a}, which we have integrated as a subroutine within BayeSED3 to enable parallel sampling of the computationally intensive parameter space.
These young populations ($t_{\rm age} < 10\,{\rm Myr}$) are modeled with the following nebular parameters: ionization parameter $\log U = -2.3$, inner radius $\log(r_{\rm in}/{\rm cm}) = 19$, hydrogen number density $\log(n_{\rm H}/{\rm cm}^{-3}) = 2$, and gas-phase metallicity $\log(Z_{\rm gas}/Z_{\rm star}) = 0$, meaning the gas metallicity matches that of the stellar population.

For SSPs older than 10\,Myr, we retain the original BC03 models, adjusting only their spectral resolution to match our optimized $R=300$ specification.
The final galaxy spectra are produced by integrating both the nebular-enhanced young stellar populations and the resolution-adjusted older populations into our composite stellar population synthesis calculations.

\subsection{Model spectral library construction} \label{ss:lib}

\begin{figure*}[ht]
    \centering
    \includegraphics[width=1.0\textwidth]{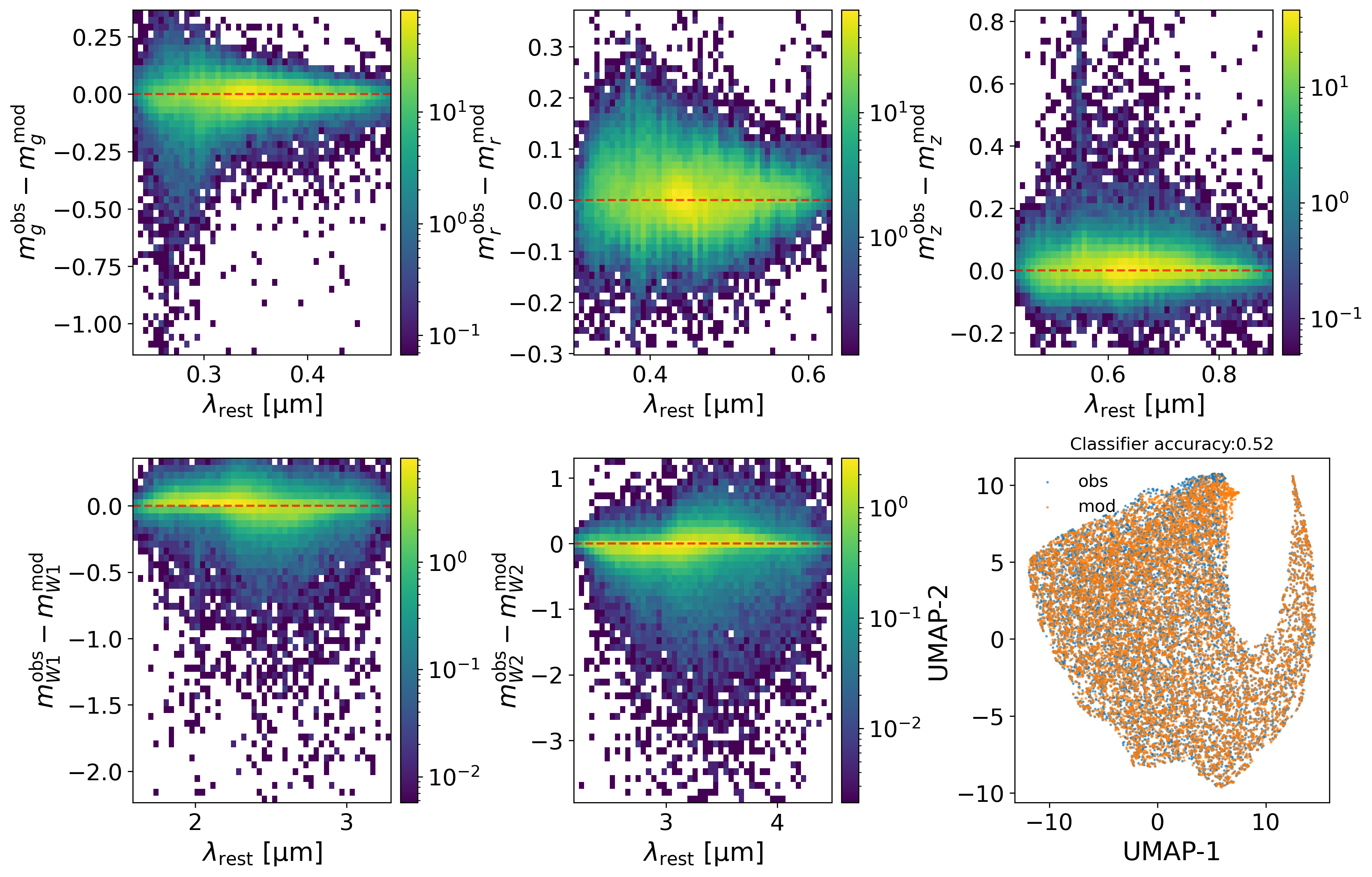}
    \caption{Residual analysis and color space validation for galaxies in the DESI DR9 photometric catalog. The five panels show magnitude residuals (observed minus model-predicted) for DESI optical bands (depths: $g$=24.5, $r$=23.9, $z$=22.9; top row) and WISE infrared bands (depths: W1=20.4, W2=19.5; bottom row). Each panel displays rest-frame wavelength on the x-axis versus magnitude residuals on the y-axis, with the density of points shown on a logarithmic color scale. Red dashed lines indicate perfect agreement (zero residuals). The tight distribution around zero residuals demonstrates the accuracy of our SED fitting across all bands and wavelengths. The sixth panel shows a UMAP embedding of standardized multi-band magnitudes with observed galaxies (blue) and model predictions (orange) overplotted. The accuracy of a simple domain classifier \textit{acc}$\approx 0.50$ indicate that the models and observations are well-mixed throughout the color space, demonstrating that our model library adequately spans the observational manifold. These results validate our model's ability to reproduce the observed photometry and generate reliable spectral templates for CSST simulations.}
    \label{fig:desi_fits_MAG}
\end{figure*}

Leveraging the stellar population synthesis and photoionization models described above, we constructed a comprehensive spectral library based on the DESI DR9 photometric catalog.
This catalog contains approximately 140 million galaxies with photometric measurements in five bands: $g$, $r$, $z$ from DESI imaging, and W1, W2 from WISE.
We utilized BayeSED3's multi-band photometric analysis capabilities to perform Bayesian SED fitting on each galaxy in the catalog.
Figure~\ref{fig:desi_fits_MAG} presents a comprehensive statistical validation of the best-fit results.
The first five panels (top row: $g$, $r$, $z$ bands; bottom row: W1, W2 bands) show magnitude residuals (observed minus model-predicted) as a function of rest-frame wavelength, demonstrating systematic accuracy with tight distributions centered on zero residuals.
The sixth panel provides a quantitative assessment of model representativeness in multi-band color space using UMAP dimensionality reduction.

We note that the residual analysis (panels 1-5) reveals systematic patterns in the model performance across different wavelength regimes, which we discuss band by band.
In the first panel ($g$-band), we observe systematic negative residuals at UV wavelengths (<0.3 $\mu$m), indicating that our models understimate the UV flux, possibly due to limitations in modeling young stellar populations, dust attenuation effects, or binary star evolution.
The second panel ($r$-band) shows generally good agreement with small residuals centered around zero, demonstrating reliable modeling in the optical continuum.
The third panel ($z$-band) exhibits some discrete positive residuals around 0.5-0.8 $\mu$m, suggesting that our models may overestimate the contributions of some emission lines.
However, no similar pattern is shown in the $r$-band data within similar rest-frame wavelength, suggesting other band specific observational issues.

Moving to the infrared bands, the fourth panel (W1 band) shows systematic negative residuals throughout the 2-3 $\mu$m range, indicating flux underestimation by our models.
This could reflect missing contributions from thermally pulsing asymptotic giant branch (TP-AGB) stars \citep{BevacquaD2025a}, which are expected to contribute significantly to NIR emission in certain stellar populations.
The fifth panel (W2 band) exhibits the strongest systematic negative residuals at 3-4 $\mu$m rest-frame wavelengths, suggesting significant flux underestimation.
This wavelength range is particularly sensitive to polycyclic aromatic hydrocarbon (PAH) features \citep{LaiT2020a} and hot dust emission from active galactic nuclei (AGN), both of which can contribute substantially to MIR photometry in star-forming and active galaxies but not be adequately represented in our current models.

To evaluate whether our model library adequately spans the observational manifold, we apply UMAP dimensionality reduction to standardized multi-band magnitudes (sixth panel).
After filtering out missing data (indicated by "-999") and subsampling the results for clarity, UMAP projects the high-dimensional color space (5 photometric bands) into a 2D space (UMAP-1 and UMAP-2), which captures the main structural patterns in the color space.
The \textit{acc}$\,\approx\,$0.5 metric indicates that a simple domain classifier cannot distinguish between models and observations better than random guessing, confirming they occupy the same regions of color space.

Despite these systematic effects across different wavelength regimes, the overall residual distributions remain well-centered around zero, indicating that our models provide reliable spectral templates for the majority of the galaxy population. These residuals highlight areas for future model improvements, particularly in incorporating TP-AGB stellar evolution in the NIR and PAH/AGN emission in the MIR.

The best-fit model spectra from this analysis serve dual purposes: they provide a foundation for CSST slitless spectroscopy simulations and establish a reference library for subsequent spectral analysis.
To efficiently manage the computational challenges posed by this large-scale analysis -- particularly the need to synthesize and store high-resolution spectra for each source -- we enhanced BayeSED3's data handling capabilities.
Specifically, we implemented support for the HDF5 file format, which enables efficient parallel read/write operations and optimized storage of the spectral library.

\subsection{CSST slitless spectroscopy simulation} \label{ss:sim}

To generate realistic mock observations for testing our analysis methodology, we employed the CESS (\citealt{WenR2024a}), a dedicated simulation tool designed to efficiently process large galaxy samples.
Using the model spectral library constructed from our DESI photometric analysis as input, CESS generates both slitless spectroscopic data and seven-band photometry (NUV, $u$, $g$, $r$, $i$, $z$, $y$) that closely match CSST's observational characteristics.
This simulation framework is part of a broader effort to develop comprehensive analysis pipelines for CSST, complementing the large-scale structure analysis pipeline described in \citet{GuY2024a}.

The simulation process begins by converting the high-resolution input spectra to CSST's spectral resolution.
This is achieved through a two-step process: first re-binning the spectra to match CSST's wavelength sampling, then applying convolution with appropriate Gaussian kernels.
To avoid artificial amplification of spectral features at boundaries, we implemented a refined weighting function for kernel integration.
The sky background contribution, primarily from earthshine and zodiacal light, is modeled using empirical measurements from space-based observations.

A key feature of CESS is its treatment of extended sources through empirical morphological modeling.
Each galaxy's 2D brightness profile is characterized by four parameters: S\'ersic index (n), effective radius ($R_{\rm e}$), position angle ($PA$), and axis ratio ($b/a$).
These parameters are used to generate two critical brightness distribution functions: one along the dispersion axis controlling the spectral self-broadening effect, and another along the spatial direction determining the extraction window width.
Our analysis reveals that while the S\'ersic index has minimal impact, the self-broadening effect is primarily driven by the effective radius, with axis ratio and position angle becoming important for edge-on galaxies.
This morphological treatment is particularly important as it enables consistent analysis with the mock galaxy catalogs developed for large-scale structure studies \citep{GuY2024a}.

To address the challenge of source confusion in slitless spectroscopy, CESS incorporates a dedicated contamination module.
This module maps the spatial distribution of spectral traces from all sources in the field, calculating overlap rates and contamination fractions by analyzing the locations and areas of zero-order images and first-order spectra.
The module is particularly valuable for assessing redshift measurement reliability in crowded environments such as galaxy clusters.
The contamination modeling is calibrated using the same mock galaxy distributions employed in large-scale structure analyses, ensuring consistency across different CSST science cases.

The final simulated data products include both 1D slitless spectra and broadband photometry, with all relevant observational effects and noise components incorporated.
This comprehensive dataset enables us to validate our joint analysis methodology and quantify the improvements achieved through combining spectroscopic and photometric information.
The simulation framework is designed to be compatible with the mock galaxy redshift surveys described in \citet{GuY2024a}, allowing for end-to-end testing of analysis pipelines from raw data to cosmological constraints (see \citealt{WenR2024a} for detailed descriptions of the simulation process).
A complete description of the mock catalog structure, including all spectroscopic and photometric data columns, is provided in Table~\ref{tab:catalog}.

\begin{deluxetable*}{lll}
\tabletypesize{\scriptsize}
\tablecaption{Column description of the CSST mock galaxy catalog\label{tab:catalog}}
\tablewidth{\textwidth}
\tablehead{
\colhead{Column} & \colhead{Unit} & \colhead{Description}
}
\startdata
ID & \nodata & Unique object identifier \\
z\_min & \nodata & Minimum of redshift prior \\
z\_max & \nodata & Maximum of redshift prior \\
E(B-V) & mag & Color excess of Milky Way \\
d/Mpc & \nodata & distance to the object \\
Photometry: mag\_[NUV,u,g,r,i,z,y] & mag & Broadband magnitudes \\
Photometry: magerr\_[NUV,u,g,r,i,z,y] & mag & Uncertainties of broadband magnitudes \\
index1 & \nodata & Internal index for cross-reference \\
z\_\{True\} & \nodata & True spectroscopic redshift in mocks \\
log(age/yr)[0,1]\_\{True\} & dex & True stellar age \\
log(tau/yr)[0,1]\_\{True\} & dex & True SF timescale \\
log(Z/Zsun)[0,1]\_\{True\} & dex & True stellar metallicity \\
Av\_2[0,1]\_\{True\} & mag & True dust attenuation \\
log(scale)[0,1]\_\{True\} & dex & True model scale \\
ageU(zform)/Gyr[0,1]\_\{True\} & \nodata & Universe age at formation \\
log(ageMW/yr)[0,1]\_\{True\} & dex & Mass-weighted age \\
log(ZMW/Zsun)[0,1]\_\{True\} & dex & Mass-weighted metallicity \\
log(CEH(0)/Zsun)[0,1]\_\{True\} & dex & Current metallicity \\
log(SFH(0)/[M\_{sun}/yr])[0,1]\_\{True\} & dex & Current SFR \\
log(SFR\_{100Myr}/[M\_{sun}/yr])[0,1]\_\{True\} & dex & SFR averaged over last 100 Myr \\
log(Mstar\_formed)[0,1]\_\{True\} & dex & Formed stellar mass over the past history \\
log(Mstar)[0,1]\_\{True\} & dex & Current stellar mass \\
log(Mstar\_liv)[0,1]\_\{True\} & dex & Living stellar mass \\
log(L\_{unabsorbed}/[erg/s])[0,1]\_\{True\} & dex & Unabsorbed luminosity \\
log(L\_{absorbed}/[erg/s])[0,1]\_\{True\} & dex & Dust-absorbed luminosity \\
RA & deg & Right ascension (R.A. in J2000) \\
Dec & deg & Declination (decl. in J2000) \\
z\_best & \nodata & Redshift used in BayeSED3 analysis of DESI DR9 photometry\\
MAG\_G & mag & DESI DR9 $G$-band magnitude \\
MAG\_R & mag & DESI DR9 $R$-band magnitude \\
MAG\_Z & mag & DESI DR9 $Z$-band magnitude \\
n & \nodata & S\'ersic index as given by CESS simulation \\
Re & arcsec & Effective radius as given by CESS simulation \\
PA & deg & Position angle as given by CESS simulation \\
baratio & \nodata & Axis ratio $b/a$ as given by CESS simulation \\
str\_mass & dex & Stellar mass used in CESS simulation \\
gu\_rms\_in\_e & \nodata & GU-band RMS noise as given by CESS simulation \\
gv\_rms\_in\_e & \nodata & GV-band RMS noise as given by CESS simulation \\
gi\_rms\_in\_e & \nodata & GI-band RMS noise as given by CESS simulation \\
gu\_snr\_mean & \nodata & GU-band mean SNR as given by CESS simulation \\
gv\_snr\_mean & \nodata & GV-band mean SNR as given by CESS simulation \\
gi\_snr\_mean & \nodata & GI-band mean SNR as given by CESS simulation \\
gu\_wave\_off & $\mu$m & GU-band wavelength offset as given by CESS simulation \\
gv\_wave\_off & $\mu$m & GV-band wavelength offset as given by CESS simulation \\
gi\_wave\_off & $\mu$m & GI-band wavelength offset as given by CESS simulation \\
index2 & \nodata & Secondary index/key \\
Nw\_u & \nodata & Number of GU wavelength samples (expected 338) \\
Nw\_v & \nodata & Number of GV wavelength samples (expected 393) \\
Nw\_i & \nodata & Number of GI wavelength samples (expected 516) \\
wu0..wu337, fu0..fu337, eu0..eu337, su0..su337 & $\mu$m, $\mu$Jy, $\mu$Jy, $\mu$m & GU arrays (0.25--0.40\,$\mu$m): wavelength, flux, error, dispersion \\
wv0..wv392, fv0..fv392, ev0..ev392, sv0..sv392 & $\mu$m, $\mu$Jy, $\mu$Jy, $\mu$m & GV arrays (0.40--0.60\,$\mu$m): wavelength, flux, error, dispersion \\
wi0..wi515, fi0..fi515, ei0..ei515, si0..si515 & $\mu$m, $\mu$Jy, $\mu$Jy, $\mu$m & GI arrays (0.60--1.00\,$\mu$m): wavelength, flux, error, dispersion \\
\enddata
\tablecomments{The complete machine-readable catalog, comprising approximately 100,000 mock galaxies with the columns described above, will be made publicly available upon publication at \url{https://doi.org/10.5281/zenodo.17221343}.}
\end{deluxetable*}

\section{Bayesian full spectrum analysis with BayeSED3} \label{sec:bayesed_dec}

In this section, we present our Bayesian framework for analyzing CSST slitless spectroscopic data.
We begin with the model selection and prior assumptions (Section \ref{ss:priors}), followed by the construction of likelihood functions for the cases with and without photometric data (Section \ref{ss:likelihood}).
We then detail our approach to model normalization through scaling factors (Section \ref{ss:scaling}) and the optimization of MultiNest runtime parameters for efficient parameter space exploration (Section \ref{sss:mutinest}).
Finally, we define comprehensive metrics for evaluating parameter estimation quality (Section \ref{ss:metrics}).

\subsection{Model Selection and Priors} \label{ss:priors}

For our Bayesian analysis, we employ the same galaxy spectral energy distribution modeling approach described in Section~\ref{ss:sedm}.
Specifically, we use the BC03 \citep{BruzualG2003a} SSP models with a Chabrier IMF, an exponentially declining SFH, and the \citet{CalzettiD2000a} DAL.
The model includes nebular emission for young stellar populations ($t_{\rm age} < 10\,{\rm Myr}$) following \citet{BylerN2017a}, with photoionization modeling using CLOUDY \citep{FerlandG2017a}.

We adopt the same priors for all model parameters as used in generating the mock data (Section~\ref{sec:csst_mock}): uniform priors for stellar population age $\log({\rm age/yr})$ over [8.0, 10.13], star formation timescale $\log(\tau/{\rm yr})$ over [6.0, 10.0], metallicity $\log(Z/Z_{\odot})$ over [-2.30, 0.70], dust extinction $A_V$ over [0.0, 4.0] mag, and flux scaling factor $\log({\rm scale})$ over [5.0, 12.0].
The ranges of these priors are chosen based on physical considerations and previous studies.
For the stellar metallicity evolution, we maintain the linear SFH-to-metallicity mapping model.
The spectral resolution is kept at $R=300$ as the optimal balance between computational efficiency and model fidelity.
This consistent treatment between mock data generation and analysis helps isolate the effects of SED modeling uncertainties in our results.

\subsection{Likelihood function} \label{ss:likelihood}

The likelihood function in our Bayesian analysis framework combines both spectroscopic and photometric data when available.
For a comprehensive treatment, we present the likelihood functions for cases with and without supplementary photometric data separately in the following subsections.

\subsubsection{Without photometry} \label{sss:likelihood_nophot}

For the analysis of slitless spectroscopic data alone, assuming Gaussian noise, the likelihood function is defined as:

\begin{equation}
\mathcal{L}(\boldsymbol{\theta}) \equiv p(\mathbf{d}|\boldsymbol{\theta},\mathbf{M},\mathbf{I}) = \mathcal{L}_{\mathrm{spec}}(\boldsymbol{\theta})
\label{eq:likelihood_def}
\end{equation}

where $\mathcal{L}_{\mathrm{spec}}(\boldsymbol{\theta})$ is given by:

\begin{equation}
\begin{split}
\mathcal{L}_{\mathrm{spec}}(\boldsymbol{\theta}) &= \prod_{i=1}^{n}\,\frac{1}{\sqrt{2\pi}\,\sigma_{i}} \\
&\quad \times \exp\left(-\frac{1}{2}\frac{(F_{\mathrm{o},i}^{\mathrm{spec}}-s\cdot f_{\mathrm{m},i}^{\mathrm{spec}}(\boldsymbol{\theta}))^2}{\sigma_{i}^2}\right)
\end{split}
\label{eq:likelihood_spec}
\end{equation}

where $F_{\mathrm{o},i}^{\mathrm{spec}}$ and $\sigma_i$ represent the observed spectral flux and its uncertainty in each wavelength bin $i$, $f_{\mathrm{m},i}^{\mathrm{spec}}(\boldsymbol{\theta})$ represents the model spectral flux with parameters $\boldsymbol{\theta}$, and $s$ is a scaling factor.
$\mathbf{M}$ represents the model assumptions and $\mathbf{I}$ represents all other relevant background information.

Our likelihood function assumes uncorrelated Gaussian noise, which is a simplification of the true noise characteristics in CSST slitless spectroscopy.
In reality, low-SNR spectra may exhibit non-Gaussian noise properties, instrumental systematics can introduce correlations between adjacent spectral elements, and spectral self-blending creates systematic effects.
We quantify the impact of spectral self-blending separately in Section~\ref{ss:disc_blend_overlap}, showing that while it increases scatter ($\sigma_{\rm NMAD}$) by $\gtrsim30\%$ across all parameters, high precision remains achievable.
Future refinements incorporating non-Gaussian noise models and covariance matrices represent important avenues for improvement, particularly for systematic error characterization in cosmological analyses.

\subsubsection{With photometry} \label{sss:likelihood_withphot}

When complementary photometric data are available, we extend the likelihood function to incorporate both spectroscopic and photometric constraints:

\begin{equation}
\mathcal{L}(\boldsymbol{\theta}) = \mathcal{L}_{\mathrm{spec}}(\boldsymbol{\theta}) \cdot \mathcal{L}_{\mathrm{phot}}(\boldsymbol{\theta})
\label{eq:likelihood_joint}
\end{equation}

where $\mathcal{L}_{\mathrm{spec}}(\boldsymbol{\theta})$ is the spectroscopic likelihood as defined in Equation~\ref{eq:likelihood_spec}, and $\mathcal{L}_{\mathrm{phot}}(\boldsymbol{\theta})$ is the photometric likelihood given by:

\begin{equation}
\begin{split}
\mathcal{L}_{\mathrm{phot}}(\boldsymbol{\theta}) &= \prod_{j=1}^{m}\,\frac{1}{\sqrt{2\pi}\,\sigma_{j}} \\
&\quad \times \exp\left(-\frac{1}{2}\frac{(F_{\mathrm{o},j}^{\mathrm{phot}}-s\cdot f_{\mathrm{m},j}^{\mathrm{phot}}(\boldsymbol{\theta}))^2}{\sigma_{j}^2}\right)
\end{split}
\label{eq:likelihood_phot}
\end{equation}

where $F_{\mathrm{o},j}^{\mathrm{phot}}$ and $\sigma_j$ are the observed photometric flux and uncertainty in band $j$, and $f_{\mathrm{m},j}^{\mathrm{phot}}(\boldsymbol{\theta})$ is the corresponding model flux.
This joint likelihood function enables us to leverage both the detailed spectral information from CSST slitless spectroscopy and the broad wavelength coverage from photometry when available.

\subsection{Determination of Scaling Factor} \label{ss:scaling}

As shown in Equations~\ref{eq:likelihood_spec} and \ref{eq:likelihood_phot}, both spectroscopic and photometric likelihoods involve a scaling factor $s$ that relates model fluxes to observed fluxes.
This scaling factor is necessary because the absolute normalization of model spectra is typically arbitrary, and we need to scale them to match the observed flux levels.
We have employed two approaches for determining this scaling factor in our analysis:

In the first approach, which was used in previous versions of BayeSED, the scaling factor $s(\boldsymbol{\theta})$ is determined using Non-negative Linear Model (NNLM) optimization by minimizing the $\chi^2$ with respect to $s$ for any given set of other model parameters $\boldsymbol{\theta}$.
The NNLM optimization approach, rather than simple least squares, ensures that the scaling factor remains positive, which is physically meaningful since we are dealing with flux measurements.
For the joint likelihood case (Equation~\ref{eq:likelihood_joint}), a single scaling factor is determined using all available data points by combining both spectroscopic and photometric measurements in the summations.
This NNLM optimization-based determination of $s(\boldsymbol{\theta})$ for each set of model parameters reduces the dimensionality of the parameter space that needs to be explored by MultiNest.

In BayeSED3, we have introduced a new approach where $s$ is treated as an additional free parameter in the Bayesian sampling along with other SED modeling parameters $\boldsymbol{\theta}$.
Similar approach have been employed by other Bayesian SED fitting codes, such as BAGPIPES and Prospector.
We adopt a log-uniform prior for $s$ to ensure positivity while allowing exploration over several orders of magnitude.
While this increases the dimensionality of the parameter space and computational cost, it allows for a more thorough exploration of scaling uncertainties and their correlations with other parameters.

\begin{figure*}
    \centering
    \begin{tabular}{cc}
        \includegraphics[width=0.48\textwidth]{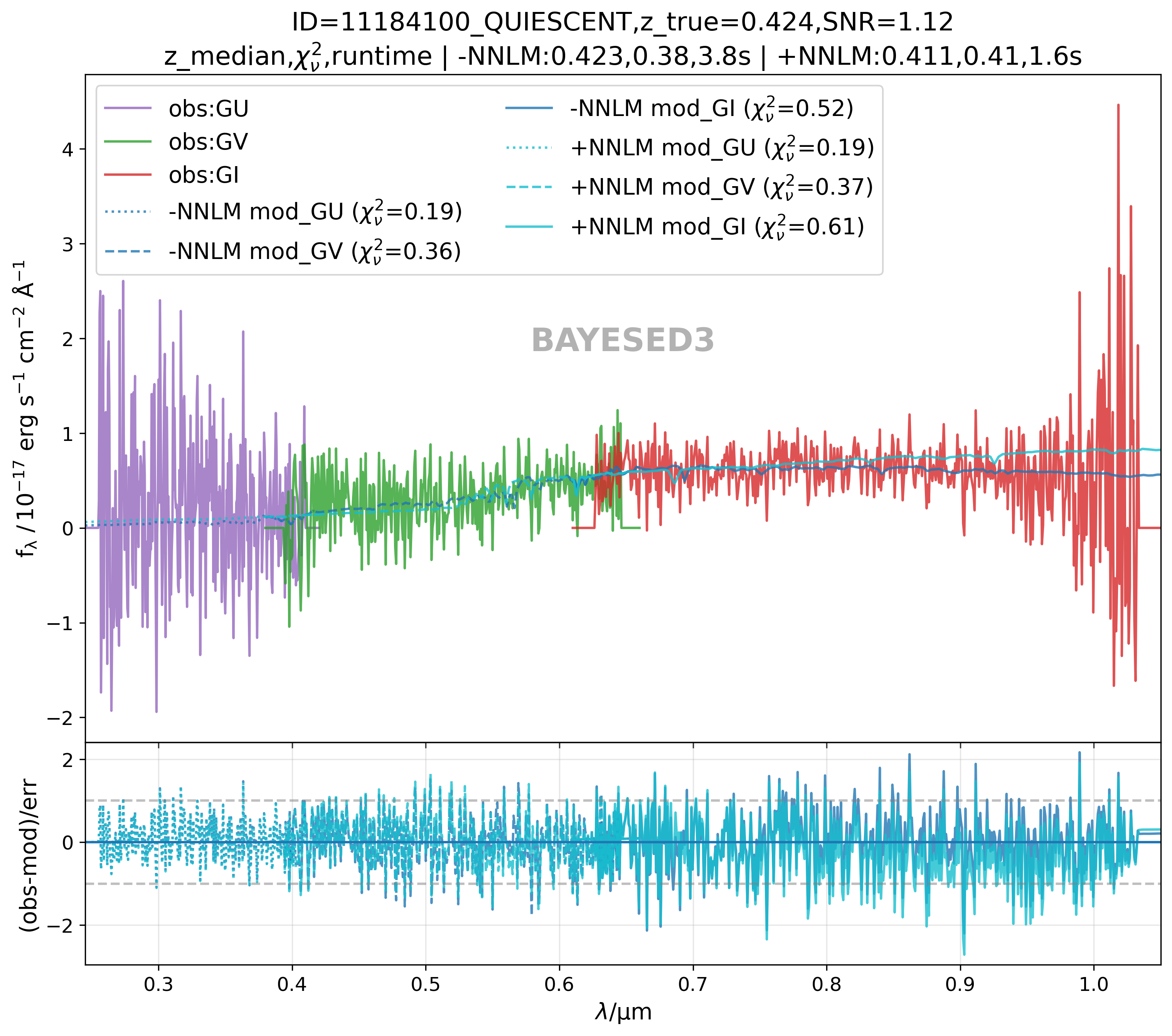} &
        \includegraphics[width=0.48\textwidth]{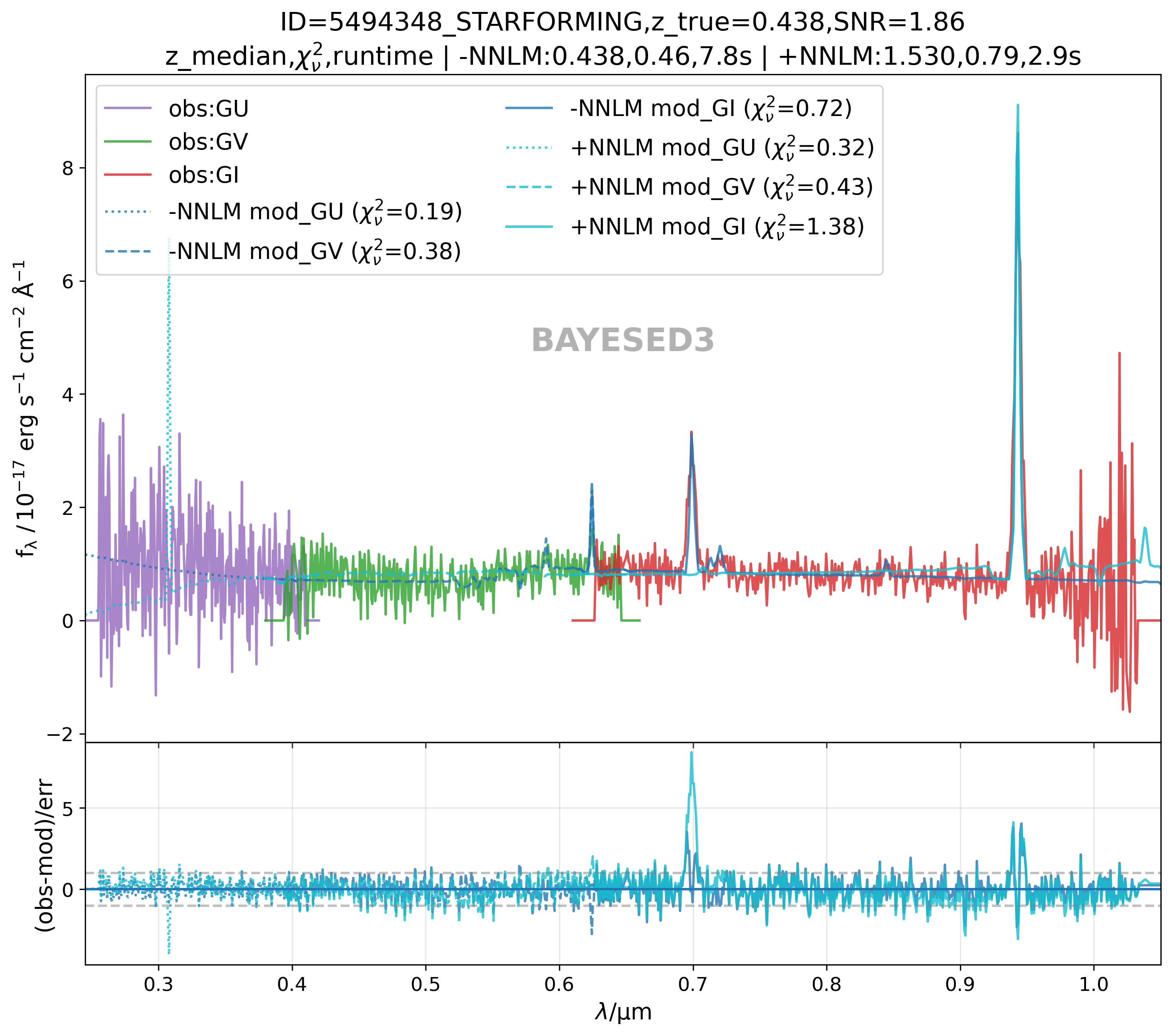} \\
        \includegraphics[width=0.48\textwidth]{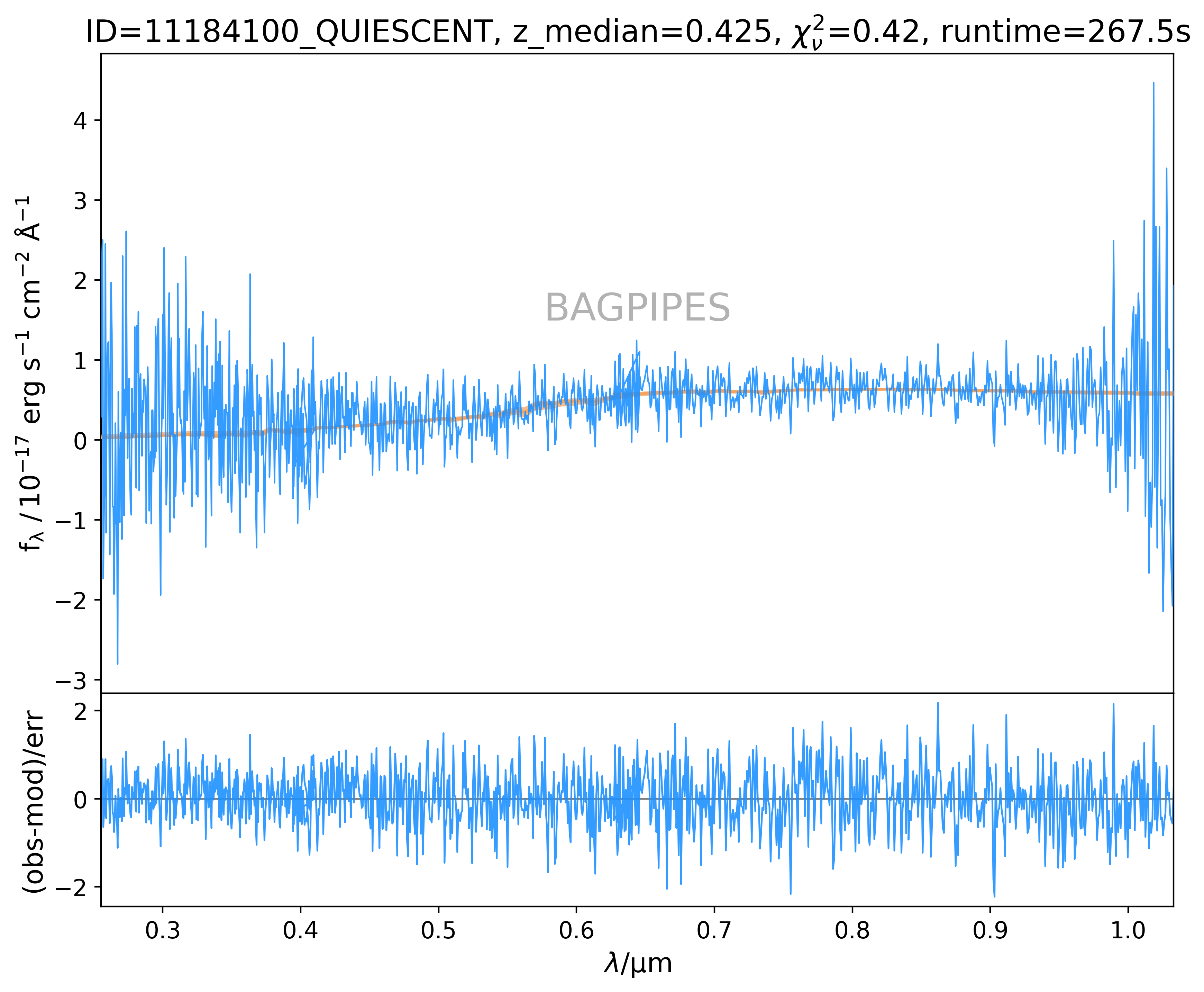} &
        \includegraphics[width=0.48\textwidth]{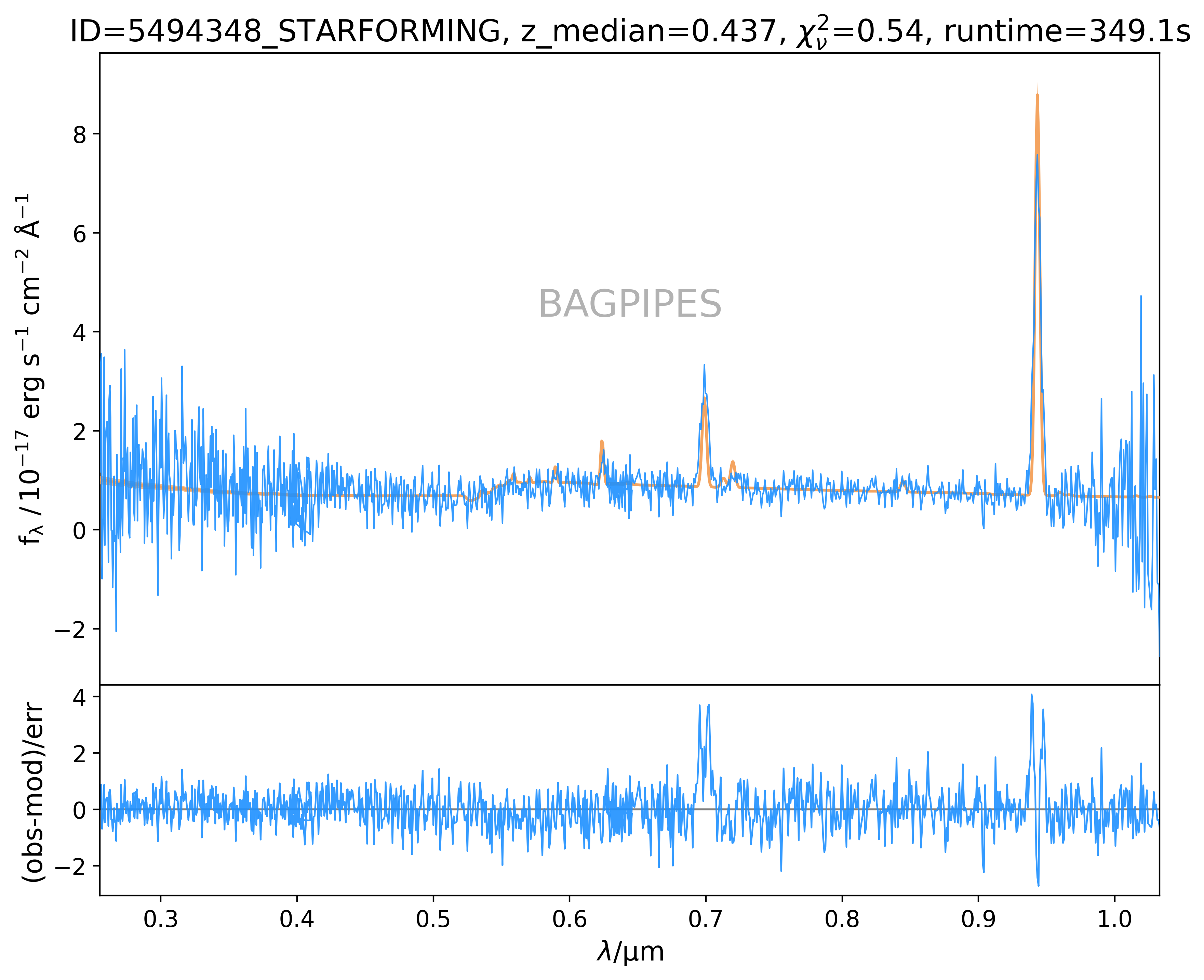} \\
        \includegraphics[width=0.48\textwidth]{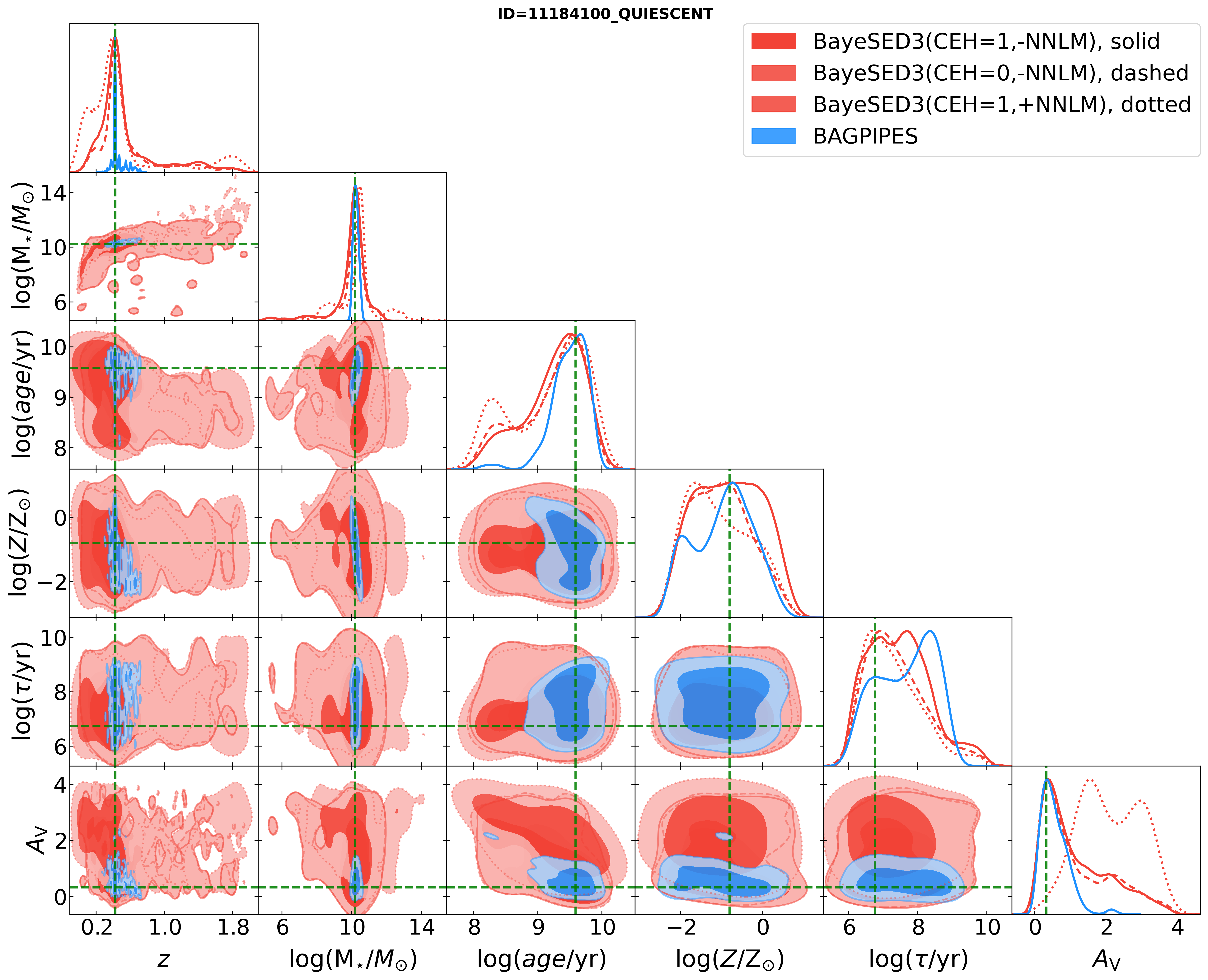} &
        \includegraphics[width=0.48\textwidth]{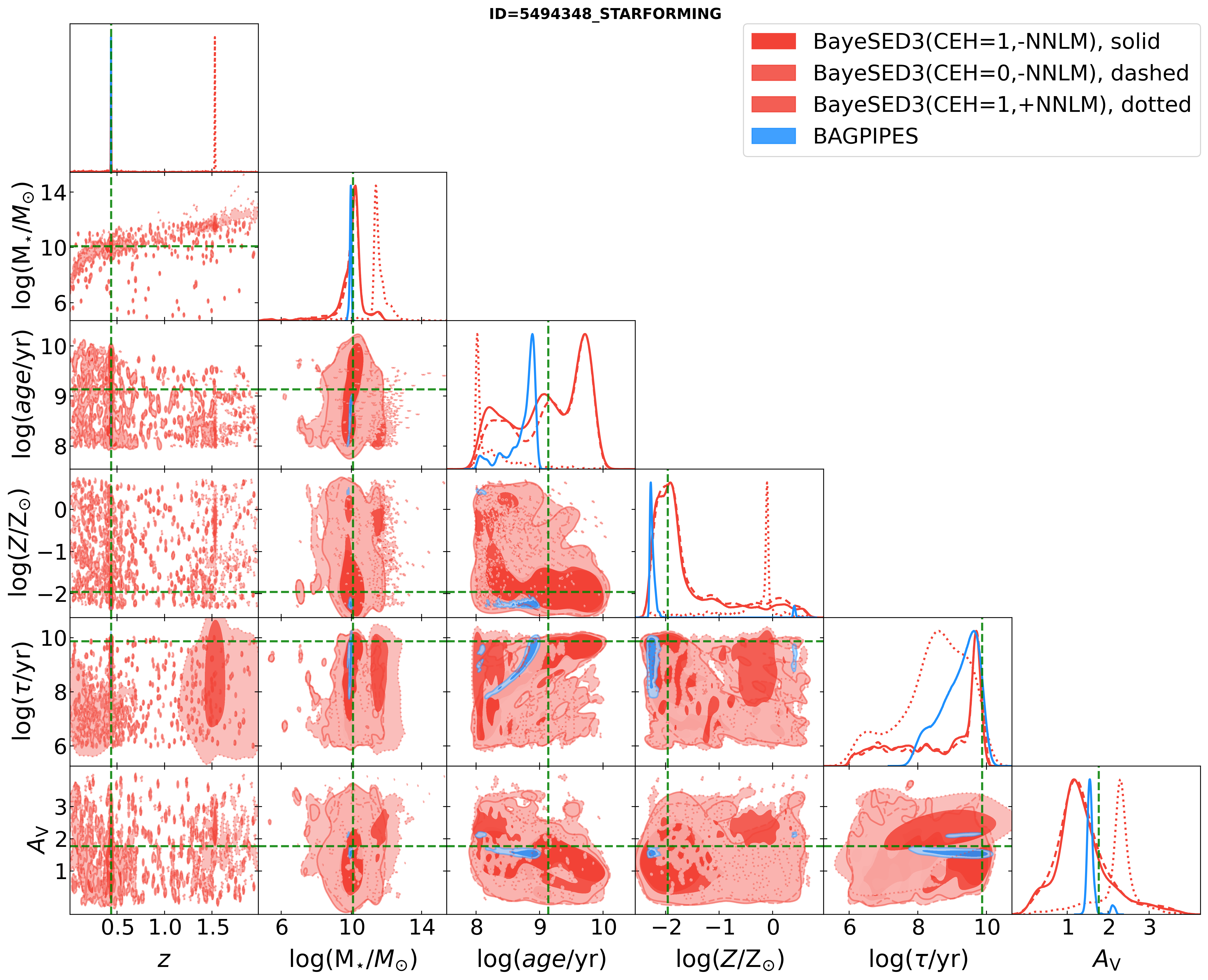}
    \end{tabular}
    \caption{Bayesian full spectrum analysis of CSST slitless spectra for a quiescent (left) and star-forming (right) galaxy. \textbf{Top row:} BayeSED3 results comparing -NNLM (Bayesian sampling) and +NNLM (optimization) approaches. For the quiescent galaxy (SNR=1.12, $z_{\rm true}=0.424$), -NNLM achieves accurate redshift ($z_{\rm median}=0.423$) while +NNLM yields a different redshift ($z_{\rm median}=0.411$). For the star-forming galaxy (SNR=1.86, $z_{\rm true}=0.438$), -NNLM perfectly recovers the redshift ($z_{\rm median}=0.438$) while +NNLM catastrophically fails ($z_{\rm median}=1.530$). \textbf{Middle row:} BAGPIPES achieves competitive quality but requires much longer runtime than BayeSED3. \textbf{Bottom row:} Corner plots comparing posterior distributions from BayeSED3 variations (CEH=1/-NNLM, CEH=0/-NNLM, CEH=1/+NNLM) and BAGPIPES for all six free parameters. Generally, the Bayesian sampling approach provides much more reliable posteriors for low-SNR slitless spectroscopy.}
    \label{fig:fitting_comparison}
\end{figure*}

Figure~\ref{fig:fitting_comparison} compares these two scaling approaches using representative quiescent and star-forming galaxies, along with BAGPIPES results for reference.
The top row shows BayeSED3 results comparing the two scaling approaches: for the quiescent galaxy (ID=11184100, SNR=1.12, $z_{\rm true} = 0.424$), the Bayesian sampling approach (-NNLM) achieves accurate redshift estimation ($z_{\rm median} = 0.423$), while the NNLM optimization approach (+NNLM) yields a slightly different result ($z_{\rm median} = 0.411$).
For the star-forming galaxy (ID=5494348, SNR=1.86, $z_{\rm true} = 0.438$), -NNLM accurately recovers the true redshift ($z_{\rm median} = 0.438$), while +NNLM shows catastrophic redshift failure ($z_{\rm median} = 1.530$).
The middle row presents BAGPIPES results, which yield competitive fitting quality ($z_{\rm median} = 0.425$ and $0.437$ for the quiescent and star-forming respectively) but require substantially longer computational times \footnote{The recent version of BAGPIPES with nautilus sampler has been employed. We found that the optimal nlive=40 for MultiNest in BayeSED3 is not suitable for BAGPIPES, therefore the default nlive=400 is adopted. We set the pool size used for parallelization to be 10 for BAGPIPES, while run BayeSED3 with "mpirun -np 10", in the same machine with 3GHz 10-Core Intel Xeon W processor. Due to the substantial computational resources required, we performed the BAGPIPES comparison on these two representative galaxies rather than the entire test sample. To facilitate similar benchmark tests with other codes, we will release the complete mock catalog (see Section~\ref{sec:csst_mock}). A python script for detailed comparison between BayeSED3 and BAGPIPES with the same input data file is publicly available at \url{https://github.com/hanyk/BayeSED3/tests/test_bayesed_bagpipes_comparison.py}.}.

The NNLM optimization approach exhibits problematic limitations for low-SNR slitless spectroscopy.
While it can be faster than the Bayesian sampling approach when it successfully converges, these quick convergences often yield incorrect solutions, highlighting that speed alone is insufficient if the solution is unreliable.
More critically, across our broader test sample, the NNLM optimization approach failed to converge within 100,000 likelihood evaluations for many cases, resulting in much longer runtimes without achieving reliable solutions.
For quiescent galaxies with low overall SNR, the NNLM optimization approach is prone to converging to incorrect solutions, likely due to getting trapped in local minima or overfitting to noise fluctuations in the low-SNR regime.
For star-forming galaxies, regions with emission lines typically have much higher SNR than the surrounding continuum spectrum, causing the NNLM optimization approach to overemphasize emission line regions while poorly fitting the continuum.
This poor convergence reliability and tendency toward incorrect solutions makes NNLM optimization impractical for large survey applications, where both computational efficiency and solution reliability are essential.
The Bayesian sampling approach provides superior performance by properly marginalizing over the scaling factor, leading to more balanced fits that appropriately weight both high-SNR and low-SNR regions across the spectrum.
This approach better handles the challenges of low-SNR slitless spectroscopy by avoiding local minima traps and preventing overfitting to noise fluctuations, as reflected in the more uniform residuals and improved $\chi^2$ values across all bands.

The corner plots in the bottom row of Figure~\ref{fig:fitting_comparison} demonstrate that BayeSED3 and BAGPIPES produce generally consistent posterior distributions, especially for the quiescent galaxy.
For this comparison, we attempted to use the same SED model with the same number of free parameters and priors as closely as possible.
It is important to note that in the low-SNR regime, posterior distributions can be highly sensitive to both prior choices and SED model assumptions.
For example, BAGPIPES does not allow the same time-evolving CEH as BayeSED3, so we ran BayeSED3 with the same constant CEH (CEH=0) as well.
For the star-forming galaxy, we observe more pronounced differences in the posterior distributions of parameters such as age and metallicity between BayeSED3 and BAGPIPES, suggesting differences in the treatment of nebular emission modeling.
These differences highlight the importance of nebular emission modeling for star-forming galaxies and represent a major subject for future work that will require more careful calibration with high-resolution spectroscopic data.

Generally, the clear advantage of pure Bayesian sampling approach (BayeSED3 without NNLM or BAGPIPES) over the optimization approach (such as NNLM) is evident in the well-constrained, reliable posterior distributions compared to the broader, less reliable constraints from optimization, confirming the superiority of the Bayesian treatment for low-SNR slitless spectroscopy.
BayeSED3's Bayesian sampling approach thus achieves the optimal balance of reliability, accuracy, and computational efficiency for large-scale spectroscopic surveys.
By combining robust Bayesian parameter estimation with optimized computational performance, BayeSED3 delivers fitting quality comparable to established codes like BAGPIPES while requiring much less computational time.
This computational advantage stems from BayeSED3's implementation in C++ with optimized SED model resolution (R=300) and carefully tuned nested sampling parameters specifically for CSST data characteristics.
The combination of methodological robustness, computational tractability, and more user friendly python interface makes BayeSED3 ideally suited for processing the millions of low-SNR slitless spectra that CSST will deliver, enabling reliable galaxy characterization at the scale required for precision cosmology and galaxy evolution studies.

\subsection{Runtime parameters of MultiNest algorithm} \label{sss:mutinest}

As the Bayesian inference engine of BayeSED, MultiNest has several runtime parameters that significantly affect the performance of parameter estimation.
Two parameters are particularly important: the total number of live points (nlive) and the target sampling efficiency (efr).
The nlive parameter determines the effective sampling resolution - how many points are being actively sampled at any given time during the nested sampling process.
The efr parameter determines the ratio of points accepted to those sampled, controlling how efficiently the algorithm explores the parameter space.

Generally, a larger nlive and lower efr lead to more accurate posteriors and evidence values but at higher computational cost.
Following \citet{HanY2023a}, we systematically tested different combinations of these parameters, exploring six values of efr (0.05, 0.1, 0.2, 0.3, 0.5, 0.8) and eight values of nlive (10, 15, 20, 25, 50, 100, 200, 400) to find the optimal balance between computational efficiency and estimation quality.

Based on these tests, we adopt nlive = 40 and efr = 0.1 for our analysis.
These values provide a good balance between computational efficiency and the quality of parameter estimation.
For the Bayesian sampling approach (treating the scaling factor as an additional free parameter), the median runtime using one processor thread on a 2.2GHz CPU is approximately 34 seconds for star-forming galaxies and 38 seconds for quiescent galaxies across our test sample.
With these settings, we achieve reliable parameter estimation while maintaining reasonable computational costs for analyzing large samples of galaxies.

\subsection{Metrics for Parameter Estimation} \label{ss:metrics}

For quantitative assessment of parameter estimation quality, we employ several key metrics that are consistently applied across redshift, stellar mass ($M_*$), and star formation rate (SFR) measurements. These metrics are calculated within different SNR bins and as functions of redshift to evaluate performance under varying observational conditions.

\subsubsection{Success and Reliability Metrics}

We define the following metrics to evaluate the success and reliability of parameter estimation:

\begin{itemize}
    \item \textbf{Completeness}: The fraction of galaxies with measurements that pass our quality criteria relative to the total test sample size:
    \begin{equation}
        {\rm Completeness} = \frac{N_{\rm acceptable}}{N_{\rm total}}
    \end{equation}
    where $N_{\rm acceptable}$ is the number of galaxies meeting both $SNR > 0$ and parameter-specific acceptance thresholds.
    
    \item \textbf{Purity}: The fraction of reliable measurements among those that pass our quality criteria:
    \begin{equation}
        {\rm Purity} = \frac{N_{\rm true\,positives}}{N_{\rm acceptable}}
    \end{equation}
    where $N_{\rm true\,positives}$ is the number of acceptable measurements with normalized error below the success threshold.
    
    \item \textbf{Success Rate}: The product of completeness and purity:
    \begin{equation}
        {\rm Success\,Rate} = \frac{N_{\rm true\,positives}}{N_{\rm total}}
    \end{equation}
    representing the overall fraction of galaxies with reliable measurements in the complete sample.
    
    \item \textbf{Outlier Fraction}: The fraction of catastrophic failures among acceptable measurements:
    \begin{equation}
        {\rm Outlier\,Fraction} = \frac{N_{\rm outliers}}{N_{\rm acceptable}}
    \end{equation}
    where $N_{\rm outliers}$ is the number of acceptable measurements with normalized error above the outlier threshold.
\end{itemize}

\subsubsection{Accuracy Metrics}

To quantify the accuracy of parameter estimation, we use:

\begin{itemize}
    \item \textbf{Bias}: The median of the normalized error distribution for acceptable measurements:
    \begin{equation}
        {\rm Bias} = {\rm median}(\Delta_{\rm acceptable})
    \end{equation}
    where $\Delta$ is defined differently for each parameter:
    \begin{itemize}
        \item For redshift: $\Delta_z = (z_{\rm pred} - z_{\rm true})/(1 + z_{\rm true})$
        \item For stellar mass: 
        \begin{align}
        \Delta_{M_*} &= \frac{\log(M_*/M_{\odot})_{\rm pred} - \log(M_*/M_{\odot})_{\rm true}}{\log(M_*/M_{\odot})_{\rm true}^{\rm max} - \log(M_*/M_{\odot})_{\rm true}^{\rm min}}
        \end{align}
        \item For SFR: 
        \begin{align}
        \Delta_{SFR} &= \\
        &\frac{\log({SFR}\,[M_{\odot}\,{\rm yr}^{-1}])_{\rm pred} - \log({SFR}\,[M_{\odot}\,{\rm yr}^{-1}])_{\rm true}}{\log({\rm SFR}\,[M_{\odot}\,{\rm yr}^{-1}])_{\rm true}^{\rm max} - \log({\rm SFR}\,[M_{\odot}\,{\rm yr}^{-1}])_{\rm true}^{\rm min}} \nonumber
        \end{align}
    \end{itemize}
    \footnote{For the test sample selected in \S\ref{ss:sample}, $\log(M_*/M_{\odot})_{\rm true}^{\rm min} \approx 5$, $\log(M_*/M_{\odot})_{\rm true}^{\rm max} \approx 13$, $\log({\rm SFR}/[M_{\odot}\,{\rm yr}^{-1}])_{\rm true}^{\rm min} \approx -3$, and $\log({\rm SFR}/[M_{\odot}\,{\rm yr}^{-1}])_{\rm true}^{\rm max} \approx 4$.}
    
    \item \textbf{$\sigma_{\rm NMAD}$}: The normalized median absolute deviation, a robust estimator of scatter that is less sensitive to outliers than the standard deviation:
    \begin{equation}
        \sigma_{\rm NMAD} = 1.48 \times {\rm median}(|\Delta|)
    \end{equation}
    where $\Delta$ is the normalized error defined for each parameter as above.
    The factor 1.48 is chosen such that $\sigma_{\rm NMAD}$ equals the standard deviation for a normal distribution, making it directly comparable to Gaussian statistics while being more robust against outliers.
\end{itemize}

For redshift estimation, achieving $\sigma_{\rm NMAD} \lesssim 0.002$-$0.005$ is required for CSST's cosmological goals \citep{MiaoH2023a,MiaoH2024a}, and we demonstrate in Sections~\ref{sec:results} and \ref{sec:disc} that our method meets these precision requirements for both star-forming and quiescent galaxies.

\subsubsection{Parameter-Specific Thresholds} \label{sss:metrics_thresholds}

For each parameter, we define specific thresholds and acceptable ranges:

\begin{itemize}
    \item \textbf{Redshift}:
    \begin{itemize}
        \item Success threshold: \\$|\Delta_z| < 0.01$ 
        \item Outlier threshold: \\$|\Delta_z| \geq 0.02$ 
        \item Acceptance threshold\footnote{The optimal value is determined through analysis in Section~\ref{sss:results_z_threshold}.}: \\$z_{\rm mean}/\sigma_z > 20$ 
    \end{itemize}
    
    \item \textbf{Stellar Mass}:
    \begin{itemize}
        \item Success threshold: \\$|\Delta_{M_*}| < 0.02$ 
        \item Outlier threshold: \\$|\Delta_{M_*}| \geq 0.04$ 
        \item Acceptance threshold\footnote{The optimal value is determined through analysis in Section~\ref{sss:results_mass_threshold}.}: \\$\langle\log M_*\rangle/\sigma_{\log M_*} > 50$ 
    \end{itemize}
    
    \item \textbf{Star Formation Rate}:
    \begin{itemize}
        \item Success threshold: \\$|\Delta_{SFR}| < 0.1$ 
        \item Outlier threshold: \\$|\Delta_{SFR}| \geq 0.2$ 
        \item Acceptance threshold\footnote{The optimal value is determined through analysis in Section~\ref{sss:results_sfr_threshold}.}: \\$\langle\log {\rm SFR}\rangle/\sigma_{\log {\rm SFR}} > 3$ 
    \end{itemize}
\end{itemize}

For all parameters, we additionally require $SNR > 0$ as a basic quality criterion.\footnote{This is a computational quality check ensuring that the SNR calculation has succeeded and returned a valid positive number, not a physical constraint. In rare cases, fitting failures or numerical issues could result in undefined or negative SNR estimates, which this filter excludes.} These metrics and thresholds are designed to provide a comprehensive assessment of parameter estimation quality while accounting for the different scales and physical meanings of each parameter. The thresholds are chosen based on typical requirements for galaxy evolution studies and the expected performance of CSST slitless spectroscopy.

\section{Results} \label{sec:results}

In this section, we present a comprehensive analysis of our method's performance in estimating three key galaxy properties: redshift, stellar mass, and star formation rate.
For each property, we follow a systematic approach: first analyzing how measurement quality varies with acceptance thresholds to determine optimal acceptance criteria, then examining detailed performance metrics separately for star-forming and quiescent galaxies.

\subsection{Redshift} \label{ss:results_z}

We now present the results of our analysis, starting with redshift estimation, which is fundamental for deriving other galaxy properties.
Accurate redshift measurements are crucial for the subsequent analysis of stellar mass and star formation rate.
This section systematically examines the performance of our redshift estimation method.
We first analyze the impact of the acceptance threshold on the overall quality of redshift measurements.
Then, we delve into the detailed performance metrics, specifically focusing on their dependence on SNR and redshift for both star-forming and quiescent galaxies.

\subsubsection{Dependence of overall performance on Acceptance Threshold} \label{sss:results_z_threshold}

\begin{figure*}
    \includegraphics[width=\textwidth]{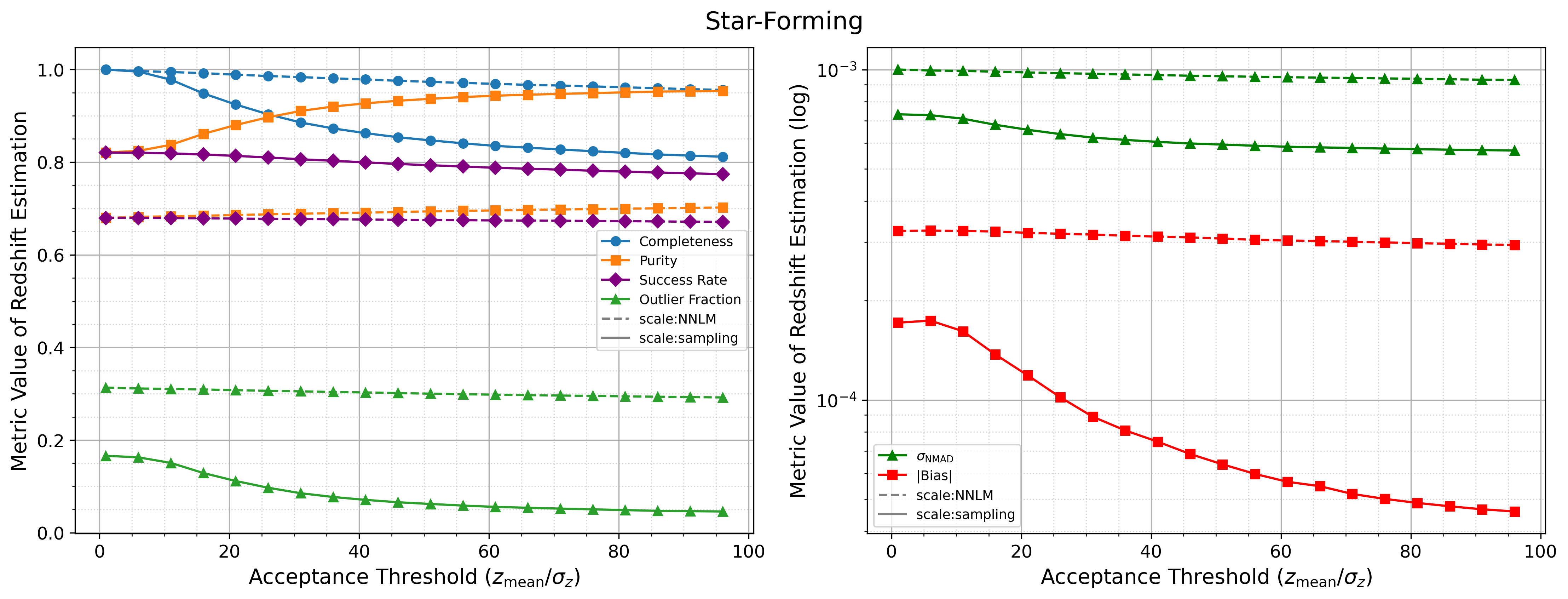}
    \includegraphics[width=\textwidth]{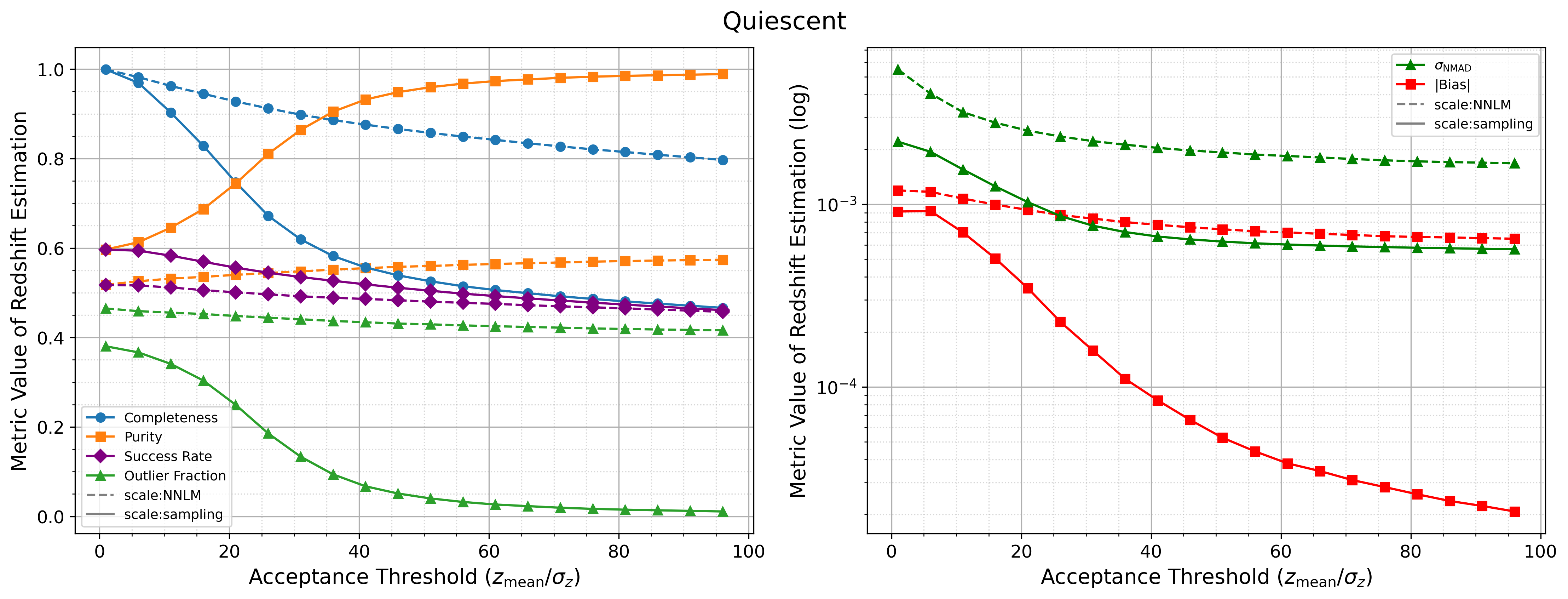}
    \caption{Quality assessment of redshift estimation as a function of acceptance threshold ($z_{\rm mean}/\sigma_z$) for star-forming (top) and quiescent (bottom) galaxies, comparing NNLM optimization (dashed lines) and Bayesian sampling (solid lines) approaches for scaling determination. Left panels show quality metrics (completeness, purity, outlier fraction, success rate); right panels show precision metrics ($\sigma_{\rm NMAD}$, $|{\rm Bias}|$). The Bayesian sampling approach consistently outperforms NNLM optimization across all metrics for both galaxy types, with particularly dramatic improvements for quiescent galaxies. An optimal threshold of $z_{\rm mean}/\sigma_z > 20$ is adopted for subsequent analysis (see Section~\ref{sss:results_z_threshold} for detailed discussion).}
    \label{fig:redshift_significance}
\end{figure*}

We first investigate how the quality of redshift measurements depends on the acceptance threshold ($z_{\rm mean}/\sigma_z$) for both star-forming and quiescent galaxies. Figure~\ref{fig:redshift_significance} shows the evolution of various quality metrics with increasing threshold values, comparing two different scaling approaches: NNLM optimization (dashed lines) and Bayesian sampling (solid lines).

For star-forming galaxies, the Bayesian sampling approach consistently outperforms the NNLM optimization approach across all metrics. Although completeness decreases faster with the Bayesian approach, it achieves significantly better quality metrics: higher purity ($80\text{--}90\%$ vs. $70\%$), higher success rate ($82\text{--}78\%$ vs. $68\%$), and lower outlier fraction ($15\text{--}5\%$ vs. $30\%$). The precision is also substantially better with the Bayesian approach, achieving lower $\sigma_{\rm NMAD}$ ($0.0007\text{--}0.00055$ vs. $0.001\text{--}0.00095$) and lower bias ($0.0002\text{--}0.00005$ vs. $0.0003$). This superior performance can be attributed to the fundamental differences between the two approaches. The Bayesian method explores the full parameter space more thoroughly, accounting for complex interdependencies between redshift and other galaxy properties. It provides more robust uncertainty estimation by sampling the posterior probability distribution rather than just finding a maximum likelihood point, resulting in fewer catastrophic failures and more reliable confidence intervals.

The results for star-forming galaxies demonstrate an important trade-off: while the Bayesian approach reduces the number of galaxies that meet the acceptance criteria at higher thresholds (shown by faster declining completeness), those that do meet the criteria have significantly more reliable measurements. This suggests that for applications requiring high precision, such as weak lensing or detailed structure evolution studies, the Bayesian approach with higher thresholds would be preferable, while for applications requiring maximum sample size with reasonable reliability, lower thresholds might be considered.

For quiescent galaxies, the quality of redshift estimation is generally lower than for star-forming galaxies regardless of the method used, reflecting the intrinsic challenge of measuring redshifts for galaxies lacking strong emission features. However, the performance gap between the two scaling approaches is even more pronounced for quiescent galaxies. The Bayesian sampling approach demonstrates remarkably superior performance, achieving higher purity ($60\text{--}100\%$ vs. $50\text{--}60\%$), higher success rate ($60\text{--}45\%$ vs. $50\text{--}45\%$), and notably lower outlier fraction ($40\text{--}0\%$ vs. $45\text{--}40\%$). The precision metrics show an even more dramatic improvement, with $\sigma_{\rm NMAD}$ decreasing from $0.0022\text{--}0.0006$ with the Bayesian approach compared to $0.0055\text{--}0.0016$ with NNLM optimization, and lower bias ($0.001\text{--}0.000$ vs. $0.0012\text{--}0.0006$).

The particularly substantial improvement for quiescent galaxies likely stems from the Bayesian approach's ability to better utilize subtle spectral features such as absorption lines and the 4000Å break. These features, which dominate quiescent galaxy spectra, contain valuable redshift information but require more sophisticated analysis than emission lines. The Bayesian method's comprehensive parameter space exploration is better suited to extract this information, accounting for complex correlations between spectral features, stellar population parameters, and redshift. The NNLM approach, focusing on optimizing a single likelihood point, struggles more with these subtleties, resulting in less reliable measurements.

The trade-off between completeness and precision is evident in both galaxy populations and with both methods. As the acceptance threshold increases, we observe a systematic improvement in measurement quality (increasing purity, decreasing outlier fraction, improving precision) at the expense of sample completeness. This suggests that adopting different thresholds may be optimal for different scientific applications: lower thresholds maximize sample size at the cost of including some less reliable measurements, while higher thresholds yield a smaller but more precisely measured sample.

Based on the comprehensive analysis of quality metrics shown in Figure~\ref{fig:redshift_significance}, we adopt an optimal acceptance threshold of $z_{\rm mean}/\sigma_z > 20$ and the Bayesian approach of scaling determination for all subsequent analyses in this section. This threshold balances sample size considerations with measurement quality, achieving a good compromise between completeness and precision for both galaxy populations.

Using the optimal acceptance threshold and scaling method, we now examine the detailed performance metrics for star-forming and quiescent galaxies (Figure~\ref{fig:redshift_metrics}), focusing on their dependencies on SNR and redshift.

\subsubsection{Dependence of performance on SNR and redshift} \label{sss:results_z_performance}

\begin{figure*}
    \includegraphics[width=\textwidth]{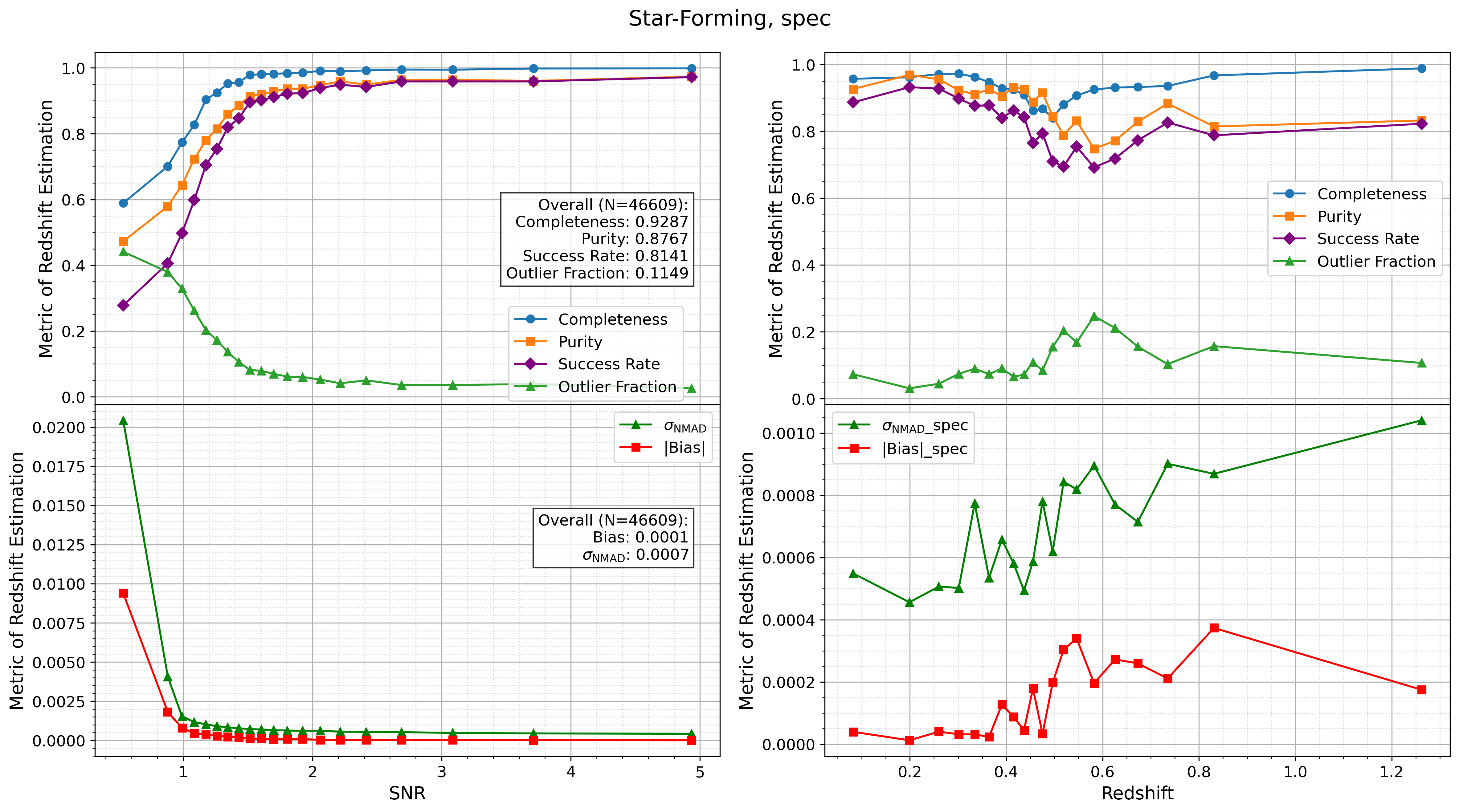}
    \includegraphics[width=\textwidth]{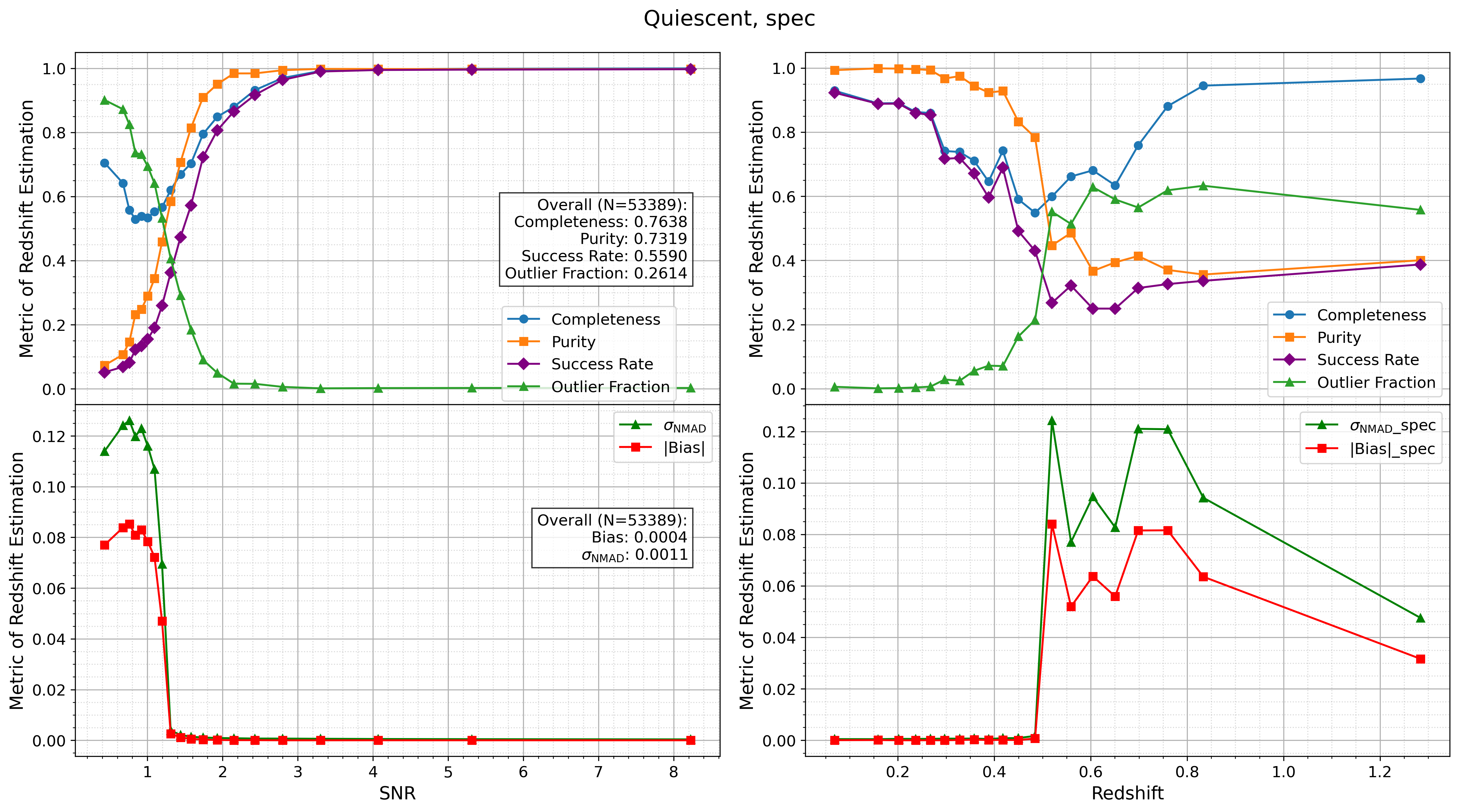}
    \caption{Performance metrics for redshift estimation using CSST slitless spectroscopy for star-forming ($N=46609$; top) and quiescent ($N=53389$; bottom) galaxies. Left panels: SNR dependence; right panels: redshift dependence. \textbf{Star-forming galaxies:} Despite challenging low SNR, performance improves dramatically with increasing SNR. Overall: 92.87\% completeness, 87.67\% purity, 81.41\% success rate, $\sigma_{\rm NMAD}=0.0007$. Performance is stable through $z\approx0.4$ but challenging at $z\approx0.4$-0.7. \textbf{Quiescent galaxies:} Clear SNR threshold at $SNR\sim2$ for reliable estimation. Overall: 76.38\% completeness, 73.19\% purity, 55.90\% success rate, $\sigma_{\rm NMAD}=0.0011$. Performance degrades significantly at $z>0.5$. See Section~\ref{sss:results_z_performance} for detailed discussion.}
    \label{fig:redshift_metrics}
\end{figure*}

\textbf{Star-forming galaxies:} For star-forming galaxies (Figure~\ref{fig:redshift_metrics}, top panels), our analysis reveals strong dependencies on both SNR and redshift. The overall statistics ($N=46609$) demonstrate excellent performance with high completeness (92.87\%), good purity (87.67\%), strong success rate (81.41\%), low outlier fraction (11.49\%), and remarkable precision ($\sigma_{\rm NMAD}=0.0007$, Bias$=0.0001$).

The SNR dependence shows that reliable redshift estimation requires $SNR > 1.5$, above which we observe excellent performance across all metrics. This threshold is particularly important given the SNR distribution of our data (See bottom middle and right panels of Figure~\ref{fig:sample_dist}), where approximately 90\% of GU and 40\% of GV measurements fall below $SNR=1$. Despite these challenging conditions, completeness improves dramatically from $\sim$60\% at $SNR=0.5$ to nearly 100\% at $SNR>2$, while purity rises from $\sim$45\% to $\sim$95\%. The success rate shows similar improvement from $\sim$30\% to $\sim$95\%, and the outlier fraction drops significantly from $\sim$40\% at low SNR to $<$5\% for $SNR>2$. The precision metrics demonstrate substantial improvement with increasing SNR - $\sigma_{\rm NMAD}$ decreases from $\sim$0.02 to $\sim$0.0007, while $|{\rm Bias}|$ reduces from $\sim$0.009 to $\sim$0.0001.

The redshift dependence reveals more complex behavior, particularly in relation to the underlying galaxy distribution (Figure~\ref{fig:sample_dist}). Performance remains stable through the peak of galaxy distribution at $z \approx 0.4$, where we have the highest number density of sources. A more challenging region emerges at $z \approx 0.4$-0.7, coinciding with the rapid decline in galaxy numbers, where completeness and success rates show modest decline and the outlier fraction peaks at $\sim$25\%. Beyond $z>0.7$, where very few galaxies exist in our sample, the metrics stabilize with completeness and success rates maintaining $\sim$90\%, though precision gradually degrades ($\sigma_{\rm NMAD}$ increases from $\sim$0.0004 to $\sim$0.001).

These results demonstrate that our method provides highly reliable redshift measurements for star-forming galaxies, particularly when $SNR>1.5$, though special attention should be paid to observations in the $z \approx 0.4$-0.7 range where performance shows some degradation.

\textbf{Quiescent galaxies:} For quiescent galaxies (Figure~\ref{fig:redshift_metrics}, bottom panels), our analysis reveals a clear demarcation in performance across different SNR regimes. For $SNR<$1, redshift estimation proves highly challenging, with completeness ranging from 50\% to 70\%, very low purity (5-35\%), and an extremely high outlier fraction (70-90\%). The precision metrics in this low SNR regime are particularly poor, with $\sigma_{\rm NMAD}$ values of 0.1-0.12 and bias of 0.07-0.08, indicating substantial systematic and random errors.

We observe a dramatic improvement in performance for measurements with $SNR>1$ and $SNR<2$. In this intermediate regime, completeness rises significantly from $\sim$55\% to $\sim$85\%, purity improves from $\sim$35\% to $\sim$95\%, and the outlier fraction drops dramatically from $\sim$70\% to $\sim$5\%. The improvement is even more pronounced for $SNR>2$, where we achieve excellent performance with completeness of 90-100\%, purity of 95-100\%, and an outlier fraction of just 0-1\%. The precision metrics in this high SNR regime are exceptional, with $\sigma_{\rm NMAD}$ of $\sim$0.001 and bias of $\sim$0.001.

The redshift dependence reveals additional challenges, particularly at higher redshifts. While performance is strong at low redshifts ($z<0.5$), we observe significant degradation beyond $z>0.5$ with increasing outlier fraction and decreasing success rate. The overall statistics demonstrate moderate completeness (76.38\%), reasonable purity (73.19\%), modest success rate (55.90\%), and a significant outlier fraction (26.14\%). When measurements can be obtained, we achieve good precision with a bias of 0.0004 and $\sigma_{\rm NMAD}$ of 0.0011. These results highlight the importance of $SNR>2$ for reliable redshift estimation of quiescent galaxies, particularly at higher redshifts where the absence of strong emission features makes measurements more challenging.

\subsection{Stellar Mass} \label{ss:results_mass}

Having analyzed redshift measurements, we now examine the performance of stellar mass estimation, which is determined jointly with redshift and other physical parameters through Bayesian full spectrum analysis. Following the same methodology as before, we examine both star-forming and quiescent galaxies across different measurement acceptance thresholds, SNR ranges, and redshifts.

\subsubsection{Dependence of overall performance on Acceptance Threshold} \label{sss:results_mass_threshold}

\begin{figure*}
    \includegraphics[width=\textwidth]{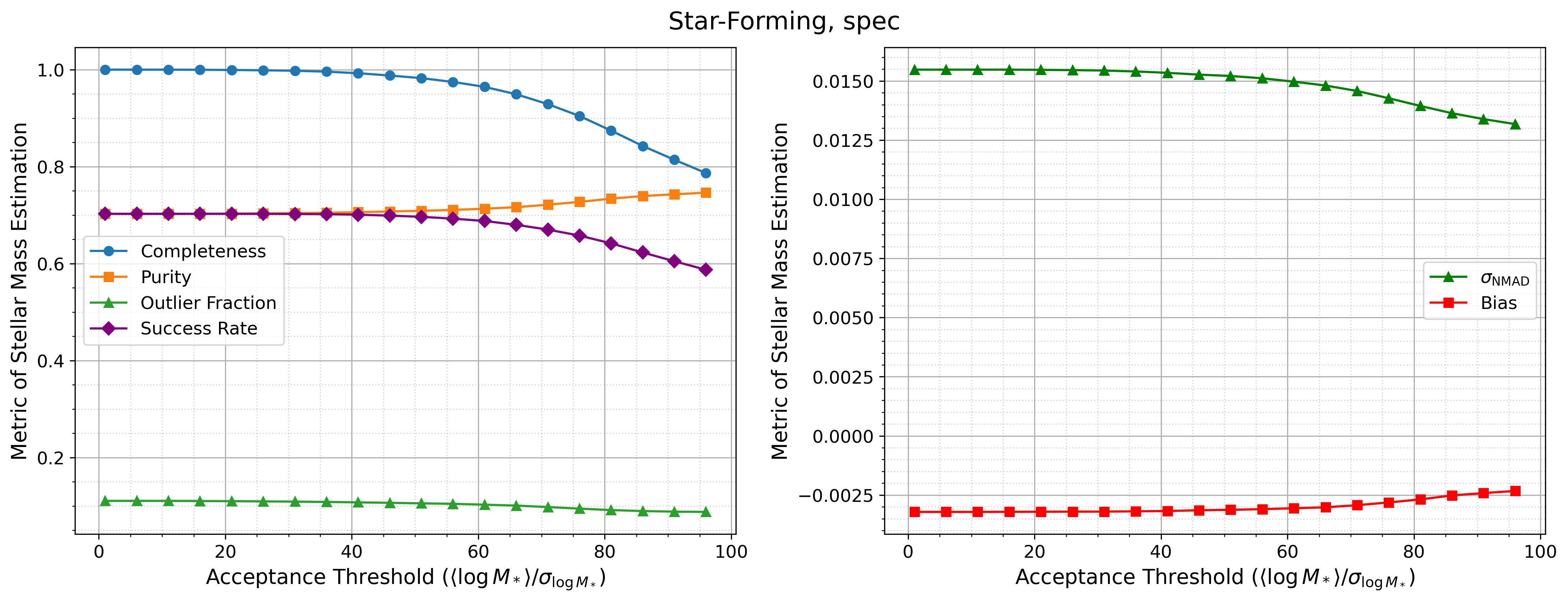}
    \includegraphics[width=\textwidth]{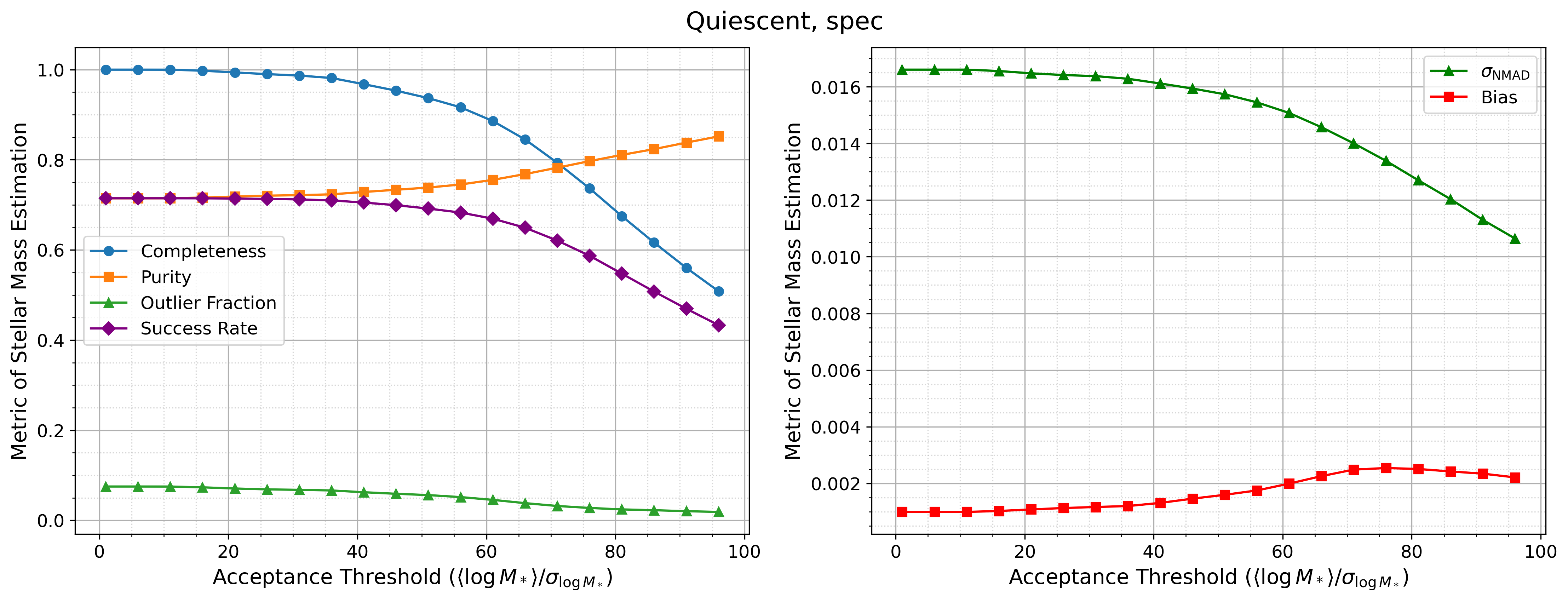}
    \caption{Quality assessment of stellar mass estimation as a function of acceptance threshold ($\langle\log M_*\rangle/\sigma_{\log M_*}$) for star-forming (top) and quiescent (bottom) galaxies using CSST slitless spectroscopy. Left panels show quality metrics (completeness, purity, outlier fraction, success rate); right panels show precision metrics ($\sigma_{\rm NMAD}$, Bias). For star-forming galaxies, purity remains stable while completeness gradually decreases. For quiescent galaxies, purity improves substantially while completeness drops more steeply. An optimal threshold of $\langle\log M_*\rangle/\sigma_{\log M_*} > 50$ is adopted, achieving $\sim$70\% success rate for both populations (see Section~\ref{sss:results_mass_threshold} for details).}
    \label{fig:mstar_significance}
\end{figure*}

We examine how the quality of stellar mass measurements varies with the acceptance threshold ($\langle\log M_*\rangle/\sigma_{\log M_*}$) for both star-forming and quiescent galaxies (Figure~\ref{fig:mstar_significance}). The analysis reveals distinct behaviors between the two galaxy populations, highlighting the importance of appropriate threshold selection for reliable mass estimation.

For star-forming galaxies, increasing the acceptance threshold leads to a gradual decline in completeness from nearly 100\% to about 80\%, while maintaining stable purity at approximately 70-75\%. The outlier fraction remains consistently low at about 10\%, indicating inherently robust mass estimation. The success rate closely follows the completeness trend, suggesting that measurement availability, rather than reliability, is the primary limitation. The precision metrics show clear improvement with increasing threshold - $\sigma_{\rm NMAD}$ decreases slightly from about 0.015 to 0.013, while the bias remains consistently low and negative at around -0.0025, demonstrating good control of systematic errors.

In contrast, quiescent galaxies exhibit notably different behavior. The completeness shows a much steeper decline, dropping from 100\% to approximately 55\% as the threshold increases. However, this is accompanied by a substantial improvement in purity, which rises steadily from 70\% to 85\%. The outlier fraction remains very low (below 8\%) across most threshold values, indicating highly reliable measurements when they can be obtained. The success rate decreases significantly with increasing threshold, primarily due to the sharp drop in completeness. The precision metrics show more dramatic improvement - $\sigma_{\rm NMAD}$ decreases from about 0.016 to 0.011, while the bias remains well-controlled below 0.003.

Based on these results, we adopt an optimal acceptance threshold of $\langle\log M_*\rangle/\sigma_{\log M_*} > 50$ for our analysis. This choice balances success rate with measurement quality improvement. At this threshold, both star-forming and quiescent galaxies maintain high success rates ($\sim$70\%), with excellent completeness ($\sim$100\% for star-forming, $\sim$95\% for quiescent galaxies) and good purity ($\sim$70\% for star-forming, $\sim$75\% for quiescent galaxies). The precision metrics at our chosen threshold ($\sigma_{\rm NMAD} \approx 0.015$ for star-forming and $\sigma_{\rm NMAD} \approx 0.016$ for quiescent galaxies) show meaningful improvement over lower thresholds while retaining large enough samples for robust statistical analyses.

\subsubsection{Dependence of performance on SNR and redshift} \label{sss:results_mass_performance}

\begin{figure*}
    \includegraphics[width=\textwidth]{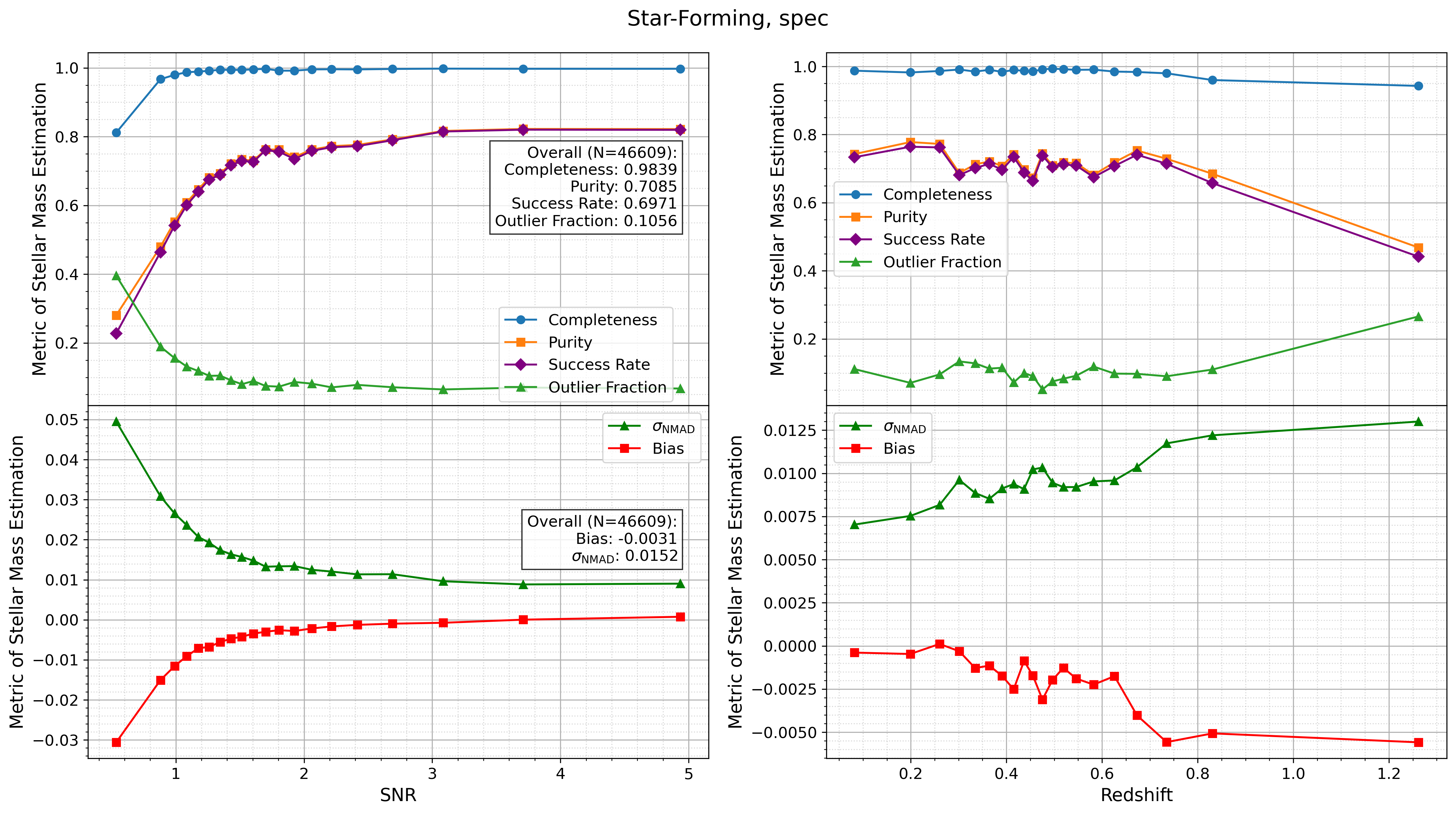}
    \includegraphics[width=\textwidth]{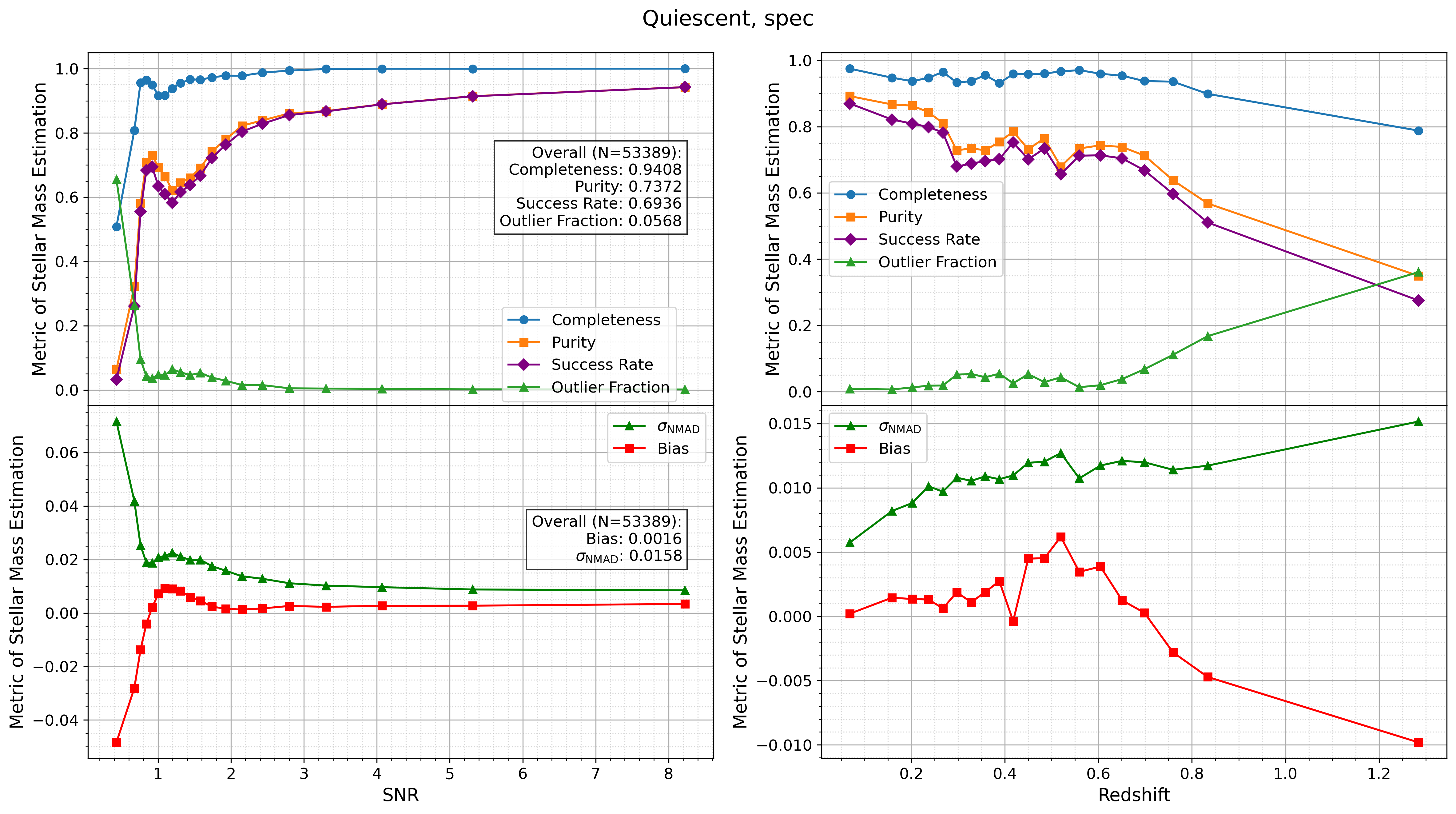}
    \caption{Performance metrics for stellar mass estimation using CSST slitless spectroscopy for star-forming (top panels, $N=46609$) and quiescent (bottom panels, $N=53389$) galaxies. Left panels: SNR dependence; Right panels: redshift dependence. \textbf{Star-forming galaxies}: Overall: 98.39\% completeness, 70.85\% purity, 69.71\% success rate, $\sigma_{\rm NMAD} = 0.0152$. Performance transition at $SNR\sim2$; stable up to $z\approx0.6$. \textbf{Quiescent galaxies}: Overall: 94.08\% completeness, 73.72\% purity, 69.36\% success rate, $\sigma_{\rm NMAD} = 0.0158$. Critical SNR threshold at $\sim$1.5; performance degrades beyond $z\approx0.6$. See Section~\ref{sss:results_mass_performance} for detailed discussion.}
    \label{fig:mstar_metrics}
\end{figure*}

For star-forming galaxies (N=46609, Figure~\ref{fig:mstar_metrics}, top panels), our analysis reveals strong dependencies on both SNR and redshift. The overall statistics demonstrate excellent performance with very high completeness (98.39\%), good purity (70.85\%), a strong success rate (69.71\%), and a moderate outlier fraction (10.56\%), with high precision indicated by $\sigma_{\rm NMAD} = 0.0152$ and bias = -0.0031. These metrics confirm our method's ability to reliably estimate stellar masses for a large fraction of star-forming galaxies.

The SNR dependence shows a clear performance transition around $SNR \approx 2$. For $SNR<2$, we see a sharp improvement in all metrics as signal strength increases. Completeness rises from 20\% to 99\%, purity from 25\% to 75\%, and the success rate climbs from 20\% to 75\%. Critically, the outlier fraction drops dramatically from 40\% to just 5\%, indicating that catastrophic failures are largely overcome with modest SNR. In the high SNR regime ($SNR>2$), performance stabilizes at an optimal level, with all metrics holding steady, suggesting that further increases in SNR yield diminishing returns.

The redshift dependence reveals two distinct phases. Up to $z \approx 0.6$, performance is remarkably stable, with completeness at ~98\% and both purity and success rate holding steady at ~65-75\%. This stability indicates the method is robust across the nearby universe. Beyond $z \approx 0.6$, we observe a gradual decline, with purity and success rate dropping to ~45\% by $z \approx 1.3$. This degradation is expected as key spectral features redshift out of the observed bands.

For quiescent galaxies (N=53389, Figure~\ref{fig:mstar_metrics}, bottom panels), our analysis also reveals distinct dependencies. The overall statistics show high completeness (94.08\%), good purity (73.72\%), a strong success rate (69.36\%), and a very low outlier fraction (5.68\%), with high precision ($\sigma_{\rm NMAD} = 0.0158$, bias = 0.0015). These metrics indicate that when we do obtain a measurement, it is highly reliable.

The SNR dependence exhibits a dramatic performance threshold around $SNR \approx 1.5$. Below this value, all metrics improve rapidly with signal strength; for instance, completeness rises from as low as 50\% to over 90\%, while the outlier fraction plummets from 65\% to 5\%. This highlights a critical SNR floor for reliable measurements. For $SNR>2$, performance is excellent and sustained, with completeness near 99\% and purity stable at ~80\%, indicating that measurements in this regime are extremely robust.

The redshift dependence shows a clear transition in performance. Up to $z \approx 0.6$, performance is strong and stable, with ~98\% completeness and purity between 65-80\%. This defines an ideal redshift range for studying quiescent galaxies with this method. Beyond $z \approx 0.6$, we see a significant decline, with purity and success rates dropping to ~30\% by $z \approx 1.3$. This decline underscores the challenges of observing older stellar populations at high redshift.

An important systematic effect emerges when examining the bias as a function of both SNR and redshift for both galaxy populations (Figure~\ref{fig:mstar_metrics}, right panels). The bias becomes systematically negative only at $SNR < 1$, and this low-SNR regime is precisely where both bias magnitude and scatter increase with redshift. This behavior suggests that our Gaussian noise assumption (Section~\ref{ss:likelihood}) may not be fully valid at very low SNR, where non-Gaussian noise properties and instrumental systematics become important. Violations of the Gaussian assumption can lead to systematic biases in likelihood evaluation, with these errors being amplified when data provide weak constraints. The apparent correlation between bias and redshift primarily reflects the underlying correlation between SNR and redshift in our sample: higher-redshift galaxies systematically have lower SNR (Figure~\ref{fig:sample_dist}), with an increasing fraction of measurements falling below $SNR=1$ at higher redshift. This finding underscores the importance of developing non-Gaussian noise models for very low SNR data, as mentioned in Section~\ref{ss:likelihood}.

In summary, our method provides reliable stellar mass estimates for both star-forming and quiescent galaxies, with performance becoming highly robust for $SNR>1.5$. The clear thresholds in both SNR and redshift help define optimal conditions for mass estimation, while the systematic bias at very low SNR highlights areas for methodological improvement.

\subsection{Star Formation Rate} \label{ss:results_sfr}

Finally, we examine the estimation of star formation rates, which complements our stellar mass analysis in characterizing galaxy populations. As with previous properties, we analyze the acceptance threshold dependence and performance metrics, though focusing primarily on star-forming galaxies where SFR measurements are most relevant.

\begin{figure*}[ht!]
    \centering
    \includegraphics[width=\textwidth]{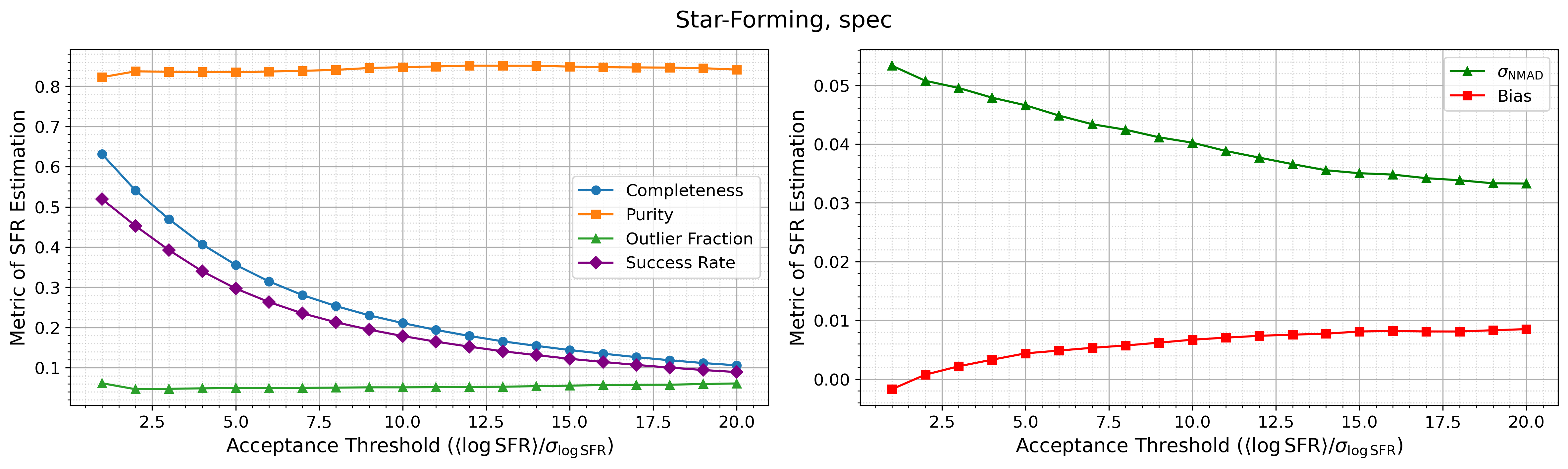} \\
    \includegraphics[width=\textwidth]{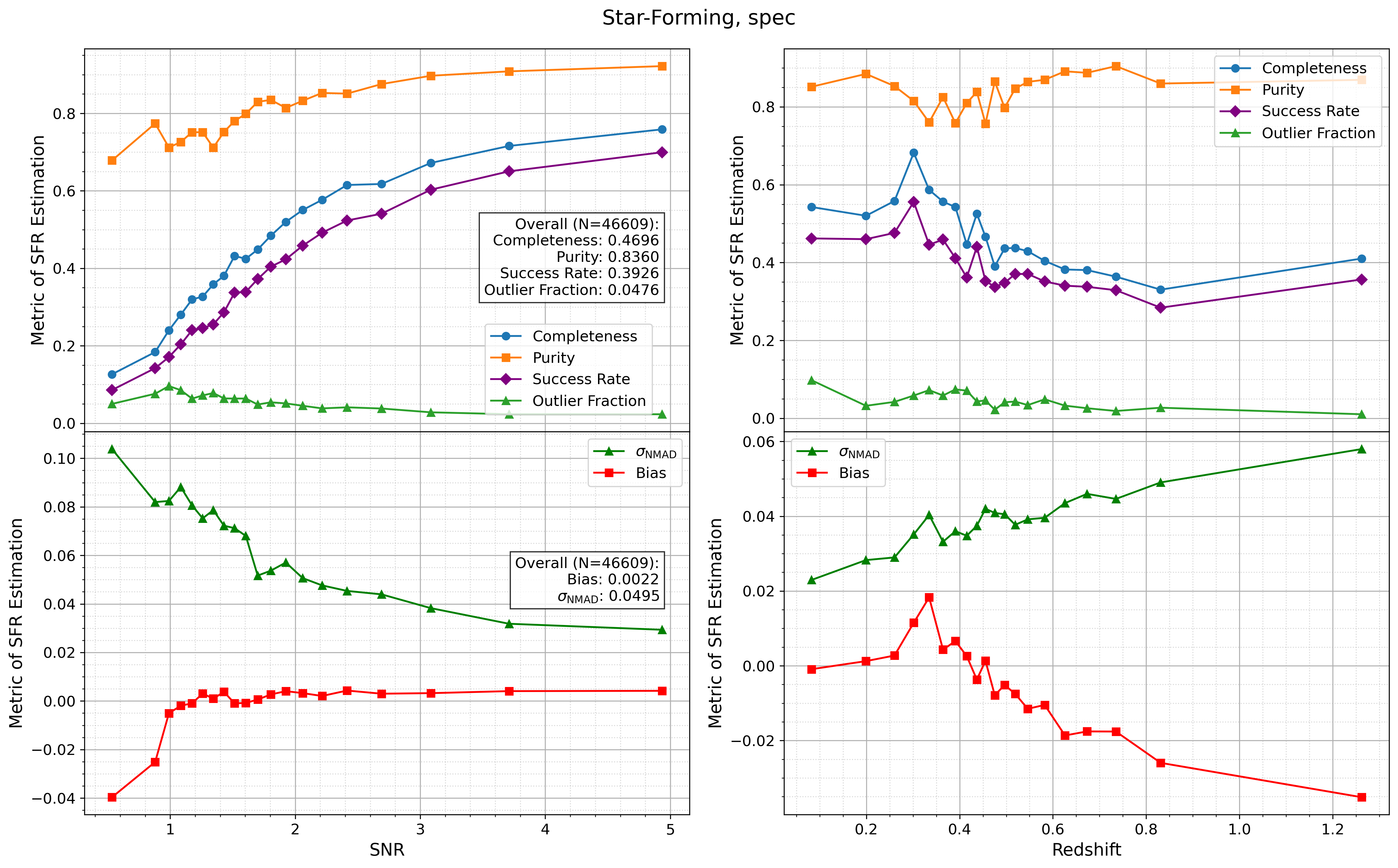}
    \caption{Performance evaluation for star formation rate (SFR) estimation of star-forming galaxies using CSST slitless spectroscopy. \textbf{Top:} Quality assessment (completeness, purity, outlier fraction, success rate) and precision metrics ($\sigma_{\rm NMAD}$, Bias) as a function of acceptance threshold ($\langle\log {\rm SFR}\rangle/\sigma_{\log {\rm SFR}}$). \textbf{Bottom:} Performance metrics ($N=46609$) showing dependence on SNR (left) and redshift (right). Performance significantly improves for $SNR>1$, with optimal results at $SNR>3$. An optimal acceptance threshold of $\langle\log {\rm SFR}\rangle/\sigma_{\log {\rm SFR}} > 3$ is adopted, achieving overall $\sigma_{\rm NMAD}$ = 0.0495 and success rate = 0.3926. See Section~\ref{sss:results_sfr_threshold} and Section~\ref{sss:results_sfr_performance} for detailed discussion.}
    \label{fig:sfr_performance}
\end{figure*}

\subsubsection{Dependence of overall performance on Acceptance Threshold} \label{sss:results_sfr_threshold}

We examine how the quality of star formation rate measurements varies with the acceptance threshold ($\langle\log {\rm SFR}\rangle/\sigma_{\log {\rm SFR}}$) for star-forming galaxies, as shown in the top panels of Figure~\ref{fig:sfr_performance}. Our analysis reveals clear trade-offs between measurement quality and sample completeness as the threshold increases.
The completeness shows a steady decline with increasing threshold, starting at approximately 65\% for low thresholds and dropping steadily to about 10\% as the threshold increases from 1 to 20. This substantial reduction in sample size indicates that very stringent uncertainty requirements significantly limit the number of acceptable measurements.

The purity demonstrates relatively stable behavior that provides insight into measurement reliability. Starting at around 82\%, it increases to a peak of approximately 85\% around threshold values of 12, then remains relatively stable around 83-85\% even at high thresholds. This pattern suggests that our SFR measurements maintain consistent reliability across different threshold selections.

The success rate closely follows the completeness trend, dropping from about 52\% to 10\%, while the outlier fraction remains relatively stable around 5-6\% across most threshold values. This low and stable outlier fraction indicates good overall measurement quality with few catastrophic failures.

The precision metrics show encouraging trends with increasing threshold. The normalized median absolute deviation ($\sigma_{\rm NMAD}$) improves from approximately 0.055 to 0.033, with most improvement occurring at lower thresholds. The bias remains near zero throughout, ranging from approximately -0.002 to 0.010, showing excellent calibration across the full range of thresholds.

As for the stellar mass case, we adopt an optimal acceptance threshold of $\langle\log {\rm SFR}\rangle/\sigma_{\log {\rm SFR}} > 3$ for our follow-up analysis. This choice balances the competing demands of sample size and measurement quality. At this threshold, we maintain reasonable completeness while achieving good precision ($\sigma_{\rm NMAD} \approx 0.035$) and minimal bias (near zero).

\subsubsection{Dependence of performance on SNR and redshift} \label{sss:results_sfr_performance}

The dependence of SFR estimation performance on SNR and redshift for star-forming galaxies ($N=46609$) is detailed in the bottom panels of Figure~\ref{fig:sfr_performance}. The overall statistics demonstrate moderate completeness (0.4696), good purity (0.8360), modest success rate (0.3926), and low outlier fraction (0.0476), with $\sigma_{\rm NMAD}$ = 0.0495 and bias = 0.0022. These metrics indicate that while SFR estimation is more challenging than stellar mass estimation, the measurements we obtain are generally reliable.

The SNR dependence exhibits three distinct regimes of performance. In the low SNR regime ($SNR<1$), we observe limited but improving performance with increasing SNR. Completeness starts at about 10\% and rises to approximately 25\%, while purity shows more encouraging values, improving from 65\% to 70\%. The success rate remains low but increases from 10\% to 20\%, and the outlier fraction fluctuates around 5-10\%. The precision metrics show significant improvement even in this challenging regime, with $\sigma_{\rm NMAD}$ decreasing from 0.10 to 0.03 and bias improving from -0.04 to near zero.

The intermediate SNR regime ($1<SNR<3$) demonstrates substantial improvement across all metrics. Completeness shows a dramatic increase from 25\% to 65\%, while purity rises from 70\% to an excellent level of approximately 90\%. The success rate shows corresponding improvement from 20\% to 60\%, and the outlier fraction decreases to a stable value of about 3-5\%. Precision continues to improve, with $\sigma_{\rm NMAD}$ reaching approximately 0.03 and bias remaining near zero.

In the high SNR regime ($SNR>3$), we observe continued improvement in most metrics. Completeness reaches approximately 75\%, purity maintains around 90\%, and the success rate climbs to around 70\%. The outlier fraction remains stable at about 2-3\%, suggesting we've reached a fundamental limit in our ability to eliminate catastrophic failures. The precision metrics maintain their excellent performance, indicating that we've achieved optimal measurement quality in this regime.

The redshift dependence reveals a moderate decline in performance with increasing distance. At low redshift ($z\approx0.2$), we achieve completeness and success rates of approximately 55\%, which decrease to about 40\% by $z\approx1.2$. Notably, purity remains consistently high at 85-90\% across all redshifts, indicating that while fewer measurements can be obtained at high redshift, those we do obtain remain reliable. The precision metrics show gradual degradation with redshift, with $\sigma_{\rm NMAD}$ increasing from about 0.025 to 0.06 and bias ranging from near zero at low redshift to -0.04 at $z\approx1.2$. This degradation likely reflects both the decreasing SNR and the reduced availability of key spectral features for SFR estimation at higher redshifts.

These comprehensive results demonstrate that our method can provide reliable SFR estimates for star-forming galaxies, particularly when $SNR>1$. The clear dependence on both SNR and redshift helps define optimal conditions for SFR estimation and understand the limitations of our methodology. While the overall completeness is lower than for stellar mass estimation, the high purity and well-controlled precision metrics make this approach valuable for statistical studies of galaxy evolution, particularly at lower redshifts where we achieve the best performance.

\section{Discussion} \label{sec:disc}

\subsection{Effects of Self-blending and Overlapping} \label{ss:disc_blend_overlap}

\begin{figure*}[ht!]
    \centering
    \begin{tabular}{c}
        \includegraphics[width=0.95\textwidth]{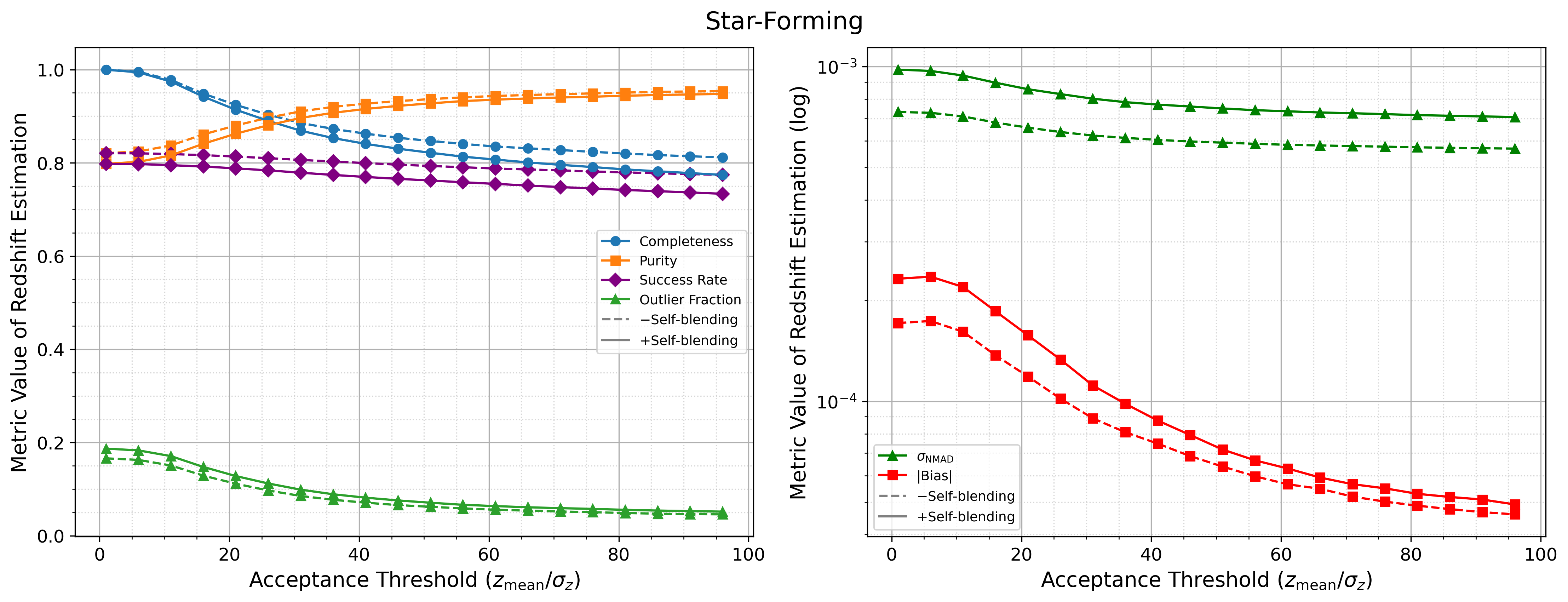} \\
        \includegraphics[width=0.95\textwidth]{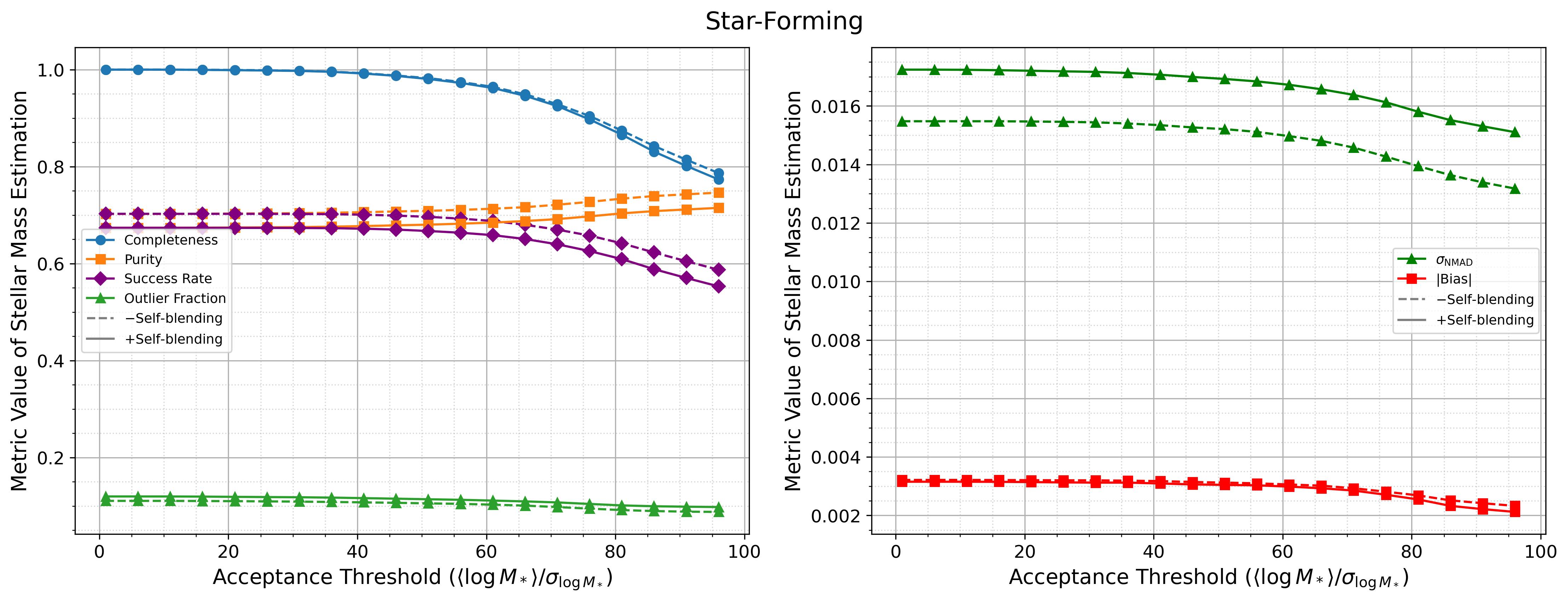} \\
        \includegraphics[width=0.95\textwidth]{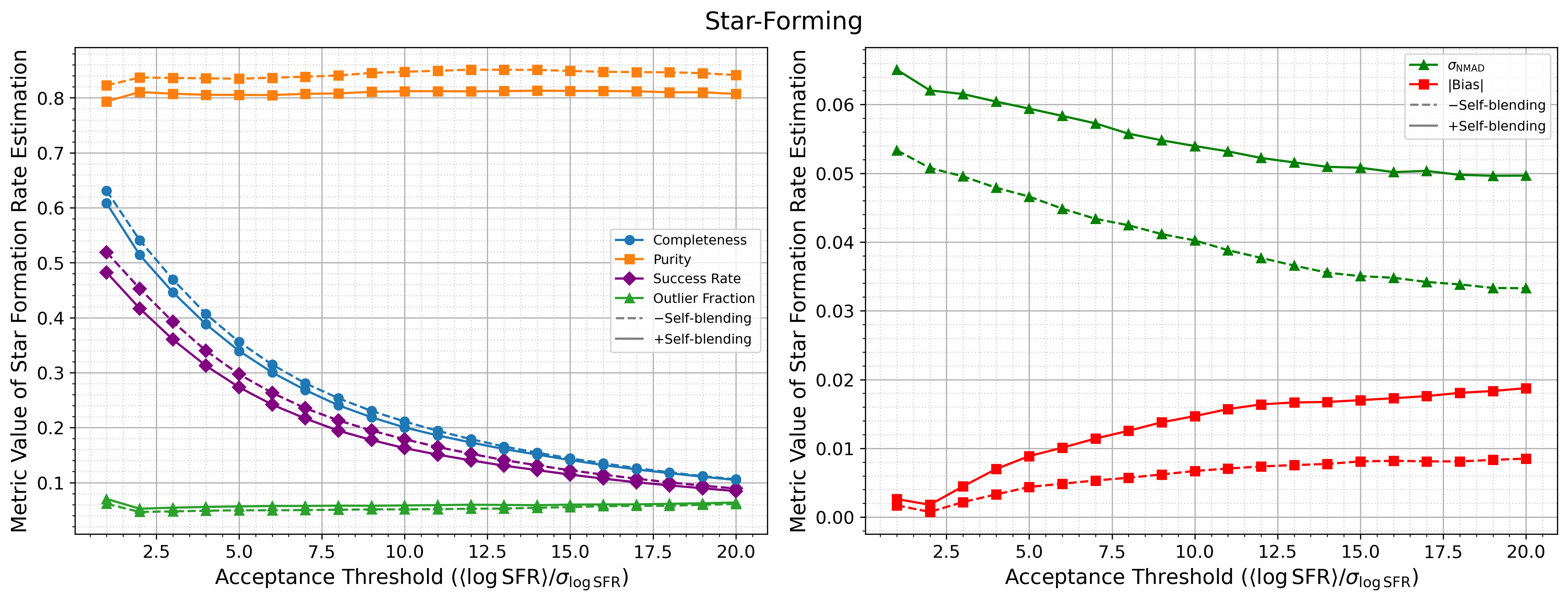} \\
    \end{tabular}
    \caption{Comparison of parameter estimation performance with (solid lines) and without (dashed lines) self-blending effects for star-forming galaxies. Each row shows results for redshift (top), stellar mass (middle), and star formation rate (bottom). Left panels show quality metrics versus acceptance threshold; right panels show precision metrics ($\sigma_{\rm NMAD}$, $|{\rm Bias}|$). Self-blending primarily impacts $\sigma_{\rm NMAD}$ by $\gtrsim$30\%, with larger effects on SFR than on stellar mass and redshift. Even with self-blending, high-quality estimates are achievable at optimal thresholds (see Section~\ref{ss:disc_blend_overlap} for details).}
    \label{fig:significance_comparison_sf}
\end{figure*}

\begin{figure*}[ht!]
    \centering
    \begin{tabular}{c}
        \includegraphics[width=0.95\textwidth]{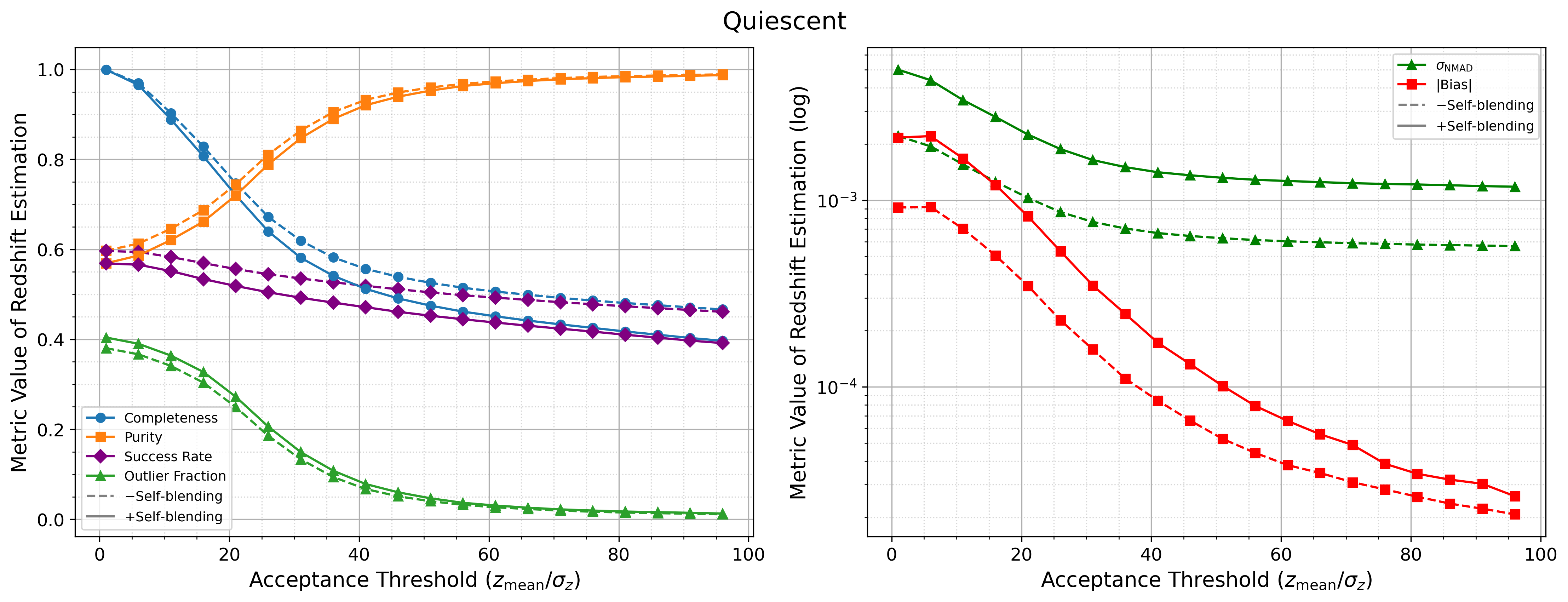} \\
        \includegraphics[width=0.95\textwidth]{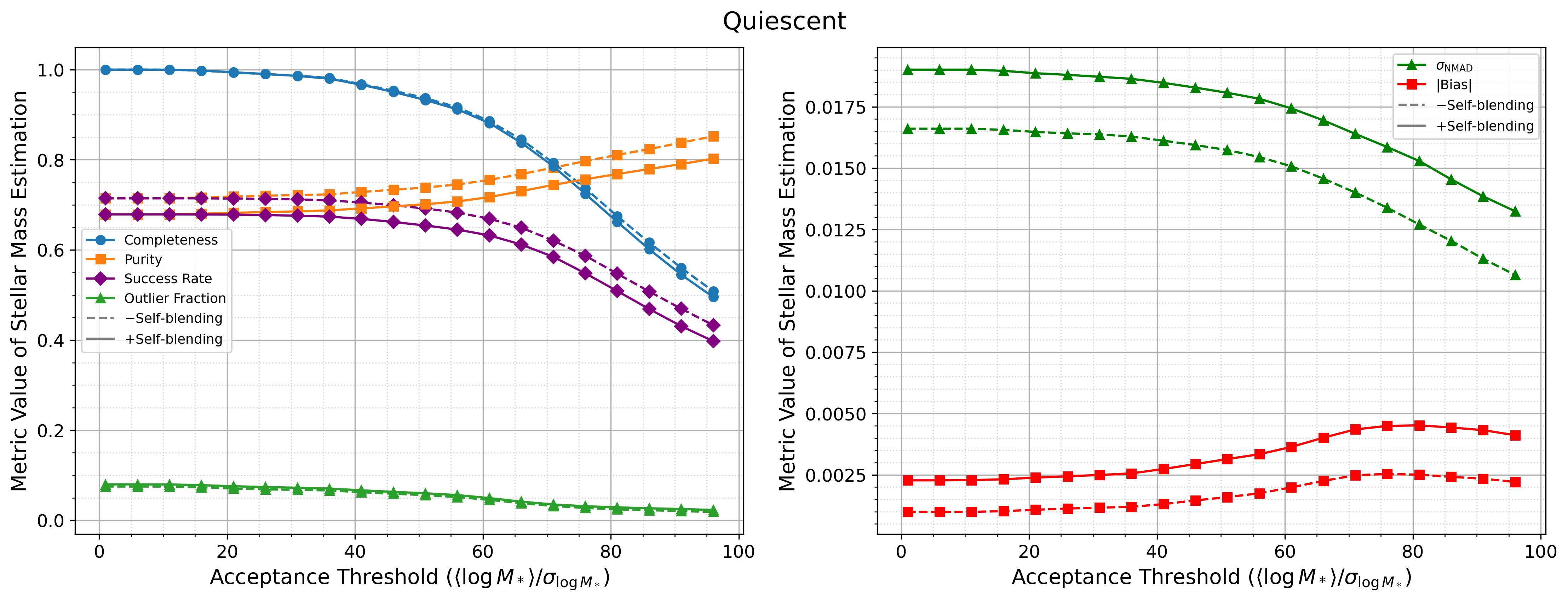} \\
    \end{tabular}
    \caption{As Figure~\ref{fig:significance_comparison_sf}, but for quiescent galaxies. Self-blending impacts $\sigma_{\rm NMAD}$ by $\gtrsim$30\% for both redshift and stellar mass estimation. Even with self-blending, reliable parameter estimates are achievable at optimal acceptance thresholds (see Section~\ref{ss:disc_blend_overlap} for details).}
    \label{fig:significance_comparison_quiescent}
\end{figure*}

In slitless spectroscopy, self-blending occurs when different parts of an extended galaxy have their dispersed spectra overlapping with each other. This inherent effect can distort spectral features and affect parameter estimation quality. To quantify the impact of spectral self-blending, we conducted a comparative analysis between simulations with and without this effect included in the CSST slitless spectroscopy mock data (see Figures~\ref{fig:significance_comparison_sf} and \ref{fig:significance_comparison_quiescent}).

Generally, self-blending primarily impacts the precision metrics, particularly $\sigma_{\rm NMAD}$ which could be increased by more than 30\% across all parameters and galaxy types, while having a less pronounced effect on quality metrics such as purity and success rate. For star-forming galaxies (Figure~\ref{fig:significance_comparison_sf}), the effect is most noticeable for star formation rate estimation, followed by stellar mass, and then redshift. At optimal acceptance thresholds, even with self-blending considered, we can achieve: for redshift (at $\langle z\rangle/\sigma_z = 30$), $\sigma_{\rm NMAD} = 0.0008$, $|{\rm Bias}| = 0.0001$, purity of $\sim95\%$, and a success rate of $\sim80\%$; for stellar mass (at $\langle\log M_*\rangle/\sigma_{\log M_*} = 60$), $\sigma_{\rm NMAD} = 0.017$, $|{\rm Bias}| = 0.003$, purity of $\sim70\%$, and a success rate of $\sim65\%$; and for SFR (at $\langle\log {\rm SFR}\rangle/\sigma_{\log {\rm SFR}} = 3$), $\sigma_{\rm NMAD} = 0.06$, $|{\rm Bias}| = 0.005$, purity of $\sim80\%$, and a success rate of $\sim35\%$. For quiescent galaxies (Figure~\ref{fig:significance_comparison_quiescent}), self-blending also impacts precision. At optimal thresholds, we find: for redshift (at $\langle z\rangle/\sigma_z = 25$), $\sigma_{\rm NMAD} = 0.0015$, $|{\rm Bias}| = 0.0005$, purity of $\sim80\%$, and a success rate of $\sim50\%$; for stellar mass (at $\langle\log M_*\rangle/\sigma_{\log M_*} = 60$), $\sigma_{\rm NMAD} = 0.017$, $|{\rm Bias}| = 0.0026$, purity of $\sim70\%$, and a success rate of $\sim65\%$. These results demonstrate that while self-blending introduces additional uncertainty, particularly affecting bias and scatter, reasonably accurate parameter estimation is still achievable.

While our current analysis focuses on self-blending, source overlapping (where spectra from different galaxies contaminate each other) presents additional challenges that require separate consideration. This effect becomes particularly significant in crowded fields and at higher redshifts where source density increases. Analysis from \citet{WenR2024a} shows that in galaxy clusters, the overlap rate (fraction of galaxies affected by contamination) increases from 55\% to 80\% as cluster richness increases from 10 to 150 members. The contamination can be classified into three levels: mild ($<$10\% flux contamination), moderate (10-40\% contamination), and severe ($>$40\% contamination). In rich clusters, while mild contamination remains relatively stable at about 35\%, the severe contamination fraction increases significantly from 10\% to 25\%, indicating substantial degradation in measurement quality. These estimates are likely conservative, as they only consider contamination from zero and first-order spectra of cataloged galaxies, not accounting for higher-order spectra, stellar sources, or intracluster light. Future work will focus on developing techniques to mitigate both self-blending and source overlapping effects in CSST slitless spectroscopy, with particular attention to dense environments where overlapping effects dominate measurement uncertainties.

\subsection{Combined analysis of all slitless spectroscopic bands and all photometric bands} \label{ss:disc_specphot_all}

CSST provides both slitless spectroscopy and photometric observations for galaxy studies. A key question is whether combining the two types of data offers significant advantages over spectroscopy alone. The complementary nature of these observations - with spectroscopy providing detailed spectral features and photometry offering broader wavelength coverage - suggests potential synergies. However, the actual benefit of combining these data depends on various factors, including the galaxy type, SNR, the specific parameters being estimated, and critically, the methodology used to combine the different data types.

In the ideal but least common scenario, all three CSST slitless spectroscopic bands (GU, GV, and GI) would be available. This optimal situation, though rare in practice due to observational constraints and limited telescope time, provides a baseline for understanding the maximum potential benefit of combining spectroscopic and photometric data. Using our Bayesian framework, we systematically evaluate the impact of adding photometric data to our spectroscopic analysis by comparing the performance of spectroscopy-only measurements with combined spectroscopy+photometry across different galaxy types and parameters.

Our analysis reveals distinct patterns in how the addition of photometric data affects parameter estimation for different galaxy populations. For star-forming galaxies, we find that combining spectroscopic and photometric data yields largely similar results to using spectroscopy alone. The quality metrics (completeness, purity, outlier fraction, success rate) and $\sigma_{\rm NMAD}$ show nearly identical behavior between the two approaches across all key parameters---redshift, stellar mass, and star formation rate. While the addition of photometric data does provide some reduction in systematic bias (e.g., for stellar mass, $|{\rm Bias}|$ decreases from $\sim$0.003 to $\sim$0.0018), the bias values are already negligibly small in both cases, indicating excellent systematic error control with spectroscopy alone. These minimal differences suggest that for star-forming galaxies, spectroscopic data alone captures most of the information necessary for reliable parameter estimation.

In contrast, quiescent galaxies show a more nuanced response to the addition of photometric data. For redshift estimation, the combined spectroscopy+photometry approach demonstrates significant improvements in overall performance, particularly when the acceptance threshold is below 40. At the optimal acceptance threshold of ${\rm z}_{\rm mean}/\sigma_{\rm z} = 20$, the combined approach achieves a redshift estimation quality characterized by $\sigma_{\rm NMAD} \sim 0.0016$, $|{\rm Bias}| \sim 0.0002$, a success rate of $\sim55\%$, purity of $\sim90\%$, and an outlier fraction of $\sim10\%$ (Figure~\ref{fig:significance_comparison_quiescent_phot}). This represents a notable improvement compared to the spectroscopy-only case (Figure~\ref{fig:significance_comparison_quiescent}), which achieves similar precision ($\sigma_{\rm NMAD} = 0.0016$) but slightly higher bias ($|{\rm Bias}| \sim 0.0005$), a lower success rate of $\sim50\%$, lower purity of $\sim80\%$, and a higher outlier fraction of $\sim20\%$ at its optimal threshold (${\rm z}_{\rm mean}/\sigma_{\rm z} = 25$). However, for stellar mass estimation, both approaches perform nearly identically across all quality metrics ($\sigma_{\rm NMAD}$, completeness, purity, success rate, outlier fraction). While adding photometric data shows a slight increase in systematic bias, the bias values remain small in both cases (typically $<$0.007 in log stellar mass), indicating good systematic error control regardless of whether photometry is included. This suggests that for quiescent galaxies, while photometric data can significantly enhance redshift determinations, spectroscopic data alone could be sufficient for stellar mass estimation. This differential impact of photometric data on different parameters highlights the importance of carefully considering which observational data to combine based on both the galaxy type and the specific parameter being estimated.

\begin{figure*}[ht!]
    \centering
    \begin{tabular}{c}
        \includegraphics[width=0.95\textwidth]{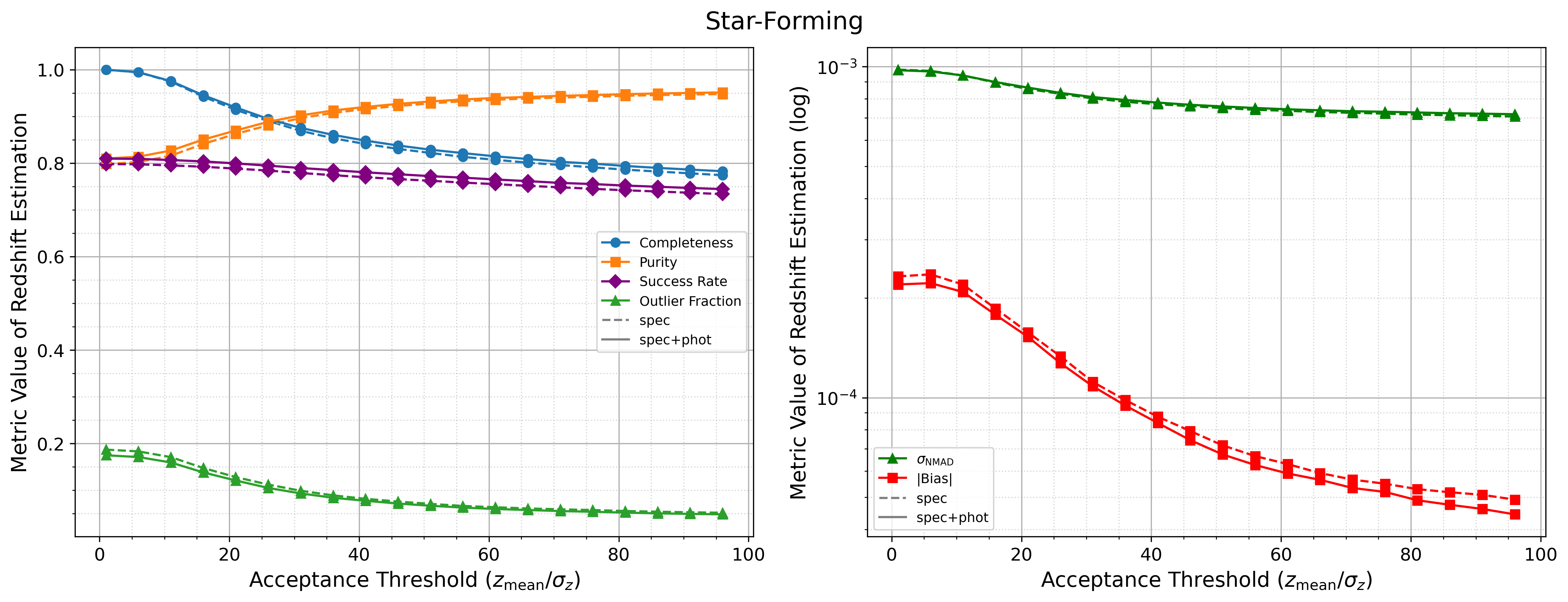} \\
        \includegraphics[width=0.95\textwidth]{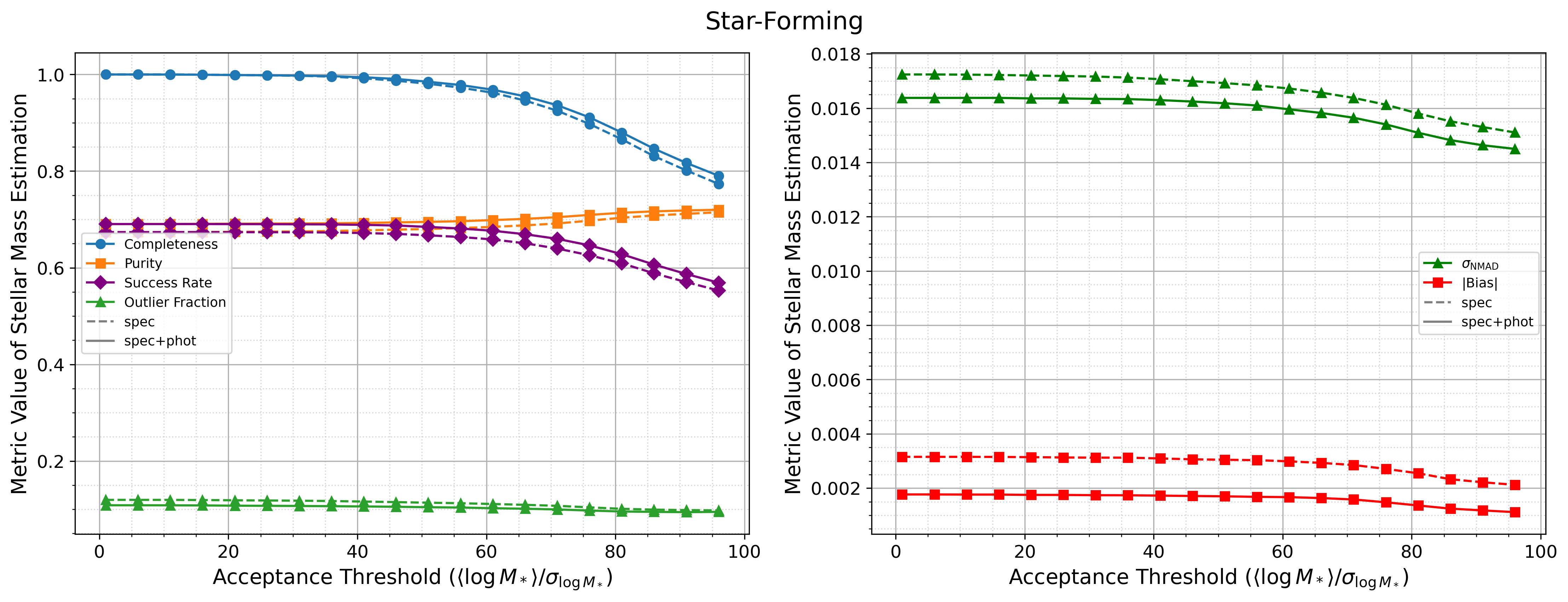} \\
        \includegraphics[width=0.95\textwidth]{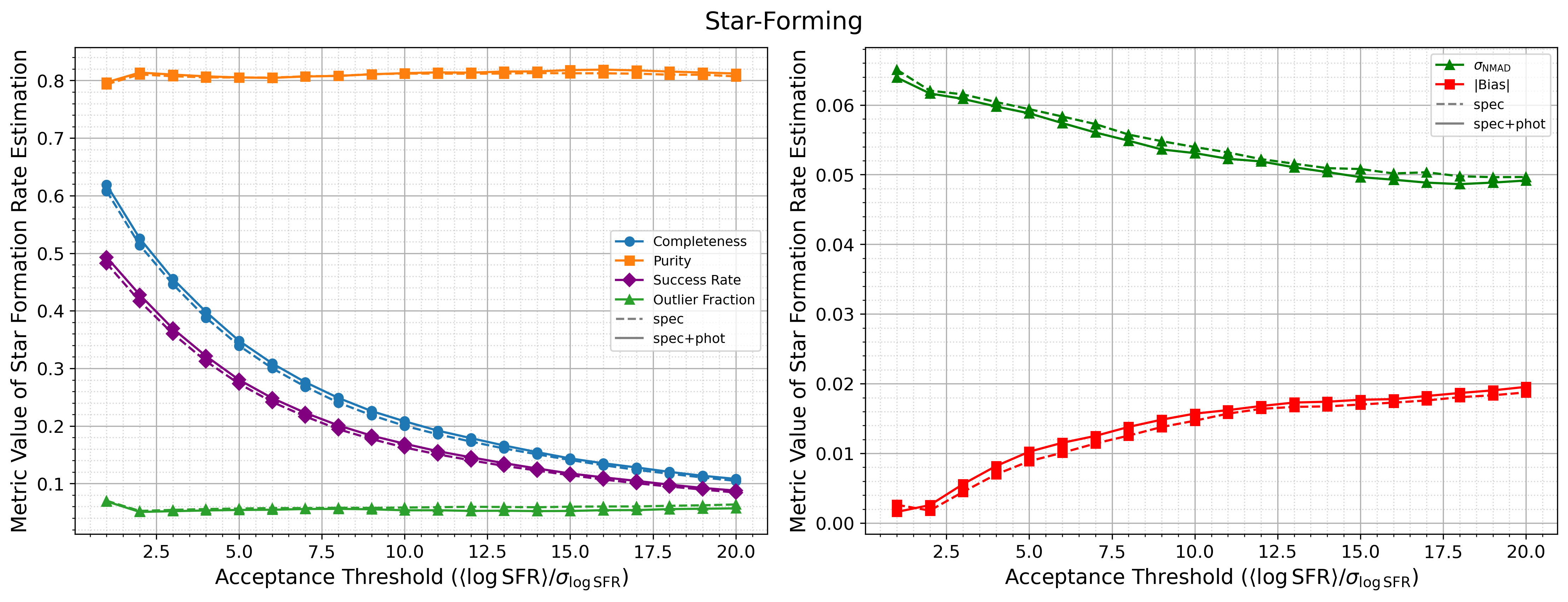} \\
    \end{tabular}
    \caption{Comparison of parameter estimation performance between spectroscopy-only (dashed lines) and combined spectroscopy+photometry (solid lines) for star-forming galaxies. Each row shows results for different parameters: redshift (top), stellar mass (middle), and star formation rate (bottom). Left panels show quality metrics (completeness, purity, outlier fraction, success rate) and right panels show precision metrics ($\sigma_{\rm NMAD}$, $|{\rm Bias}|$). The two approaches show nearly identical behavior across all parameters and metrics. While adding photometric data provides some reduction in systematic bias (e.g., stellar mass $|{\rm Bias}|$ decreases from $\sim$0.003 to $\sim$0.0018), the bias values are already negligibly small in both cases. Overall, these results suggest that spectroscopic data alone captures most of the information needed for reliable parameter estimation of star-forming galaxies.}
    \label{fig:significance_comparison_sf_phot}
\end{figure*}

\begin{figure*}[ht!]
    \centering
    \begin{tabular}{c}
        \includegraphics[width=0.95\textwidth]{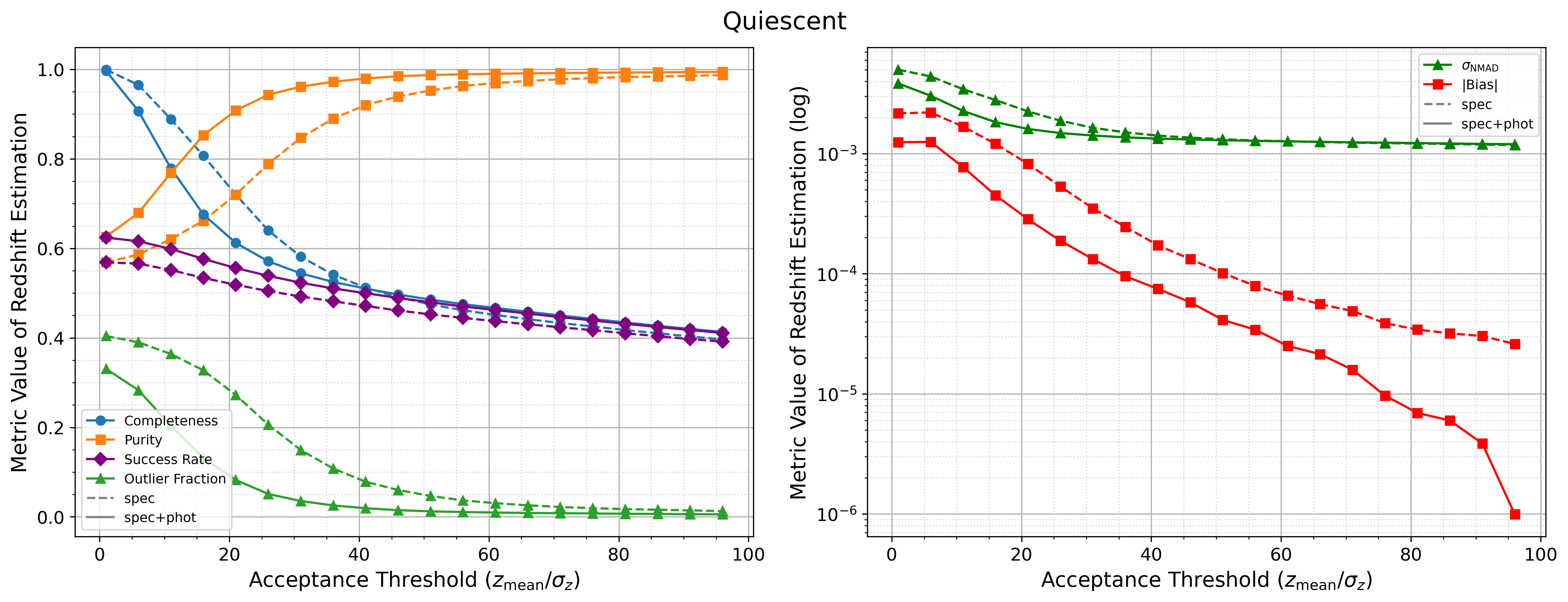} \\
        \includegraphics[width=0.95\textwidth]{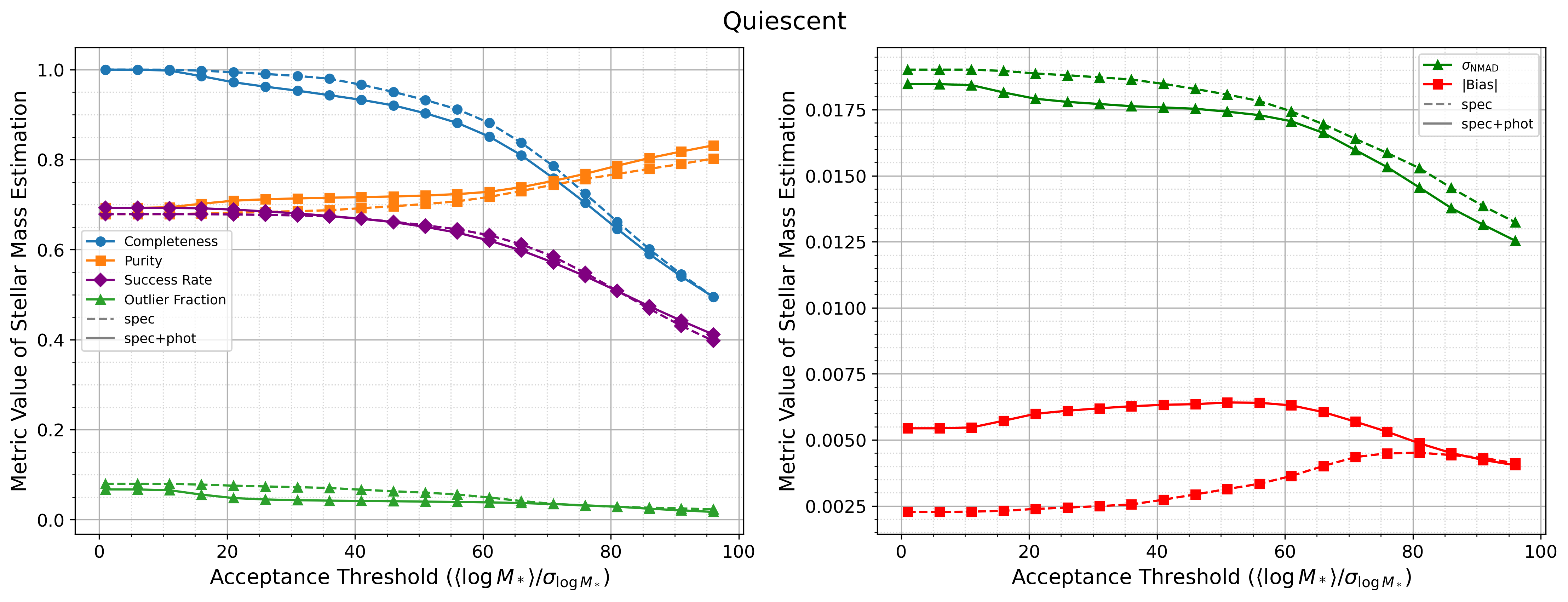} \\
    \end{tabular}
    \caption{As Figure~\ref{fig:significance_comparison_sf_phot}, but for quiescent galaxies. For redshift estimation (top row), combined spectroscopy+photometry significantly improves performance, especially at acceptance thresholds $<$40. For stellar mass estimation (bottom row), both approaches perform nearly identically. While photometric data provides modest benefits for redshift, spectroscopic data alone appears sufficient for stellar mass determination in quiescent galaxies (see Section~\ref{ss:disc_specphot_all} for details).}
    \label{fig:significance_comparison_quiescent_phot}
\end{figure*}

\subsection{Combined analysis of two slitless spectroscopic bands and all photometric bands} \label{ss:disc_specphot_part2}

Building on our analysis of the full three-band spectroscopic dataset, we now examine more realistic observational scenarios where only two of the three CSST bands (GU, GV, and GI) are available. Such intermediate-coverage scenarios occur with moderate frequency in actual surveys due to observational constraints and telescope time limitations. Each two-band combination offers different wavelength coverage and SNR characteristics, potentially influencing the added value of photometric data. By systematically evaluating all three possible combinations (GU+GV, GU+GI, and GV+GI), we provide practical guidance for observational strategies under these common conditions.

\begin{figure*}[ht!]
    \centering
    \begin{tabular}{c}
        \includegraphics[width=0.95\textwidth]{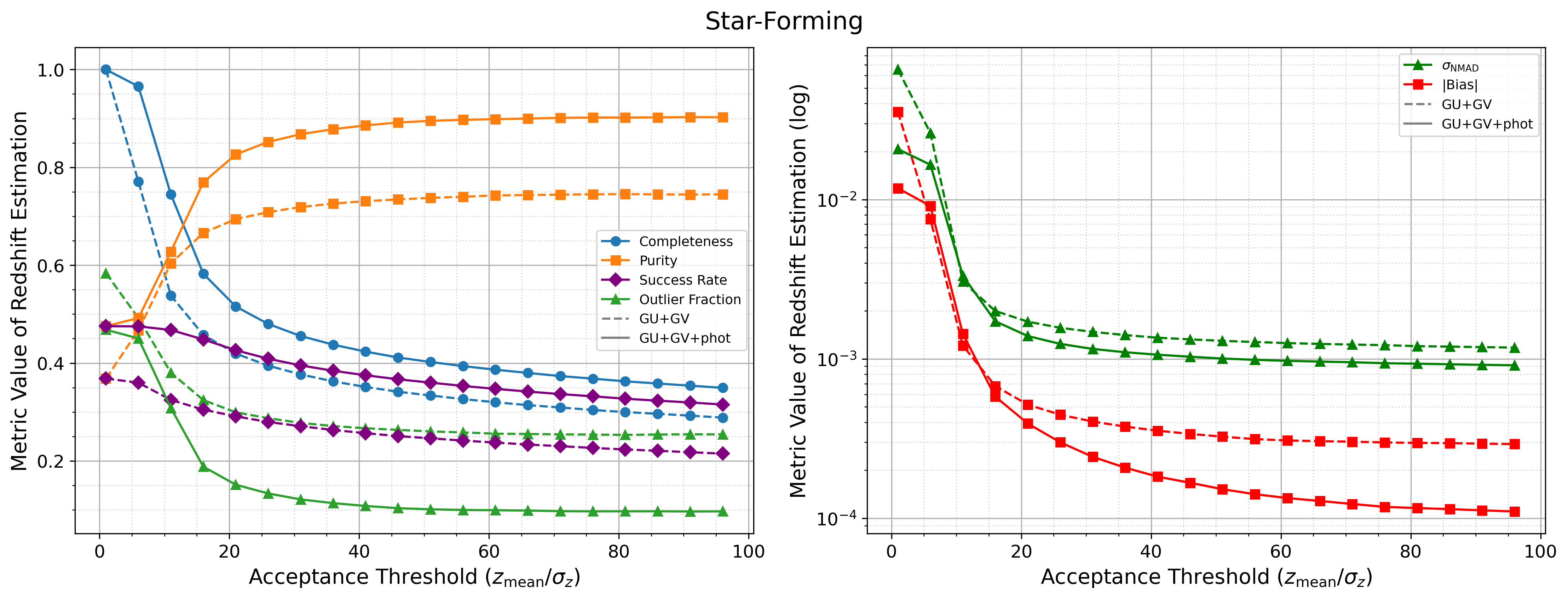} \\
        \includegraphics[width=0.95\textwidth]{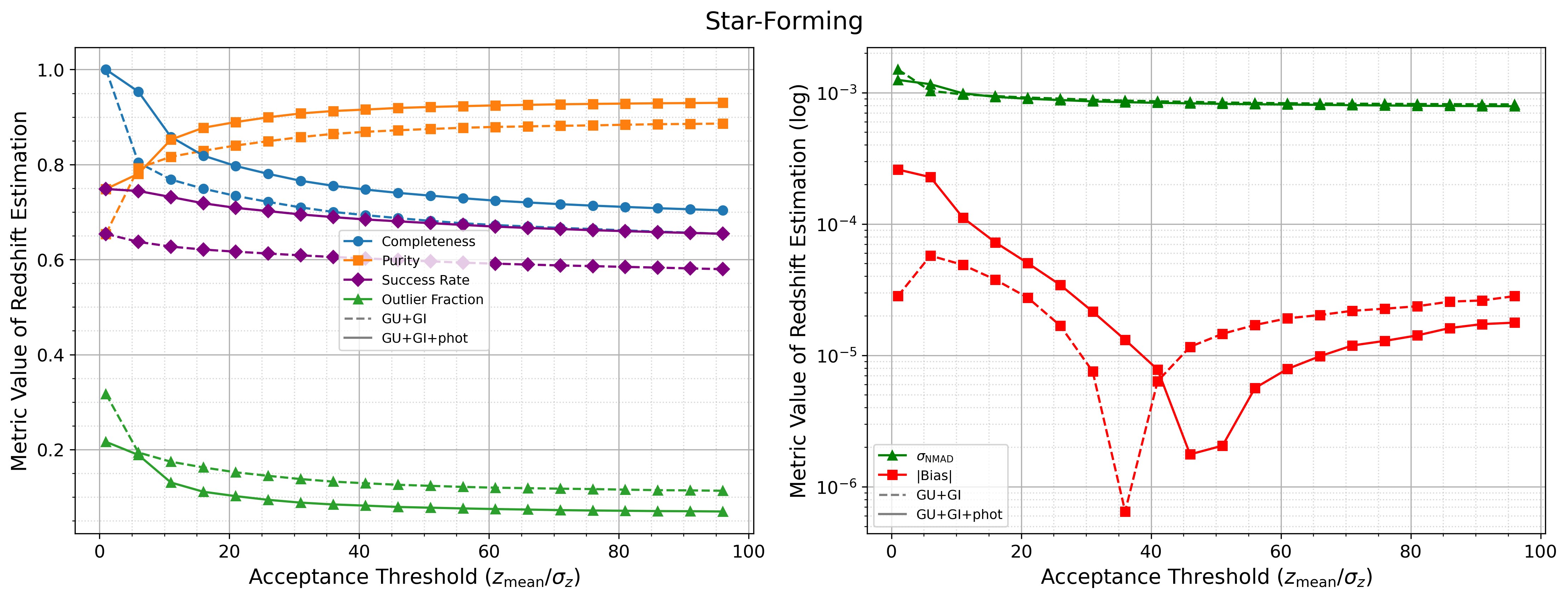} \\
        \includegraphics[width=0.95\textwidth]{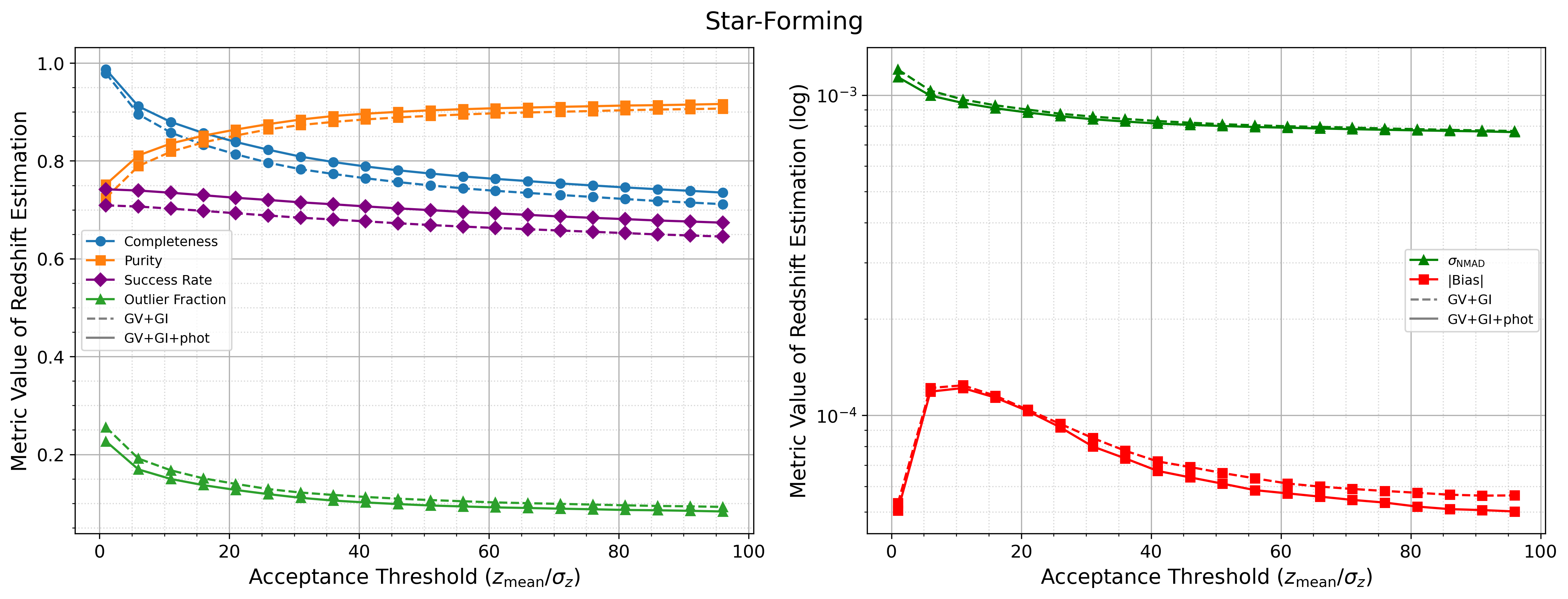}
    \end{tabular}
    \caption{Comparison of redshift estimation performance between spectroscopy-only (dashed lines) and combined spectroscopy+photometry (solid lines) for star-forming galaxies using two-band combinations: GU+GV (top), GU+GI (middle), and GV+GI (bottom). Left panels show quality metrics; right panels show precision metrics ($\sigma_{\rm NMAD}$, $|{\rm Bias}|$). Adding photometric data provides substantial improvement with two-band spectroscopy. The GV+GI and GU+GI combinations achieve the best performance when combined with photometry (see Section~\ref{ss:disc_specphot_part2} for detailed results).}
    \label{fig:significance_comparison_sf_phot_part2}
\end{figure*}

\begin{figure*}[ht!]
    \centering
    \begin{tabular}{c}
        \includegraphics[width=0.95\textwidth]{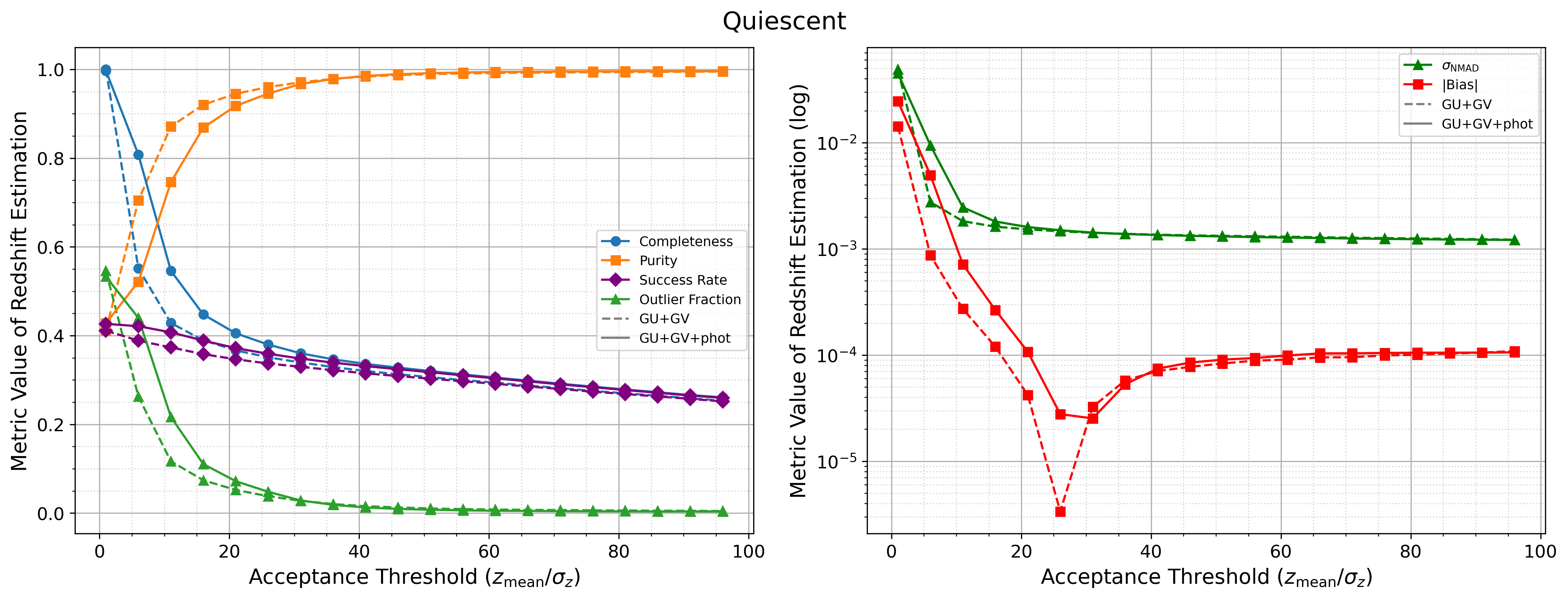} \\
        \includegraphics[width=0.95\textwidth]{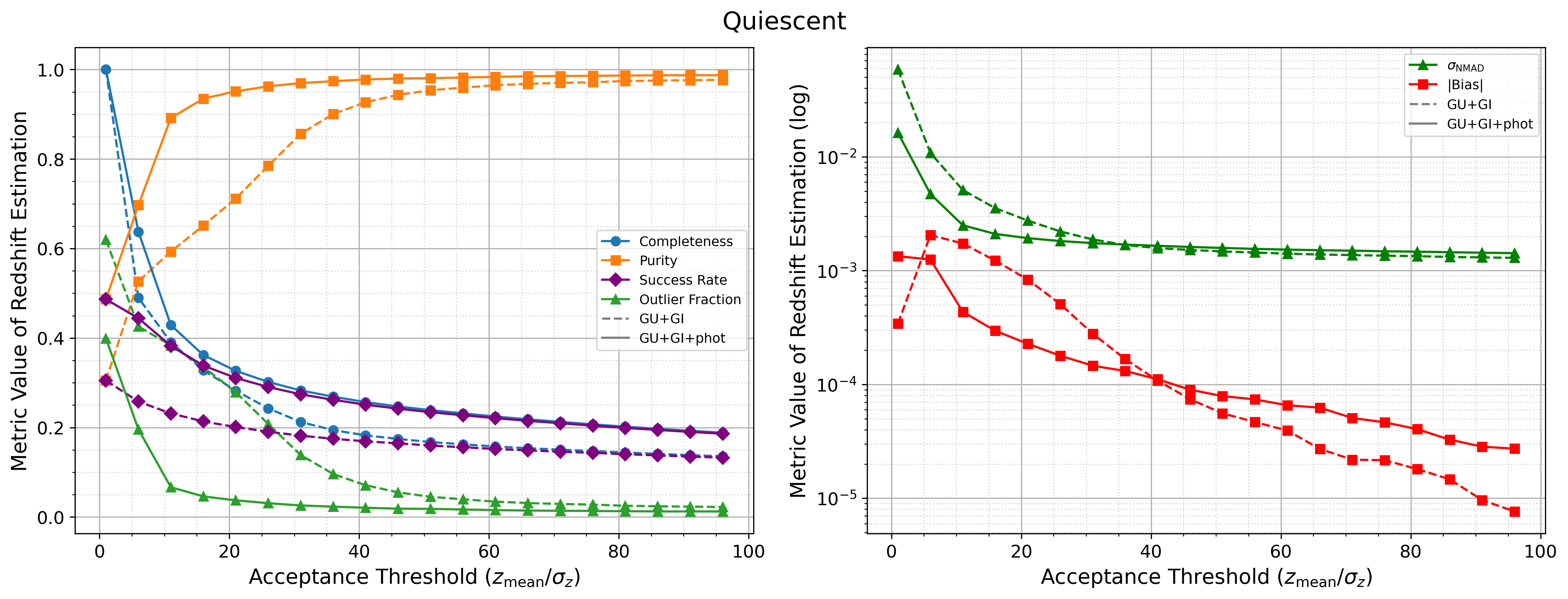} \\
        \includegraphics[width=0.95\textwidth]{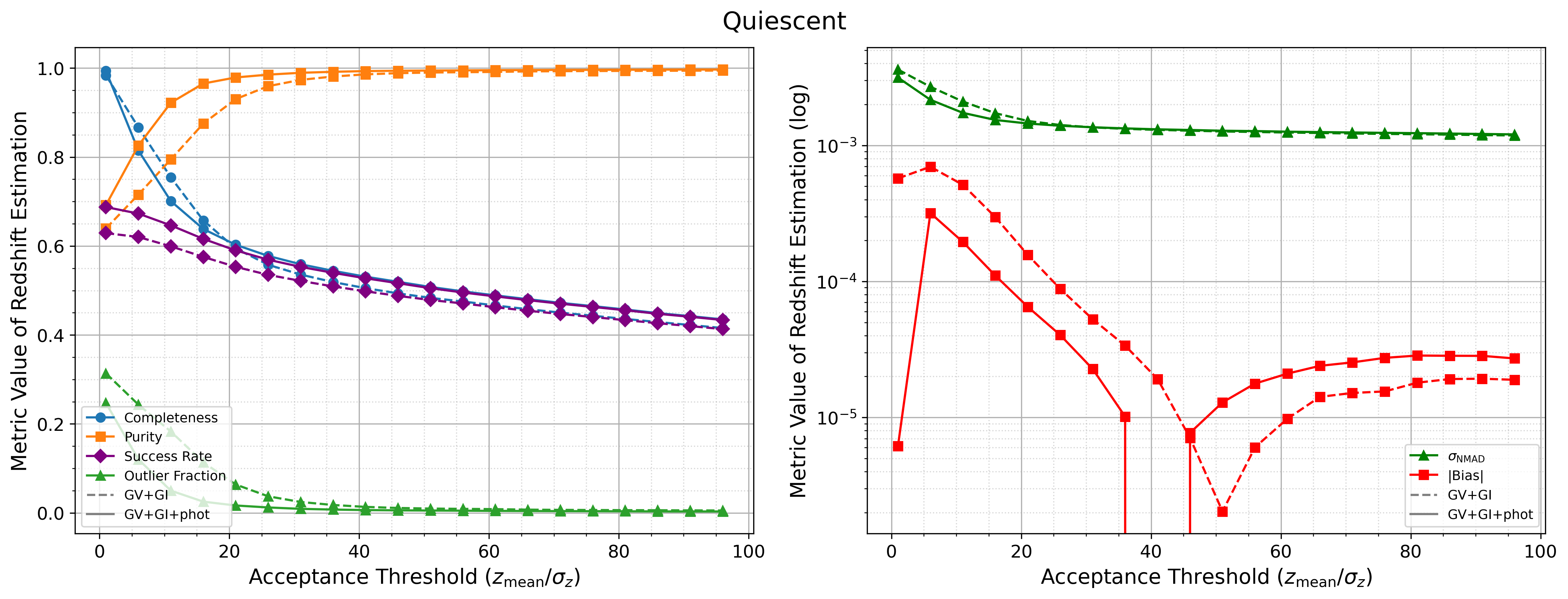}
    \end{tabular}
    \caption{As Figure~\ref{fig:significance_comparison_sf_phot_part2}, but for quiescent galaxies. Adding photometric data provides more substantial improvement with two-band spectroscopy than with three-band spectroscopy (cf. Figure~\ref{fig:significance_comparison_quiescent_phot}). The GV+GI combination shows the best performance when combined with photometry (see Section~\ref{ss:disc_specphot_part2} for detailed results).}
    \label{fig:significance_comparison_quiescent_phot_part2}
\end{figure*}

We analyzed redshift estimation performance for both star-forming and quiescent galaxies using each two-band combination, with and without the addition of photometric data (Figures~\ref{fig:significance_comparison_sf_phot_part2} and \ref{fig:significance_comparison_quiescent_phot_part2}). Generally, adding photometric data provides more obvious improvements when only two spectroscopic bands are available compared to the three-band case.

For star-forming galaxies (Figure~\ref{fig:significance_comparison_sf_phot_part2}), the performance varies depending on the band combination. With the GU+GV combination, combining with photometry allows achieving redshift estimates with a success rate $\gtrsim$45\%, purity $\gtrsim$75\%, outlier fraction $\lesssim20\%$, $\sigma_{\rm NMAD}\lesssim0.002$, and $|{\rm Bias}|\sim0.0006$ at the optimal acceptance threshold of 15. The GU+GI combination, when enhanced with photometry, yields better results: a success rate of $\sim$75\%, purity of $\sim80\%$, outlier fraction $\lesssim$20\%, $\sigma_{\rm NMAD}\sim0.001$, and $|{\rm Bias}|\lesssim0.0002$ at the optimal acceptance threshold of 5. The GV+GI combination, also combined with photometry, performs similarly well, achieving a success rate of $\sim75\%$, purity $\gtrsim$80\%, outlier fraction $\lesssim$20\%, $\sigma_{\rm NMAD}\sim0.001$, and $|{\rm Bias}|\lesssim0.0002$ at an optimal threshold of 5.

For quiescent galaxies (Figure~\ref{fig:significance_comparison_quiescent_phot_part2}), adding photometric data to two-band spectroscopy also provides notable improvements compared to the three-band case (Figure~\ref{fig:significance_comparison_quiescent_phot}). The GU+GV combination with photometry achieves a success rate of $\sim40\%$, purity $\gtrsim$85\%, outlier fraction $\sim10\%$, $\sigma_{\rm NMAD}\lesssim0.002$, and $|{\rm Bias}|\sim0.0003$ at an optimal threshold of 15. The GU+GI combination with photometry yields a success rate of $\sim$35\%, purity $\sim90\%$, outlier fraction $\lesssim$5\%, $\sigma_{\rm NMAD}\sim0.002$, and $|{\rm Bias}|\lesssim0.0003$ at the same optimal threshold of 15. The GV+GI combination stands out, achieving the best performance among the two-band options when combined with photometry: success rate $\gtrsim60\%$, purity $\gtrsim$90\%, outlier fraction $\lesssim$5\%, $\sigma_{\rm NMAD}\lesssim0.002$, and $|{\rm Bias}|\sim0.0001$ at an optimal threshold of 5.

These results highlight that even with only two spectroscopic bands, combining them with CSST's photometric data enables robust redshift estimation, with the specific performance depending on the chosen band combination and galaxy type. The GV+GI combination appears particularly effective for both star-forming and quiescent galaxies when photometric data is included.


\subsection{Combined analysis of one slitless spectroscopic band and all photometric bands} \label{ss:disc_specphot_part1}

In real observational surveys, the most common scenario is having access to only one of the three spectroscopic bands (GU, GV, or GI). Understanding the performance implications of each individual band and the potential complementary value of photometric data allows for optimized observational strategies in these limited-data scenarios, which represent the majority of actual survey conditions. This section analyzes how combining single spectroscopic bands with photometry affects parameter estimation, providing practical guidance for the most frequently encountered observational situations.

\begin{figure*}[ht!]
    \centering
    \begin{tabular}{c}
        \includegraphics[width=0.95\textwidth]{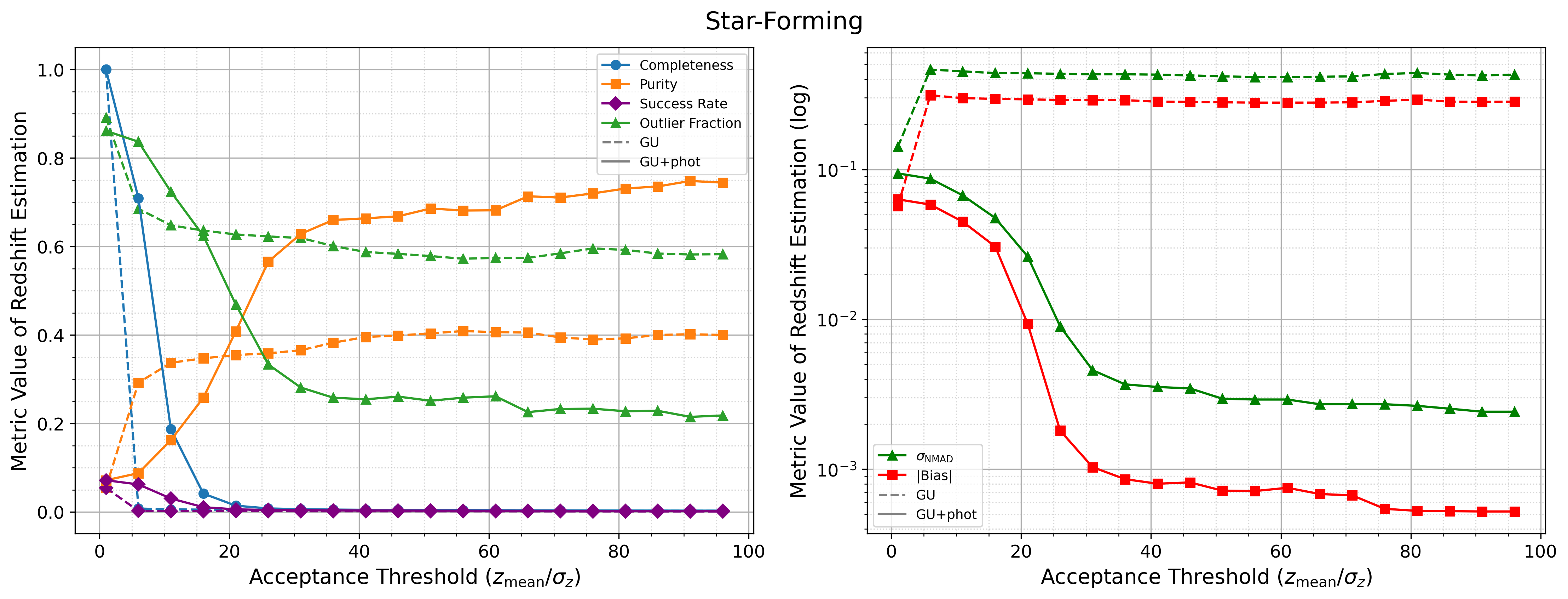} \\
        \includegraphics[width=0.95\textwidth]{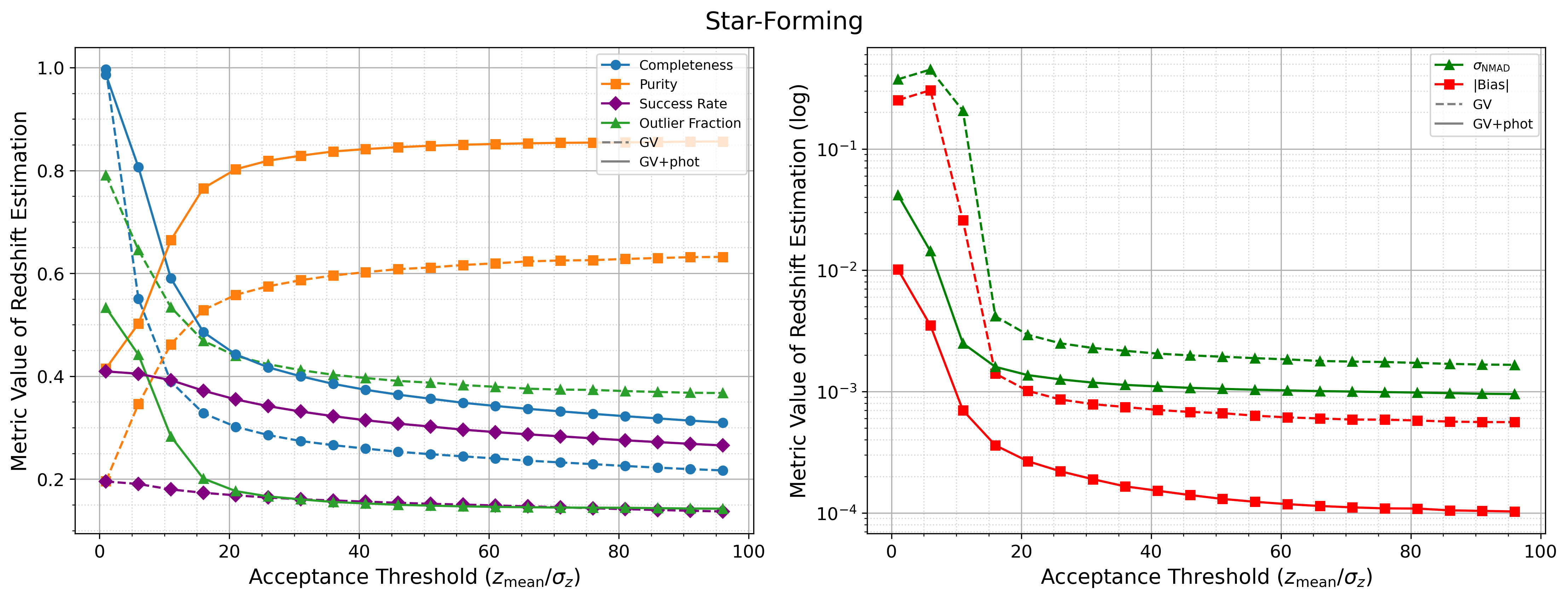} \\
        \includegraphics[width=0.95\textwidth]{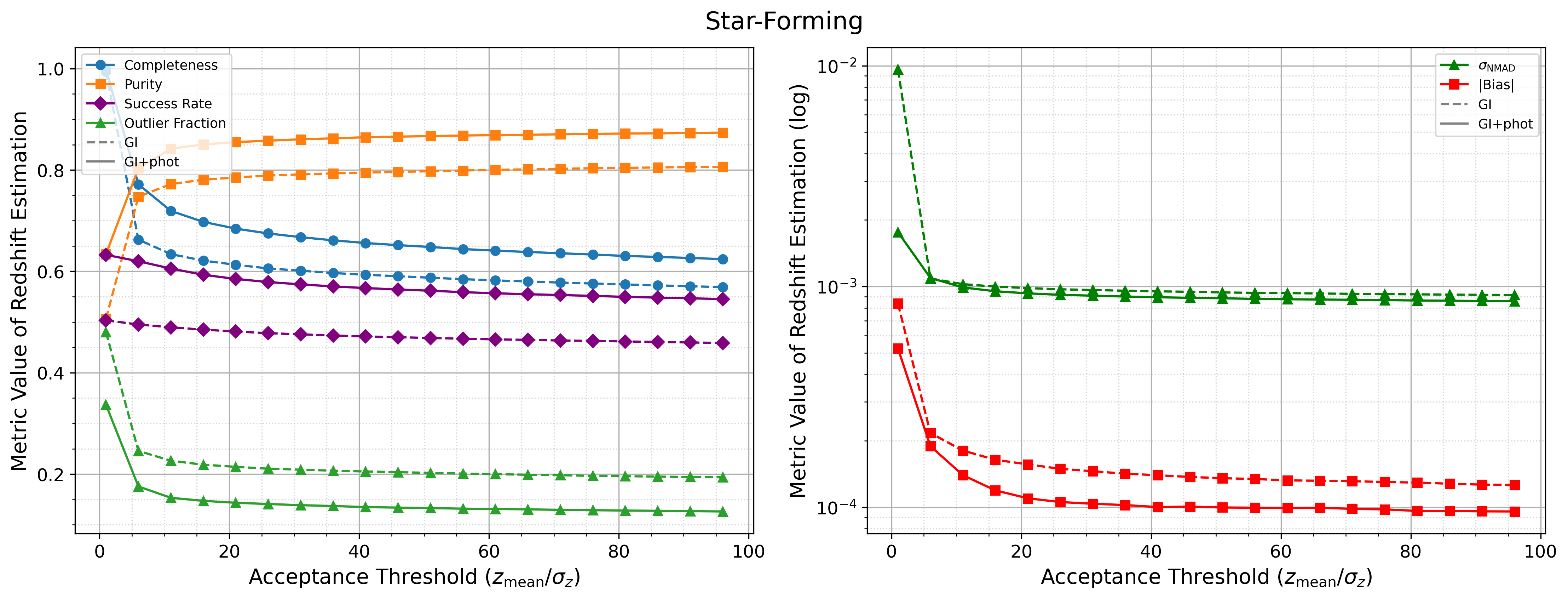}
    \end{tabular}
    \caption{Comparison of redshift estimation performance between single spectroscopic band alone (dashed lines) versus combined with photometry (solid lines) for star-forming galaxies: GU (top), GV (middle), and GI (bottom) bands. Adding photometric data provides substantial improvement with single-band spectroscopy. The GI band combined with photometry shows the best performance, followed by GV. The GU band alone yields insufficient performance even when combined with photometry (see Section~\ref{ss:disc_specphot_part1} for detailed results).}
    \label{fig:significance_comparison_sf_phot_part1}
\end{figure*}

\begin{figure*}[ht!]
    \centering
    \begin{tabular}{c}
        \includegraphics[width=0.95\textwidth]{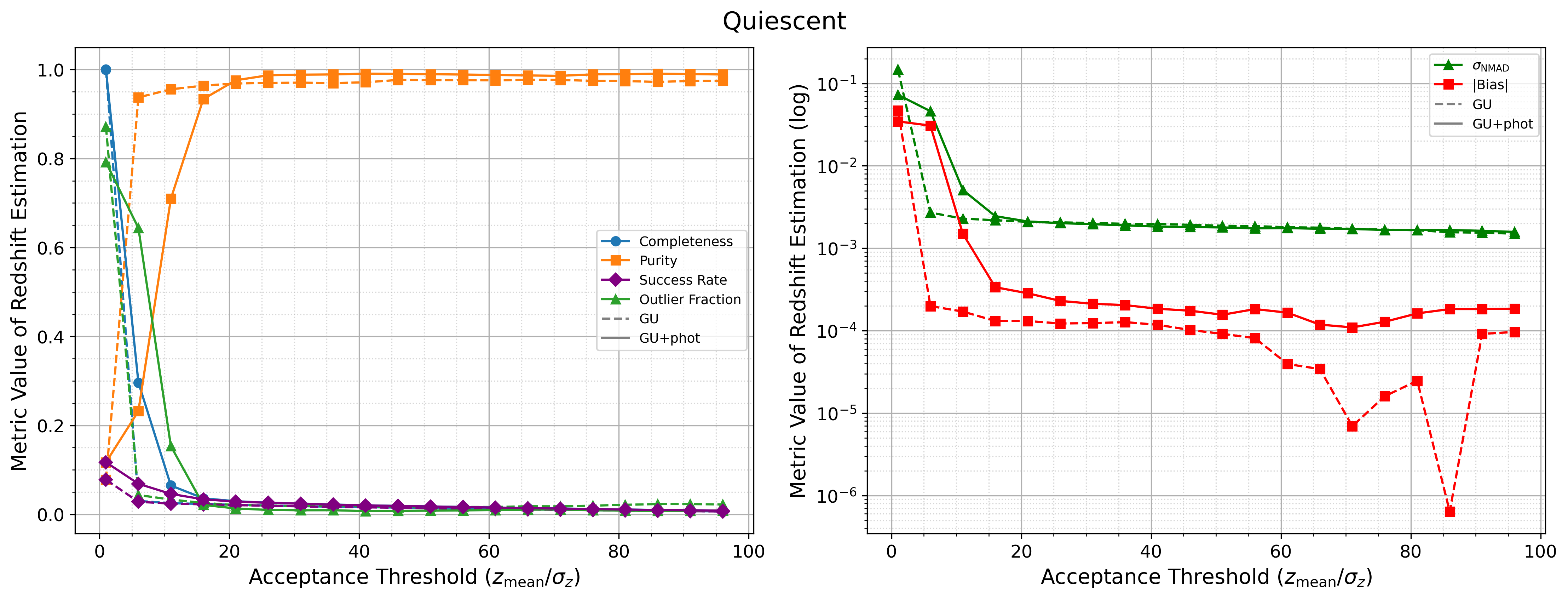} \\
        \includegraphics[width=0.95\textwidth]{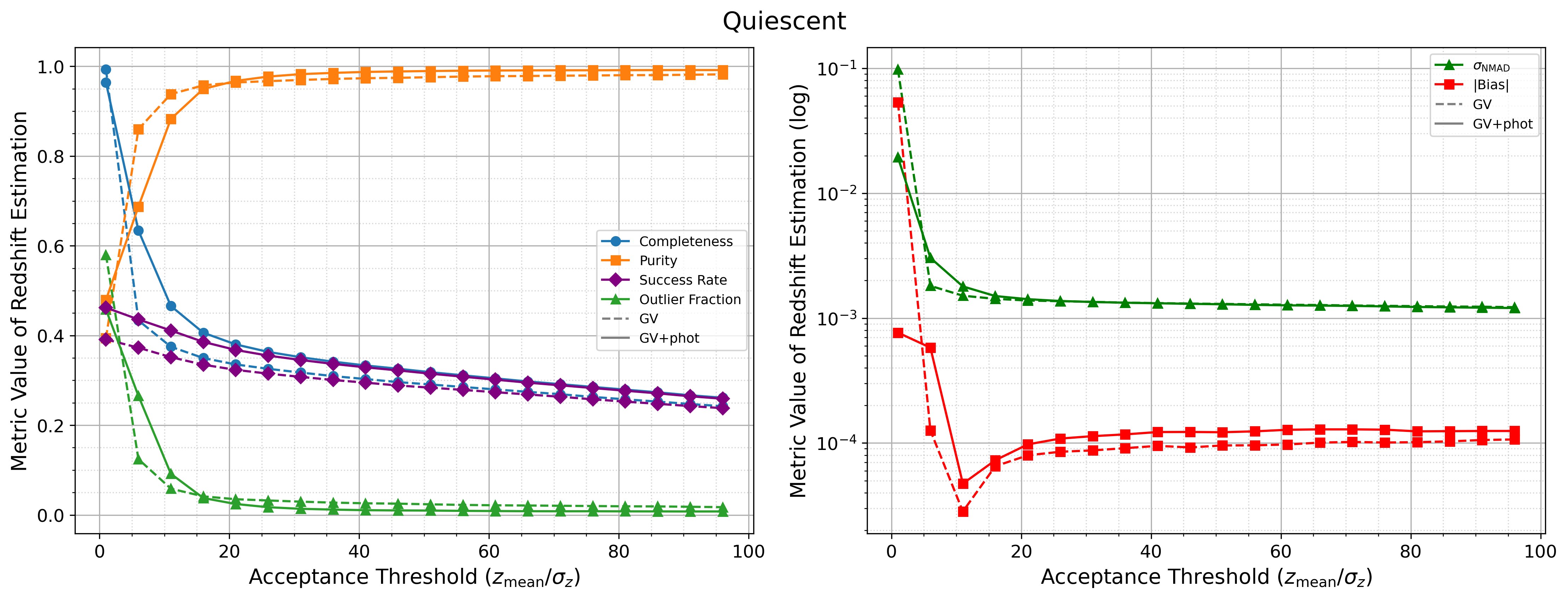} \\
        \includegraphics[width=0.95\textwidth]{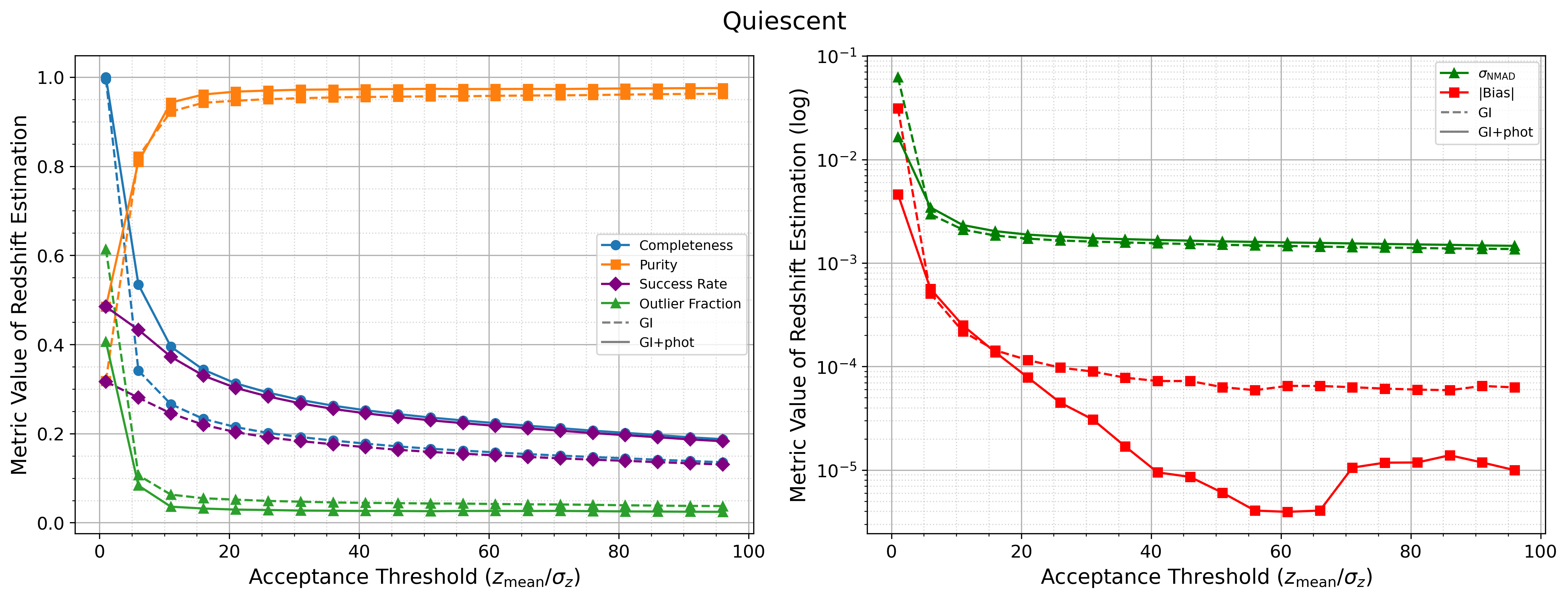}
    \end{tabular}
    \caption{As Figure~\ref{fig:significance_comparison_sf_phot_part1}, but for quiescent galaxies. Adding photometric data provides more substantial improvement with single-band spectroscopy than with multi-band spectroscopy. The GI and GV bands combined with photometry achieve reliable performance, while the GU band remains challenging even with photometry (see Section~\ref{ss:disc_specphot_part1} for detailed results).}
    \label{fig:significance_comparison_quiescent_phot_part1}
\end{figure*}

Our detailed analysis reveals significant variations in redshift estimation performance across the three CSST spectroscopic bands, with distinct patterns between star-forming and quiescent galaxies (Figures~\ref{fig:significance_comparison_sf_phot_part1} and \ref{fig:significance_comparison_quiescent_phot_part1}). For both galaxy types, we find that adding photometric data generally provides more substantial improvements when only one spectroscopic band is available compared to multi-band spectroscopy, though the magnitude of this improvement varies considerably depending on which band is used.

For star-forming galaxies (Figure~\ref{fig:significance_comparison_sf_phot_part1}), the GU band alone proves extremely challenging for accurate redshift estimation, with poor performance metrics even when supplemented with photometric data. This limitation stems from the band's significantly lower SNR compared to the other CSST bands, making it particularly difficult to detect key emission features in star-forming galaxies. In contrast, the GV band demonstrates considerably better performance, achieving high-quality redshift estimation with success rates exceeding 35\%, purity above 75\%, and outlier fractions around 20\% when combined with photometry, at an optimal acceptance threshold of 15. The precision metrics are particularly impressive, with $\sigma_{\rm NMAD}$ values of approximately 0.002 and absolute bias around 0.0003.
The GI band emerges as the most effective single band for star-forming galaxies, delivering substantially superior performance compared to either GU or GV. When combined with photometric data, GI-based redshift estimation achieves success rates greater than 60\%, purity exceeding 80\%, and outlier fractions below 20\% at an optimal acceptance threshold of 5. The precision metrics are excellent, with $\sigma_{\rm NMAD}$ around 0.001 and $|{\rm Bias}|$ less than 0.0005, making GI the clearly preferred single-band option for star-forming galaxies.

For quiescent galaxies (Figure~\ref{fig:significance_comparison_quiescent_phot_part1}), the pattern of band effectiveness follows a similar hierarchy, but with generally lower overall performance compared to star-forming galaxies. The GU band alone remains challenging even with photometric data, reflecting the difficulty in capturing the characteristic spectral features of quiescent galaxies in this wavelength range. The GV band shows moderate performance when combined with photometry, achieving success rates around 40\%, purity up to 90\%, and outlier fractions of approximately 10\% at an optimal acceptance threshold of 10. The precision metrics are respectable, with $\sigma_{\rm NMAD}$ values around 0.002 and absolute bias less than 0.0001.
As with star-forming galaxies, the GI band provides the best single-band performance for quiescent galaxies. When supplemented with photometric data, GI-based redshift estimation achieves success rates exceeding 40\%, purity above 90\%, and outlier fractions below 5\%. The precision metrics are excellent, with $\sigma_{\rm NMAD}$ values below 0.002 and absolute bias around 0.0002. 

These findings have important implications for observational strategies in CSST and similar surveys. When resource constraints limit spectroscopic coverage to a single band, our results strongly favor prioritizing the GI band for both galaxy populations. If GI observations are not feasible, the GV band provides a viable alternative, particularly for star-forming galaxies, while the GU band alone appears insufficient for reliable redshift estimation regardless of galaxy type. Importantly, our results demonstrate that combining even a single spectroscopic band (particularly GV or GI) with photometric data can yield redshift measurements of sufficient quality for many scientific applications, providing a practical approach for enhancing scientific return in resource-limited survey scenarios.

\subsection{Caveats and Limitations} \label{ss:disc_caveats}

A key aspect of our methodology in this paper is the use of identical models for both mock data generation and subsequent fitting. This deliberate choice allows us to isolate and quantify the impact of observational effects, such as low Signal-to-Noise Ratio (SNR), instrumental systematics, and spectral self-blending, on parameter estimation without confounding factors from model uncertainties. By eliminating model mismatches, our analysis provides a clean assessment of the intrinsic performance limits of the Bayesian full spectrum analysis framework under realistic CSST observational conditions.

However, it is crucial to acknowledge that this approach does not directly reveal systematic biases that might arise from model assumptions themselves---e.g., due to varying stellar population synthesis (SPS) models, star formation histories (SFHs), or dust attenuation laws (DALs). 

Our specific model choices carry potential systematic uncertainties that warrant careful consideration.
For SPS models, BC03 does not include binary stellar evolution, which models like BPASS \citep{EldridgeJ2017a} show can significantly affect UV light from stripped stars and blue stragglers, stellar mass estimates for young populations, and interpretation of some emission line ratios.
Different SPS models (e.g., FSPS, BPASS, Maraston) can yield systematic differences in mass estimates.
Our adoption of an exponentially declining SFH, while computationally efficient, has been shown to introduce biases for galaxies with rising or bursty star formation histories \citep{CarnallA2019a,LowerS2020u}.
Similarly, the Calzetti dust law may not capture the full diversity of dust attenuation curves observed in real galaxies, particularly for edge-on or heavily obscured systems\citep{SalimS2020a}.

In a previous study (\citealt{HanY2023a}), we extensively explored these model uncertainties for the CSST imaging survey, utilizing Horizon-AGN hydrodynamical simulation-based mock galaxy samples. We found that for photometric data, model uncertainties (e.g., in SFHs and DALs) can significantly impact parameter estimation, sometimes even more than observational noise, especially for parameters like stellar mass and SFR. The optimal model complexity was found to be survey-dependent, and Bayesian model comparison proved effective in identifying the best model for a given survey's discriminative power. Additionally, recent work by \citet{JonesG2025a} leveraging EAGLE simulations with synthetic SDSS spectral and VISTA photometric observations has further demonstrated that variations in stellar spectral library, initial mass function (IMF), and metallicity prescriptions can cause inferred galaxy properties (mass, age, SFR) to vary significantly, often more than observational uncertainties. This highlights the critical need for careful propagation of SED modeling uncertainties. The high level of precision required for subsequent cosmological analyses necessitates a thorough understanding and mitigation of all potential sources of error and bias.

Future work will extend the investigation of these model-related uncertainties to spectroscopic data, employing different sets of modeling assumptions for mock generation and fitting. This will enable us to quantify the impact of model mismatches, assess the sensitivity of our metrics to varying SPS models, SFHs, and DALs, and develop strategies for robust parameter estimation in the presence of model ambiguities. Such comprehensive analysis will be critical for fully realizing the cosmological potential of CSST data.

Additionally, we note that our simulation framework does not account for wavelength calibration errors, which are a common source of systematic uncertainty in spectroscopic observations. In real CSST data, wavelength calibration uncertainties could introduce additional systematic biases in redshift estimation and other derived parameters. Future work should investigate the impact of wavelength calibration errors on our parameter recovery performance and develop appropriate mitigation strategies.

\section{Summary and Conclusions} \label{sec:summary}

In this paper, we extended the Bayesian spectral energy distribution fitting code BayeSED3 to perform full spectrum analysis of low-resolution ($R \sim 200$), low SNR slitless spectroscopic data anticipated from the upcoming CSST survey. Motivated by forecasts highlighting CSST's potential for transformative cosmological studies, which critically depend on precise redshift measurements (requiring $\sigma_{\mathrm{NMAD}} \lesssim 0.002-0.005$), our primary goal was to develop and validate a robust methodology for estimating key galaxy properties like redshift ($z$), stellar mass ($M_*$), and star formation rate (SFR) from these challenging data, thereby enhancing the scientific return from CSST's unique combination of wide-field photometry and spectroscopy.

To achieve this, we first developed a realistic simulation framework. Starting with a parent sample drawn from the DESI Legacy Survey, we created intrinsic model spectra using BC03 stellar populations (Chabrier IMF, exponentially declining SFH, \citealt{CalzettiD2000a} DAL) at a resolution of $R=300$, incorporating nebular emission (both continuum and lines) for young stellar populations ($<10$\,Myr) calculated using \textsc{Cloudy} following the \citet{BylerN2017a} framework. We validated our spectral library against the DESI DR9 photometric catalog ($\sim$140 million galaxies) through residual analysis across five photometric bands ($g$, $r$, $z$, W1, W2) as a function of rest-frame wavelength. This analysis identified systematic model limitations: UV flux underestimation in the $g$-band (possibly due to limitations in modeling young stellar populations, dust attenuation effects, or binary star evolution), discrete residuals in the $z$-band (suggesting emission line modeling issues), NIR flux underestimation in W1 (likely missing TP-AGB star contributions), and significant MIR flux underestimation in W2 (indicating missing PAH features and AGN emission). While the MIR band extends beyond CSST's wavelength coverage , the UV and NIR worth more attentions in future works. Despite these systematic effects, the overall residual distributions remain well-centered around zero, and UMAP dimensionality reduction analysis confirms that our model library adequately spans the observational manifold in multi-band color space (classifier accuracy $\approx$0.5), validating our approach for generating reliable spectral templates.

These validated intrinsic spectra were then processed through CESS to generate realistic mock observations. CESS simulates instrumental effects (spectral resolution and wavelength sampling), sky background contributions, and galaxy morphology effects that lead to spectral self-blending. The resulting mock dataset includes both 7-band photometry and 3-band (GU, GV, GI) slitless spectra, characterized by a challenging low median SNR of approximately 1.65.

Our analysis methodology relies on the BayeSED3 framework to perform Bayesian inference on these mock CSST data. A key methodological advancement within BayeSED3 is the treatment of the model normalization (the scaling factor $s$) as a free parameter sampled by the MultiNest algorithm, rather than being optimized separately via NNLM. Through detailed comparison using representative quiescent and star-forming galaxies, we demonstrate that the NNLM optimization approach exhibits problematic limitations for low-SNR slitless spectroscopy: it frequently fails to converge within reasonable computational limits, often converges to incorrect solutions (e.g., catastrophic redshift failures for quiescent galaxies), and tends to overemphasize high-SNR emission line regions while poorly fitting the continuum in star-forming galaxies. While NNLM optimization is more suitable for high-SNR photometric and spectroscopic data where the parameter space is better constrained, it proves unreliable for CSST's low-SNR slitless spectra. In contrast, the Bayesian sampling approach provides superior performance by properly marginalizing over the scaling factor, leading to more balanced fits and reliable parameter estimation. Comparison with BAGPIPES confirms that BayeSED3 achieves comparable fitting quality while requiring substantially less computational time, making it ideally suited for processing millions of low-SNR slitless spectra (as discussed in Section \ref{ss:scaling} and illustrated in Figure~\ref{fig:fitting_comparison}). 

We optimized the MultiNest runtime settings (nlive=40, efr=0.1) to balance computational efficiency with estimation accuracy and defined a comprehensive suite of performance metrics (completeness, purity, success rate, outlier fraction, bias, $\sigma_{\rm NMAD}$), along with specific parameter acceptance thresholds, to quantitatively evaluate the recovery of $z$, $M_*$, and SFR (Section \ref{ss:metrics}).
Our main findings, derived from the analysis of the combined GU+GV+GI spectroscopic data as a baseline, and subsequently exploring more realistic scenarios, are as follows:

\begin{itemize}
    \item \textbf{Baseline Performance (Three-Band Spectroscopy):}
        \begin{itemize}
            \item \textit{Redshift ($z$):} Excellent redshift precision is achievable, significantly exceeding the requirements for cosmological analysis despite the low median SNR. In the idealized case without self-blending, for star-forming galaxies, we obtain $\sigma_{\mathrm{NMAD}} = 0.0007$ with a success rate exceeding 81\% (defined as purity $\times$ completeness for $|\Delta_z| < 0.01$) using an acceptance threshold of $z_{\rm mean}/\sigma_z > 20$. For quiescent galaxies, performance is understandably lower but still robust and meets the precision requirement, with $\sigma_{\mathrm{NMAD}} = 0.0011$ and a success rate $>55\%$. When realistic self-blending effects are included, optimal thresholds increase to $z_{\rm mean}/\sigma_z > 30$ (star-forming) and $> 25$ (quiescent), achieving $\sigma_{\mathrm{NMAD}} = 0.0008$ and $0.0015$ respectively, with success rates of $\sim$80\% and $\sim$50\% (see Table~\ref{tab:redshift_summary}). Performance strongly depends on SNR, with  $SNR> 1.5$ (star-forming) and  $SNR> 2$ (quiescent) being crucial for high reliability (Section~\ref{sss:results_z_performance}).
            \item \textit{Stellar Mass ($M_*$):} Stellar masses are recovered with high fidelity. Using an acceptance threshold of $\langle\log M_*\rangle/\sigma_{\log M_*} > 50$, we achieve $\sigma_{\mathrm{NMAD}} \approx 0.015$ dex for star-forming galaxies and $\sigma_{\mathrm{NMAD}} \approx 0.016$ dex for quiescent galaxies, with success rates of $\sim$70\% for both populations. Similar to redshift,  $SNR>1$ is essential for reliable $M_*$ estimation (Section~\ref{sss:results_mass_performance}).
            \item \textit{Star Formation Rate ($SFR$):} SFRs for star-forming galaxies can be reliably recovered, especially at higher SNR. With an acceptance threshold of $\langle\log {\rm SFR}\rangle/\sigma_{\log {\rm SFR}} > 3$, we find $\sigma_{\mathrm{NMAD}} \approx 0.05$ dex, although the overall success rate is lower ($\sim$39\%) compared to redshift and mass. Performance significantly improves for  $SNR>1$  and is optimal for  $SNR>3$ (Section~\ref{sss:results_sfr_performance}).
        \end{itemize}

    \item \textbf{Impact of Self-Blending Effects:} Spectral self-blending, an inherent characteristic of slitless spectroscopy for extended sources, primarily impacts the precision metric $\sigma_{\mathrm{NMAD}}$, which could be increased by $\gtrsim30\%$ across all parameters. However, even when these effects are included in our simulations, we demonstrate that high precision remains achievable. For instance, redshift precision reaches $\sigma_{\mathrm{NMAD}} = 0.0008$ for star-forming galaxies and $0.0015$ for quiescent galaxies, while stellar mass precision is maintained at $\sigma_{\mathrm{NMAD}} \approx 0.017$ dex for both types (as detailed in Section \ref{ss:disc_blend_overlap}). The challenge of overlapping spectra from distinct sources, particularly in crowded fields, remains an area for future investigation.

    \item \textbf{Synergistic Benefits of Spectroscopy + Photometry:} Combining CSST's slitless spectroscopy with its 7-band photometry significantly enhances parameter estimation, particularly for quiescent galaxies and in scenarios where fewer spectroscopic bands are available.
        \begin{itemize}
            \item \textit{With all three spectroscopic bands (GU+GV+GI) + photometry:} For quiescent galaxies, redshift estimation improves notably (e.g., success rate $>$55\%, purity $\sim$90\%, outlier fraction $\sim$10\% at the optimal threshold), while $M_*$ estimation shows little change. For star-forming galaxies, adding photometric data offers minimal benefit when all three spectroscopic bands are already utilized (Section \ref{ss:disc_specphot_all}).
            \item \textit{With two spectroscopic bands + photometry:} Reliable results that meet the redshift precision target are achievable. The GV+GI band combination performs best in this scenario, yielding redshift success rates of $\sim$75\% for star-forming galaxies and $>60\%$ for quiescent galaxies, with a precision of $\sigma_{\mathrm{NMAD}} \lesssim 0.002$ (Section \ref{ss:disc_specphot_part2}).
            \item \textit{With only one spectroscopic band + photometry:} This represents the most common survey scenario, and even here, scientifically valuable results meeting the redshift precision requirement are possible. The GI band is demonstrably superior when used as a single spectroscopic band, achieving redshift success rates of $>60\%$ for star-forming galaxies and $>40\%$ for quiescent galaxies, with a precision of $\sigma_{\mathrm{NMAD}} \lesssim 0.002$. The GV band serves as a viable alternative, providing success rates of $>35\%$ (star-forming) and $\sim$40\% (quiescent) with similar precision ($\sigma_{\mathrm{NMAD}} \lesssim 0.002$). The GU band alone, even when combined with photometry, proves insufficient for reliable redshift estimation (Section \ref{ss:disc_specphot_part1}).
        \end{itemize}
\end{itemize}

A consolidated overview of the redshift estimation performance, highlighting the impact of different data combinations and observational scenarios discussed above, is presented in Table~\ref{tab:redshift_summary}.

\begin{table*}[htbp]
\centering
\small
\setlength{\tabcolsep}{4pt}
\caption{Summary of redshift estimation performance at optimal acceptance thresholds for different CSST data combinations. All results include realistic self-blending effects from spectral dispersion of extended sources.}
\label{tab:redshift_summary}
\begin{tabular}{lcccccccc}
\hline
\hline
Data Set & Galaxy & Threshold & Completeness & Success Rate & Purity & Outlier Fraction & $\sigma_{\rm NMAD}$ & $|$Bias$|$ \\
 & & $z_{\rm mean}/\sigma_z$ & (\%) & (\%) & (\%) & (\%) & & \\
\hline
\multicolumn{9}{c}{\textit{Spectroscopy Only}} \\
\hline
GU+GV+GI & Star-forming & $> 30$ & $\sim$88.9 & $\sim$80 & $\sim$90 & $\sim$10 & $\sim$0.0008 & $\sim$0.0001 \\
GU+GV+GI & Quiescent & $> 25$ & $\sim$62.5 & $\sim$50 & $\sim$80 & $\sim$20 & $\sim$0.0015 & $\sim$0.0005 \\
\hline
\multicolumn{9}{c}{\textit{Spectroscopy + Photometry}} \\
\hline
GU+GV+GI+Phot & Star-forming & $> 30$ & $\sim$88.9 & $\sim$80 & $\sim$90 & $\sim$10 & $\sim$0.0008 & $\sim$0.0001 \\
GU+GV+GI+Phot & Quiescent & $> 20$ & $\sim$61.1 & $\sim$55 & $\sim$90 & $\sim$10 & $\sim$0.0016 & $\sim$0.0002 \\
\hline
GU+GV+Phot & Star-forming & $> 15$ & $\sim$60 & $>$45 & $>$75 & $<$20 & $<$0.002 & $\sim$0.0006 \\
GU+GV+Phot & Quiescent & $> 15$ & $\sim$47 & $\sim$40 & $>$85 & $\sim$10 & $<$0.002 & $\sim$0.0003 \\
\hline
GU+GI+Phot & Star-forming & $> 5$ & $\sim$94 & $\sim$75 & $\sim$80 & $<$20 & $\sim$0.001 & $<$0.0002 \\
GU+GI+Phot & Quiescent & $> 15$ & $\sim$39 & $\sim$35 & $\sim$90 & $<$5 & $\sim$0.002 & $<$0.0003 \\
\hline
GV+GI+Phot & Star-forming & $> 5$ & $\sim$94 & $\sim$75 & $>$80 & $<$20 & $\sim$0.001 & $<$0.0002 \\
GV+GI+Phot & Quiescent & $> 5$ & $\sim$67 & $>$60 & $>$90 & $<$5 & $<$0.002 & $\sim$0.0001 \\
\hline
GU+Phot & Star-forming & $> 5$ & $\lesssim10$ & $\lesssim10$ & $\lesssim10$ & $\gtrsim80$ & $\sim0.1$ & $\gtrsim0.05$ \\
GU+Phot & Quiescent & $> 5$ & $\lesssim10$ & $\lesssim10$ & $\lesssim25$ & $\gtrsim60$ & $\sim0.06$ & $\gtrsim0.03$ \\
\hline
GV+Phot & Star-forming & $> 15$ & $\sim$47 & $>$35 & $>$75 & $\sim$20 & $<$0.002 & $\sim$0.0003 \\
GV+Phot & Quiescent & $> 10$ & $\sim$44 & $\sim$40 & $<$90 & $\sim$10 & $\sim$0.002 & $<$0.0001 \\
\hline
GI+Phot & Star-forming & $> 5$ & $\sim$75 & $>$60 & $>$80 & $<$20 & $\sim$0.001 & $<$0.0005 \\
GI+Phot & Quiescent & $> 10$ & $\sim$44 & $>$40 & $>$90 & $<$5 & $<$0.002 & $\sim$0.0002 \\
\hline
\hline
\end{tabular}
\tablecomments{"Phot" indicates the inclusion of CSST's seven-band photometry (NUV, $u$, $g$, $r$, $i$, $z$, $y$). }
\end{table*}

In conclusion, our comprehensive Bayesian framework, BayeSED3, specifically adapted for full spectrum analysis with detailed nebular emission modeling and robust Bayesian treatment of model scaling, provides a powerful and validated tool for extracting reliable physical parameters from CSST's challenging low-resolution, low-SNR slitless spectroscopic data. The validation against DESI DR9 photometry confirms that our models provide reliable templates for the majority of the galaxy population, despite systematic limitations in certain wavelength regimes (UV, NIR, and MIR). The demonstrated superiority of Bayesian sampling over NNLM optimization ensures robust parameter estimation even in the low-SNR regime where optimization-based methods fail. Our findings demonstrate that this methodology can meet and often exceed the precision requirements for key cosmological probes, with the strategic combination of spectroscopy and CSST's seven-band photometry significantly enhancing performance in data-limited scenarios. These results affirm that CSST, despite the inherent difficulties of slitless spectroscopy and spectral blending, is poised to achieve precise characterization of galaxy properties, thereby enabling transformative studies in both galaxy evolution and cosmology.

\begin{acknowledgments}
We especially thank the anonymous referee for careful reading of our manuscript and providing us a detailed report that is very helpful in improving the manuscript. 
We thank Prof.  Xu Zhou,  Xuebin Wu,  Ran Li, and  Linhua Jiang for helpful discussions about CSST surveys and the synergistic use of slitless spectra and photometry.

This work is supported by the National Key Research and Development Program of China (2021YFA1600401, 2021YFA1600404, 2021YFA1600400, 2023YFA1608100, 2023YFA1607804, 2023YFA1607800), the National Science Foundation of China (NSFC, Grant No. 11773063, 12233005, 12288102, 12173037, 12233008), the China Manned Space Program with grants nos. CMS-CSST-2025-A08,  CMS-CSST-2025-A20, CMS-CSST-2025-A04 and CMS-CSST-2025-A06, Office of Science and Technology, Shanghai Municipal Government (grant Nos. 24DX1400100, ZJ2023-ZD-001), National Astronomical Observatories Chinese Academy of Sciences No. E4TG2001, the International Centre of Supernovae (ICESUN), Yunnan Key Laboratory of Supernova Research (No. 202505AV340004).

YKH gratefully acknowledges the support of the Light of West China program of the CAS, the Yunnan Ten Thousand Talents Plan Young \& Elite Talents Project, the Natural Science Foundation of Yunnan Province (No. 202201BC070003), and the PHOENIX Supercomputing Platform jointly operated by the Binary Population Synthesis Group and the Stellar Astrophysics Group at Yunnan Observatories, Chinese Academy of Sciences (CAS).
XYK acknowledge the support of the Yunnan Revitalization Talent Support Program.

This work made use of the following software: Astropy \citep{Astropy-Collaboration2013a,Astropy-Collaboration2018m}, Matplotlib \citep{HunterJ2007x}, NumPy \citep{HarrisC2020a}, h5py \citep{ColletteA2013a}, BayeSED3 \citep[][available at \url{https://github.com/hanyk/BayeSED3}]{HanY2012a,HanY2014a,HanY2019a,HanY2023a} , and CESS \citep[][available at \url{https://github.com/RainW7/CESS}]{WenR2024a} .

The DESI Legacy Imaging Surveys consist of three individual and complementary projects: the Dark Energy Camera Legacy Survey (DECaLS), the Beijing-Arizona Sky Survey (BASS), and the Mayall z-band Legacy Survey (MzLS). DECaLS, BASS and MzLS together include data obtained, respectively, at the Blanco telescope, Cerro Tololo Inter-American Observatory, NSF's NOIRLab; the Bok telescope, Steward Observatory, University of Arizona; and the Mayall telescope, Kitt Peak National Observatory, NOIRLab. NOIRLab is operated by the Association of Universities for Research in Astronomy (AURA) under a cooperative agreement with the National Science Foundation. Pipeline processing and analyses of the data were supported by NOIRLab and the Lawrence Berkeley National Laboratory (LBNL). Legacy Surveys also uses data products from the Near-Earth Object Wide-field Infrared Survey Explorer (NEOWISE), a project of the Jet Propulsion Laboratory/California Institute of Technology, funded by the National Aeronautics and Space Administration. Legacy Surveys was supported by: the Director, Office of Science, Office of High Energy Physics of the U.S. Department of Energy; the National Energy Research Scientific Computing Center, a DOE Office of Science User Facility; the U.S. National Science Foundation, Division of Astronomical Sciences; the National Astronomical Observatories of China, the Chinese Academy of Sciences and the Chinese National Natural Science Foundation. LBNL is managed by the Regents of the University of California under contract to the U.S. Department of Energy. The complete acknowledgments can be found at \url{https://www.legacysurvey.org/acknowledgment/}.
\end{acknowledgments}

\bibliography{hanyk.bib}{}
\bibliographystyle{aasjournal}

\end{document}